\input epsf.tex
 \documentstyle[12pt]{article}
\newcommand{\bmat}{\left(\begin{array}}
\newcommand{\emat}{\end{array}\right)}
\def\NPB{Nucl. Phys. B}

\def\yzero{\smash{\hbox{$y\kern-4pt\raise1pt\hbox{${}^\circ$}$}}}

\def\a{\alpha}
\def\b{\beta}
\def\g{\gamma}
\def\d{\delta}
\def\beqa{\begin{eqnarray}}
\def\eeqa{\end{eqnarray}}

\def\om{\omega}

\def\vt{\vartheta}

\def\-{\hphantom{-}}
\def\ov{\overline}
\def\s2{\frac{1}{\sqrt2}}

\def\oh{\frac{1}{2}}
\def\beq{\begin{equation}}
\def\eeq{\end{equation}}
\def\beqa{\begin{eqnarray}}
\def\eeqa{\end{eqnarray}}

\def\IF{\relax{\rm I\kern-.18em F}}
\def\II{\relax{\rm I\kern-.18em I}}
\def\IP{\relax{\rm I\kern-.18em P}}
\def\IC{\relax\hbox{\kern.25em$\inbar\kern-.3em{\rm C}$}}
\def\IR{\relax{\rm I\kern-.18em R}}

\def\b{{\beta}}

\def\ent{{\bf Z}}
\def\C{{\bf C}}

\def\OR{\Omega {\cal R}}

\def\ent{{\bf Z}}

\def\OR{\Omega {\cal R}}

\def\ca{{\cal A}}
\def\lam{\lambda}

\def\ep{\epsilon}
\def\arr{\arrowvert}

\def\Dsl{\,\raise.15ex\hbox{/}\mkern-13.5mu D} %this one can be subscripted
\def\IZ{Z\kern-.4em  Z}

\catcode`\@=12
\begin{document}

%----------------------------------------------------------------------%
%  numbering equations with section number
%----------------------------------------------------------------------%
\makeatletter
\@addtoreset{equation}{section} \makeatother
\renewcommand{\theequation}{\thesection.\arabic{equation}}
%----------------------------------------------------------------------%
%  title page
%----------------------------------------------------------------------%
\pagestyle{empty}
%\vspace{1.0cm}
%\rightline{FTUAM-02-10; IFT-UAM-CSIC-02-09}
%\vspace{2.5cm}
%----------------------------------------------------------------------%
%  Resetting of counters
%----------------------------------------------------------------------%
%\setcounter{page}{1}
%\pagestyle{plain}
\pagestyle{empty}
%\vspace{1.0in}
\rightline{FTUAM-02-20}
\rightline{IFT-UAM/CSIC-02-17}
\rightline{\today}
%\vspace{2.5cm}
\vspace{0.5cm}
\setcounter{footnote}{0}

\begin{center}
{\LARGE{\bf Exact Standard model structures from Intersecting D5-Branes}}
\\[7mm]
%\medskip
{\Large{{  Christos ~Kokorelis} }
\\[2mm]}
\small{ Dep/to de F\'\i sica Te\'orica C-XI and 
Instituto de F\'\i sica 
Te\'orica C-XVI}
,\\[-0.3em]
{ Universidad Aut\'onoma de Madrid, Cantoblanco, 28049, Madrid, Spain}
\end{center}
\vspace{3mm}

%%%%%%%%%%%%%%%%%%%%%%%%%%%%%%%%%%%%%%%

\begin{center}
{\small \bf ABSTRACT}
\end{center}
\begin{center}
\begin{minipage}[h]{14.5cm}
We discuss the appearance of non-supersymmetric
compactifications with exactly the 
Standard Model (SM)
 at low energies, in the context of IIB orientifold
constructions with
 D5 branes intersecting at angles
on the $T^4$ tori, of the orientifold
of $T^4 \times (\C /Z_N)$.
 We discuss constructions where the Standard Model
embedding is considering within
four, five and six stacks of D5 branes.
The appearance of the three generation observable 
Standard Model at low energies is 
accompanied by a gauged baryon number, thus ensuring
automatic
proton stability.
Also, a compatibility with a low scale
of order TeV is
ensured by having a two dimensional space transverse to all
branes.
The present models complete the discussion of some recently
constructed
four stack models of D5 branes with the SM at low energy. 
By embedding the four, five and six stack Standard Model
configurations into quiver diagrams, deforming them
around the
QCD intersection numbers,
we find
a rich variety of vacua that may have exactly the
Standard Model at low energy. Also by using brane recombination on
the U(1)'s, we show that the five and six vacua flow into their
four stack counterparts. Thus 
string vacua with five and six stack deformations
are continuously connected to the four stack vacua.
 \end{minipage}                 
\end{center}

\newpage
%----------------------------------------------------------------------%
%  Resetting of counters
%----------------------------------------------------------------------%
\setcounter{page}{1}
\pagestyle{plain}
\renewcommand{\thefootnote}{\arabic{footnote}}
\setcounter{footnote}{0}
%----------------------------------------------------------------------%
%  Paper begins
%----------------------------------------------------------------------%

\section{Introduction}

One of the main objectives of current string theory research, in the absence
of a dynamical mechanism that can select a particular string vacuum, is 
the search for particular string vacua which can give us at low energies, the
observable chiral spectrum of the Standard Model (SM) and gauge interactions.
In this sense, semirealistic four dimensional (4D) models have been examined
both in $N=1$ heterotic compactifications and in orientifold
constructions \cite{see}.

In this work, we will examine 
standard model compactifications in the context
of some recent constructions \cite{ibanez1} 
which use
intersecting branes \cite{inter1, inter2} and give 
4D non-supersymmetric (non-SUSY) models. 
At present non-SUSY models are considered a
necessity in the search
of realistic 4D string models. This is to be
contrasted with past research, where the
majority of the relevant for
phenomenology models considered, were preserving 
${\cal N} = 1$ supersymmetry in four dimensions. 
Past model building studies in the context of $N=1$ weak coupling 
heterotic orbifold 
compactifications (HOC), gave rise to semirealistic
supersymmetric model generation that included at low energy
the
MSSM particle content, accompanied by a variety of exotic matter 
representations and gauge group factors. Additional problems included, 
among others, the fact that it was not possible 
to reconcile the observed discrepancy
between the unification of gauge couplings in the MSSM at
$10^{16}$ GeV \cite{see} and the string scale at 
HOC which is of order $10^{17}-10^{18}$ GeV, even though  the latter
discrepancy 
was attributed  
to the string one loop corrections of the $N=1$ 
gauge coupling constants \cite{kokos1}. Thus
the failure of obtaining realistic string compactifications   
in the context of heterotic strings directed research towards 
orientifold constructions and type
IIB constructions with
D-branes at singularities. See for
example \cite{ibar, cvet, bere, kep, allo5, allo5a, allol1,
allo6}.  In general these compactifications are accompanied by
extra matter.

However, recent results, that made use of the fact that 
in type I compactifications (IC) the string scale is a
free parameter, suggested 
that it is possible in IC to lower the string scale, thus
solving the
hierarchy problem, by having some
compact directions transverse to all stacks of branes, in the
TeV region
even without SUSY \cite{anto}. That alone changed the way
that we view
non-SUSY models, thus suggesting us that non-SUSY models with a
string scale in the TeV range is a strong alternative 
to SUSY models.  In this context, the only question remaining 
is to find realistic compactification examples, giving at
low energies
exactly the observable SM. As we will see in a moment,
the latter was shown
that is possible
in the context of intersecting branes.

In intersecting
brane models the fermions get
localized 
in the intersections between branes \cite{ber}.
In these constructions the
introduction of a quantized NS-NS B 
field \cite{flux} effectively produces
semirealistic models with 
three generations \cite{inter2}. These
backgrounds are T-dual to models with
magnetic deformations \cite{carlo}.
For additional non-SUSY constructions and recent developments 
in the context 
of intersecting branes,
see \cite{allo2, allo3, allo4, allo4a, allo5b, allo7, allo8,
allo7a,  
allo8a,  kane, nanop}.
For some new directions in the context of D6 intersecting branes in
backgrounds of
Calabi-Yau 3-folds see \cite{allo9}. For backgrounds with intersecting 
D6-branes on non-compact Calabi-Yau 3-folds see \cite{allo9a}.
For a $N=1$ SUSY construction in the context of intersecting branes
and its phenomenology see \cite{shiu}.

Quite recently, a class of models, with four stacks
of branes, was presented
that gives at low
energies exactly the SM content \cite{luis1}. 
These models were based on
D6-branes intersecting at angles on an orientifolded six-torus
compactification \cite{inter1, inter2}.  
These models have been generalized, to the same
D6-brane backgrounds, to classes of
models with
five- \cite{kokos2} and six-stacks of branes \cite{kokos3} at the 
string scale that give exactly the SM at low energies.  The models
of \cite{luis1, kokos2, kokos3} share some interesting common
features such as a gauged baryon, with a stable proton,
and
lepton number symmetries \footnote{The corresponding gauge bosons  
become massive through $B \wedge F $ couplings, arising on a generalized
Green-Schwarz mechanism \cite{allo3, luis1, iru, sa}. Thus the 
associated gauge symmetries
  survive as global
symmetries to low energies.}
 small neutrino masses and a remarkable
Higgs sector that,
that for some choices of brane angle parameters, is similar
of MSSM.
However, we should note that, while the non-SUSY four
stack model
has a variety of sectors where the non-SUSY chiral
fields of the SM get localized, models
that have five-\cite{kokos2}, six-stacks \cite{kokos3}
of D6 branes
have one additional unusual feature.
They have some sectors that preserve $N=1$ SUSY,
even though the model is
overall a non-SUSY one. The usefulness of $N=1$ SUSY
sectors stems from the
fact that their presence guarantees the presence of singlet scalars, 
superpartners of right handed neutrinos, $\nu_R$'s, necessary for the 
breaking of extra 
$U(1)$ symmetries. In this respect the five, six stack models,
uniquely
predict the existence of SUSY particles from non-SUSY SM's. 
 
Moreover, the models of \cite{luis1}, \cite{kokos2}, \cite{kokos3} 
have been extended
to describe the first examples of string GUT models
that give exactly the SM at low
energies \cite{kokos4}. The latter classes of models
are based on the
Pati-Salam \cite{pati} gauge group $SU(4)
\otimes SU(2)_L \otimes SU(2)_R$ and have
 a four stack structure at the
string scale. They
maintain
essential
features of \cite{luis1, kokos2, kokos3} and in particular the fact that
proton is
stable, as the baryon number is an unbroken gauged symmetry, and small
neutrino masses. Also the GUT four-stack classes of
models share the unusual features of the five-, six- stack
classes of
SM's of \cite{kokos2, kokos3} respectively, namely
they allow SUSY sectors to exist. These models provide us with 
a nice realization
of the see-saw mechanism.
It is quite interesting to note that,
even though it was generally believed that in D6 brane orientifolded six
torus models it was not possible to find an apparent explanation
for lowering the string scale in the TeV region, in the classes of
GUT models
of \cite{kokos4} this issue was solved differently. In particular, the models
predict the existence of light weak doublets with mass of order
$\upsilon^2/M_s$, that necessarily needs the string scale to be
less than 650 GeV. 
The latter results are particularly encouraging as
they represent strong predictions for D-brane scenarios and are directly
testable at present or future accelerators.

In this work, we will discuss compactifications of
intersecting
D5 branes, that use four-, five- and six-stacks of
D5-branes on an orientifolded $T^4 \otimes {C/Z_N}$
\cite{ibanez1}. These classes of models may
have only the SM at low energy.

First we discuss again the appearance of exactly
the SM at low energy from four stacks
of branes.
For each of the, four stack, 
quivers of \cite{ibanez1},
that give rise to a
different class of exactly the SM at low
energy \footnote{denoted as ai-type, $(i=1,..,4)$, and seen
in figure (2)},
we present `alternative'
parametric solutions to the twisted RR tadpole cancellation conditions. 
Moreover, we discuss four new `reflected'
quivers
\footnote{, denoted as of Qi-type, $(i=1,..,4)$, and
 seen in figure (1),} that give rise to the
SM at low energy. These `reflected' quivers are obtained from
the ai-type quivers by keeping the $Z_N$ quiver 
transformation properties of the nodes fixed, while simultaneously
 interchanging  
the brane/orientifold content of the nodes with $Z_N$
transformation properties different from unity.
Moreover, we show that `reflected' quivers give SM vacua that are equivalent
and in a one to one correspondence to the quivers of \cite{ibanez1}. 
The comparison of the different SM vacua is most easily seen, 
between the quiver vacua $ai$ coming from the alternative
RR tadpole solutions
and the ones associated with the `reflected'
quivers $Qi$.
We also discuss the
embedding of the SM into five and six stacks of D5 $Z_3$ quivers.
These classes of models may aslo have the SM at low energy and thus may be
of phenomenological interest.

The new classes of SM's exhibit some general features:

\begin{itemize}

\item The models are non-SUSY and in the low energies we have apart
from the SM spectrum, and gauge interactions, the natural 
appearance of some scalars fields in various colour, doublet, 
and singlet representations. All scalars receive one loop corrections.
 Also present are some singlet scalars 
that receive a vev and thus break the extra anomaly 
free $U(1)$, beyond hypercharge.

\item Baryon number (B) is a gauged $U(1)$ symmetry, thus proton
is stable. The corresponding gauge bosons become massive from a dimensional 
reduction mechanism \cite{ibanez1} resembling 
previously discussed constructions \cite{allo3}.
 
\item The placement of SM chiral fermions to the different 
intersections inside the four-, five- and six stack D5-brane
structure, is exactly the one used in the four-, five and six stack
counterpart D6-models of \cite{luis1, kokos2, kokos3} respectively.

\item The models may have a low string scale $M_s \sim 1$ TeV,
in the way suggested in \cite{anto}, as there is a transverse compact 
two dimensional manifold transverse to all D5 branes, whose size 
may vary, providing
for the hierarchy difference between $M_{Planck}$ and $M_s$.  

\item The four, five, six stack quivers have exactly the observable 
SM at low energies.
They are classified according to their brane
transformation properties under the $Z_N$ orbifold action, schematically
seen, by placing the branes and their orientifolded
images in various
quiver diagrams in the spirit of \cite{quiv1, quiv2, quiv3}.    
As a particular example, in this work we examine in
detail
the SM embedding in $Z_3$ quiver diagrams.
The five stack and six stack $Z_3$ quivers provide us
with more SM-like examples at low energies. In the latter
constructions e.g. A1, C1, C2 qivers additional anomaly free $U(1)$'s
orthogonal
to hypercharge become massive through the presence
of singlet scalars
getting a vev.

\item Using brane recombination we are able to show
that five and six stack string vacua with intersecting
D5-branes `flow' to their associated four stack
quivers (see sect. 7).

\end{itemize}

The paper is organized as follows.
In the next section we describe the rules, as well the constraints coming
from RR tadpole cancellation conditions,
for constructing the intersecting at angles D5-brane models in an 
orientifolded 
$T^4 \times C/Z_N$ IIB background. 
  In section 3 we describe
the exact form of the chiral SM fermionic structure 
configurations that all the SM's 
reproduce at low energies. 
We describe the SM embedding in a four-, five- and six stack D5 structure,
which follow from past discussions in 
\cite{luis1}, \cite{kokos2}, \cite{kokos3}, respectively.
In section 4, we discuss the four stack $Z_3$ `reflected'
quiver embedding 
of the three generation (3G) SM. 
We start by discussing the Q1-quiver, presenting in detail
the cancellation of the mixed $U(1)$ gauge anomalies by a
dimensional reduction scheme which is equivalent to the cancellation
of the field theory anomaly by its Green-Schwarz amplitude.
This mechanism \cite{ibanez1} is an extension of a similar mechanism
used in the context of toroidal models with branes at angles
in \cite{allo3}. 
Also, we discuss the rest of quivers, of Q2, Q3, Q4 type, giving 
only the SM at low energies.  
Next we discuss the alternative RR twisted tadpole cancellation
conditions to the a-type quivers of \cite{ibanez1}.
We show that the reflected quivers $Qi$ give equivalent
vacua at low energy to their `images' $ai$-type quivers.                                 
In section 5, we discuss the five stack SM embedding
in $Z_3$ quivers. We discuss a number of quivers characterized as 
belonging to the A1, $\overline{A1}$, A2, $\overline{A2}$-quivers. 
For the case of A1-quiver we examine in detail 
the Higgs sector
giving our emphasis on the definition of the geometrical quantities
that characterize the geometry of the Higgs sector of the model.
Also we discuss the scalar sector of the A1-quiver class of
 SM's.
In section 6, we discuss two examples of six stack
SM-like quivers providing
more examples with the SM 
at low energy.
In section 7, we make use of brane recombination
to show the equivalence of six, five and four stack vacua.
Our conclusions together with some comments are
presented in section 8.
In appendix I, we list the
RR tadpole solutions of
all the SM embeddings in five stack quivers not examined
explicitly in the main body of the paper.
In appendix II we list the 
RR tadpole solutions of
all the SM embeddings in six stack quivers.

\section{Exact Standard model structures from Intersecting branes}

In this work, we are going to describe new type IIB compactification vacua
that have at low energy just the observable Standard Model gauge group and 
chiral content. The proposed three generation SM's
make use of four, five and six-stacks
of D5 branes, intersecting at angles, at the string scale. 
They are based on the following compactification \cite{ibanez1}
scheme

\beq
IIB / \left(\frac{T^4 \times C/Z_N}{\{1 + {\Omega}{\cal R} \}}  \right) \ ,
\label{con1}
\eeq 
where $\Omega$ is the worldvolume 
parity and the presence of the parity $\cal R$ involution is 
associated to the reflections ${\cal R}_i$ of the i-th
coordinate, namely ${\cal R} = {\cal R}_{(5)}{\cal R}_{(7)} 
{\cal R}_{(8)}{\cal R}_{(9)}$. In a compact form the action of $\cal R$ 
on the coordinates could be expressed alternatively by using complex 
coordinates $Z_i = X_{2 i + 2} + X_{2 i + 3}$, $i = 1, 2, 3$. In this case,
the presence of $\cal R$ is more elegantly expressed as
\beqa
{\cal R} :&Z_i \rightarrow {\bar Z}_i, \ i= 1, 2;\ \
           Z_3 \rightarrow {\bar Z}_3 .
\label{invol1}
\eeqa

The presence of tadpoles induced by the presence
of the orientifold planes is cancelled by the introduction of 
N-stacks of D5 branes \footnote{Effectively, we will use 
later $N = 3$.}. The latter
wrap 2-cycles across the 
four dimensional 
tori, while they are located at the tip of the
orientifold singularity.
We note that under T-duality these constructions are dual to D7-branes
with fluxes across the compact $T^4$ dimensions.

The general picture involves D$5_a$-branes wrapping 1-cycles 
$(n^i_a, m^i_a)$, $i=1,2$ along each of the ith-$T^2$
torus of the factorized $T^4$ torus, namely 
$T^4 = T^2 \times T^2$.
Thus we allow the four-torus to wrap factorized 2-cycles, so we can 
unwrap the 2-cycle into products of two 1-cycles, one for each $T^2$.
The definition of the homology of the 2-cycles as
\beq
[\Pi_a] =\ \prod_{i=1}^2(n^i_a [a_i] + m^i_a[b_i])
\label{homo1}
\eeq
defines consequently the 2-cycle of the orientifold images as
\beq
[\Pi_{a^{\star}}] =\ \prod_{i=1}^2(n^i_a [a_i] - m^i_a[b_i]).
\label{homo2}
\eeq
We note that because of the $\Omega {\cal R}$ symmetry 
 each D$5_a$-brane
1-cycle, must be accompanied by its $\Omega {\cal R}$
orientifold image
partner $(n^i_a, -m^i_a)$; $n, m \in Z$.
In addition, because of   
the presence of discrete NS B-flux \cite{flux}, 
the tori involved are not
orthogonal but tilted. In this way 
the wrapping numbers become the effective tilted wrapping numbers, 
\beq
(n^i, m ={\tilde m}^i + b^i \cdot {n^i}/2);\; 
n,\;{\tilde m}\;\in\;Z, \  b^i = 0, 1/2
\label{na2}
\eeq
Thus semi-integer values are allowed for the m-wrapping numbers.

Let us now discuss the effect
of the orbifold action on the open string sectors.
The $\ent_N$ orbifold 
twist in the third complex dimension is
generated by the twist vector
$v = \frac{ 1}{N} (0,0,-2,0)$, which is fixed
by the requirements of modular invariance and for
the variety to be spin. The $\ent_N$ action is 
embedded in the $U(N_a)$ degrees of freedom emanating from the
$a^{th}$ stack of D5-branes, through the unitary matrix in the 
form
\beq
\g_{\om,a} = {\it diag} \left( {\bf 1}_{N_a^0},\ \alpha {\bf 1}_{N_a^1}, 
\ldots,\ \alpha^{N-1} {\bf 1}_{N_a^{N-1}} \right),
\label{Chan}
\eeq
with $\sum_{i=0}^{N-1} N_a^i = N_a$ and 
$\a \equiv {\rm exp} (2\pi i/N)$.

As we have already said, in the presence of
$\OR$ orientifold action, we need to include 
sectors where 
its brane is accompanied by is orientifold image.
Lets us denote the $\OR$ image of the brane D5$_a$-brane
by ${\OR}$D5$_a$ or alternatively as D5$_{a^{\star}}$. 

Thus for example if the D5$_a$-brane data are given by 
\beqa
& (n_a^1,m_a^1) \otimes (n_a^2,m_a^2) \nonumber \\
& \g_{\om,a} = {\it diag} \left( {\bf 1}_{N_a^0},\
\a {\bf 1}_{N_a^1},
\ldots,\ \a^{N-1} {\bf 1}_{N_a^{N-1}} \right),
\label{brana_a}
\eeqa
then the D5$_{a^*}$ data are given by
\beqa
& (n_a^1,-m_a^1) \otimes (n_a^2,-m_a^2) \nonumber \\
& \g_{\om,a^*} = {\it diag} \left( {\bf 1}_{N_a^0},\ \a^{N-1} 
{\bf 1}_{N_a^{1}}, \ldots,\ \a {\bf 1}_{N_a^{N-1}} \right),
\label{brana_a*}
\eeqa

The closed string sector,
can be computed using orbifold techniques. However, as its
spectrum will be non-chiral and non-supersymmetric, it will 
give rise to 
four dimensional gravitation and thus will be of no interest to us. 
As far as the twisted closed string sector is concerned it will
rise to some tachyons from the NSNS sectors.
Thus from the point of view of the low energy spectrum of the
theory the closed string sector is not interesting.
We thus focus our attention to the open string sector.

The open string spectrum is computed by
taking into account the combined geometric plus the Chan-Paton $Z_n$
action \cite{ibanez1}.
There are a number of different sectors that should be taken into account
when computing the chiral spectrum of the models.
They include the sectors $D5_a - D5_b$,
$D5_a - D5_{b^*}$ and $D5_a - D5_{a^*}$.
E.g. because the $D5_a-D5_b$ sector is not
constrained by the
$\OR$ projection, its spectrum is computed
as in orbifold compactifications.
Thus we introduce
the twist vector $v_\vt = (\vt_{ab}^1, \vt_{ab}^2, 0, 0)$,
where $\pi \vt_{ab}^i$ is the angle between the branes 
on the $i^{th}$ torus.
Also the states localized in the $ab$ intersection are
characterized by the four-dimensional vectors 
$r + v_\vt$ which enter the mass formula
\beq
\a' M_{ab}^2 = {Y^2 \over 4\pi\a^\prime} + N_{bos}(\vt) 
+ {(r + v_\vt)^2 \over 2} -\oh + E_{ab}.
\label{mass2}
\eeq
In (\ref{mass2}),  $Y$ is the transverse distance between the
branes $a$, $b$;
$N_{bos}(\vt)$ is the bosonic oscillator contribution and $E_{ab}$ is
the vacuum energy
\beq
E_{ab} = \sum_{i=1}^3 \oh |\vt^i| (1 - |\vt^i|)\ .
\label{vacio}
\eeq

The complete massless and tachyonic states reads:
{\small\beq
\begin{array}{cccc}
{\rm\bf Sector} & {\rm\bf State} & {\rm\bf \ent_N \ phase} & {\rm\bf \a' Mass^2}
\vspace{2mm}\\
{\rm NS} & (-1+\vartheta^1,\vt^2,0,0) & 1 & -\oh(\vt^1 - \vt^2) \nonumber \\
         & (\vt^1,-1+\vartheta^2,0,0) & 1 & \oh(\vt^1 - \vt^2)\nonumber 
\vspace{2mm}\\
{\rm R}  & (-\oh+\vartheta^1,-\oh+\vartheta^2,-\oh,+\oh) &
e^{2\pi i\frac {1}{N}} & 0\\
         &  (-\oh+\vartheta^1,-\oh+\vartheta^2,+\oh,-\oh) &
e^{-2\pi i\frac{1}{N}} & 0
\end{array}
\label{sector5ab}
\eeq}
where $\vt^i \equiv \vt^i_{ab}$ and we have set 
 $0 < \vt^i < 1$, $i = 1,2$. 
As is it clear from above, one of the NS states will
be necessarily tachyonic while the other will 
have a positive mass. Also, when $|\vt^1| = |\vt^2|$ the associated state
becomes massless. 
The final spectrum will be found again by projecting
to the invariant subspace. We note that the number of chiral fermions localized
in the intersection between the $a$-, $b$-branes, is given by 
the intersection number  
\beq
I_{ab} \equiv [\Pi_a]\cdot[\Pi_b]  
= (n_a^1 m_b^1- m_a^1\ n_b^2)
(n_a^2 m_b^2- m_a^2 n_b^2),
\label{interfive}
\eeq
where its sign denotes the chirality of the corresponding fermion.
The full spectrum appearing in the $ab$-sector is given by
{\small\beq
\begin{array}{rl}
{\rm\bf Tachyons} & \quad \sum_{a<b} \sum_{i=1}^N \; I_{ab}\times
(N_a^i,{\ov N}_b^i) \nonumber \\
{\rm\bf Left\; Fermions} & \quad \sum_{a<b} \sum_{i=1}^N \;
I_{ab}\times(N_a^i,{\ov  N}_b^{i+1}) \nonumber \\
{\rm\bf Right\; Fermions} & \quad \sum_{a<b} \sum_{i=1}^N \;  I_{ab}\times
(N_a^i,{\ov  N}_b^{i-1})
\end{array}
\label{espectro5ab}
\eeq}

The complete spectrum for the D5 branes intersecting at angles
in the backgrounds (\ref{con1}) reads:
{\small\beq
\begin{array}{l}
{\rm\bf Complex\; Scalars} \\ \sum_{a<b} \sum_{i=1}^N \; 
[\; \arr I_{ab}\arr (N_a^i,{\ov N}_b^i) + 
\arr I_{ab^*}\arr (N_a^i,N_b^{-i})\;]\\
\sum_a [\;2 \arr m_a^1 m_a^2\arr (\arr n_a^1 n_a^2\arr + 1) ({\bf A}_a^0)
+ 2 \arr m_a^1 m_a^2\arr (\arr n_a^1 n_a^2\arr - 1) ({\bf S}_a^0)\;] 
\vspace{3mm}\\
{\rm\bf Left\; Fermions} \\ \sum_{a<b} \sum_{i=1}^N \;
[\;I_{ab}(N_a^i,{\ov  N}_b^{i+1}) + I_{ab*}(N_a^i,N_b^{-i-1})\;]  \\
\sum_a \sum_{j,i=1}^N \; \d_{j,-i-1}
[\;2 m_a^1 m_a^2 (n_a^1 n_a^2 + 1) ({\bf A}_a^j)
+ 2 m_a^1 m_a^2 (n_a^1 n_a^2 - 1) ({\bf S}_a^j)\;] 
\vspace{3mm}\\
{\rm\bf Right\; Fermions} \\ \sum_{a<b} \sum_{i=1}^N \;
[\;I_{ab}(N_a^i,{\ov  N}_b^{i-1}) + I_{ab*}(N_a^i,N_b^{-i+1})\;]  \\
\sum_a \sum_{j,i=1}^N \; \d_{j,-i+1}
[\;2 m_a^1 m_a^2 (n_a^1 n_a^2 + 1) ({\bf A}_a^j)
+ 2 m_a^1 m_a^2 (n_a^1 n_a^2 - 1) ({\bf S}_a^j)\;] 
\end{array}
\label{espectro5ab*}
\eeq}

Any vacuum derived from the previous intersection constraints  
is subject additionally to constraints coming from RR tadpole
cancellation conditions. The latter are equivalent to the
Gauss law and imply the vanishing of the total
RR charge in the compact $T^4$ space.
That demands cancellation of
D5-branes charges \footnote{Taken together with their
orientifold images $(n_a^i, - m_a^i)$  wrapping
on two cycles of homology
class $[\Pi_{\alpha^{\star}}]$.}, wrapping on two cycles with
homology $[\Pi_a]$ and O5-plane 6-form
charges wrapping on 2-cycles with homology $[\Pi_{O_5}]$. 
Note that the RR tadpole cancellation conditions
in terms of cancellations of RR charges in homology are expressed as 
\beq
c_k^2 \ \sum_a \left([\Pi_a] \ {\rm Tr} \gamma_{k,a} 
+ [\Pi_{a^*}] \ {\rm Tr} \gamma_{k,a^*} \right)
= [\Pi_{O5}] \ 16 \b^1\b^2 \
{\rm sin} \left(\frac{\pi k}{N} \right),
\label{tadpoleO5}
\eeq
where $[\Pi_{O5}]$ describes the 2-cycle of ${\bf T^4}$ the O5-plane wraps,
$\b^i = 1 - b^{(i)}$ parametrizes the NS B-field background and $c_k$
where $c_k^2 = {\rm sin \ } \frac{2\pi k}{N}$ is a weight for each
$k^{th}$ twisted sector usually arising in $\ent_N$ orientifold 
compactifications \cite{GJ}. Also,
$16 \b^1\b^2$ can be seen as the number of O$5$-planes,
e.g. $4 \b^1\b^2$, times their relative charge to the 
D$5$-brane, e.g. $-4$.

In explicit form the tadpoles are given by \cite{ibanez1}
{\beq
\begin{array}{l}
c_k^2 \ \sum_a n_a^1 n_a^2 \ 
\left({\rm Tr} \gamma_{k,a} + {\rm Tr} \gamma_{k,a^*} \right)
= 16 \ {\rm sin} \left(\frac{\pi k}{N} \right) \\
c_k^2 \ \sum_a m_a^1 m_a^2 \ 
\left({\rm Tr} \gamma_{k,a} + {\rm Tr} \gamma_{k,a^*} \right) = 0\\
c_k^2 \ \sum_a n_a^1 m_a^2 \ 
\left({\rm Tr} \gamma_{k,a} - {\rm Tr} \gamma_{k,a^*} \right) = 0\\
c_k^2 \ \sum_a m_a^1 n_a^2 \ 
\left({\rm Tr} \gamma_{k,a} - {\rm Tr} \gamma_{k,a^*} \right) = 0
\label{tadpoleO5b}
\end{array}
\eeq}
The presence of a non-zero term in the first tadpole condition should be 
interpreted as a negative RR charge induced by the presence of an O5-plane. 
We should note that, alternatively, the first of 
twisted tadpole conditions can be
rewritten as
\beq
\sum_a{n_a^1 n_a^2 \left({\rm Tr} \gamma_{2k,a} 
+ {\rm Tr} \gamma_{2k,a^*}\right)} =
{16 \over {\alpha^k + \alpha^{-k}}}.
\label{decomp}
\eeq 

%
%
%
%
%
%
%
%

%%%%%%%%%%%%%%%%%%%%%%%%%%%%%%%%%%%%%%%%%%%%%%%%%%%%%
%%%%%%%%%%%%%%%%%%%%%%%%%%%%%%%%%%%%%%
%%%%%%%%%%%%%%%%%%%%%%%%%%

\section{The D5 Standard Model Configuration's}

In this section we will describe the
embedding of the SM chiral spectrum
into 
configurations of four-, five, and six-
D5-brane stacks.
Our search is facilitated 
using the four, five of six stack embedding of the SM chiral fermions
that have appeared before in the context of orientifolded intersecting 
D6 branes \cite{luis1, kokos2, kokos3} respectively into
$Z_3$, quiver diagrams.
We describe the basic characteristics of the classes of SM 
vacua produced,following their derivation from the various quivers.

\subsection{The Standard Model configurations}

In this section, we describe the
general characteristics of the different SM
configurations that
we will make use in this work.
We will describe models based on a
four stack 
\beq
U(3) \otimes U(2) \otimes U(1)_c \otimes U(1)_d \ ,
\label{four}
\eeq
a five stack 
\beq
U(3) \otimes U(2) \otimes U(1)_c
\otimes U(1)_d \otimes U(1)_e \ ,
\label{five}
\eeq
and a six-stack structure
\beq
U(3) \otimes U(2) \otimes U(1)_c \otimes U(1)_d \otimes U(1)_e \otimes U(1)_f
\label{six}
\eeq
at the string scale.
The localization of fermions into particular intersections,
as seen in table (1), have been considered before
in \cite{ibanez1} for the four stack D5-models.
The
complete accommodation
of the fermion structure of the chiral SM fermion content 
in the
five- and six-stack SM models in the different open string sectors can be seen
in tables (2) and (3). 
\newline
Several comments are in order:

\begin{table}[htb] \footnotesize
\renewcommand{\arraystretch}{1}
\begin{center}
\begin{tabular}{|c|c|c|c|c|c|c|c|c|}
\hline
Matter Fields & & Intersection & $Q_a$ & $Q_b$ & $Q_c$ & $Q_d$ & Y
\\\hline
 $Q_L$ &  $(3, 2)$ & $I_{ab}$ & $1$ & $-1$ & $0$ & $0$ &  $1/6$ \\\hline
 $q_L$  &  $2(3, 2)$ & $I_{a b^{\ast}}$ &  
$1$ & $1$ & $0$ & $0$  & $1/6$  \\\hline
 $U_R$ & $3({\bar 3}, 1)$ & $I_{ac}$ & 
$-1$ & $0$ & $1$ & $0$ & $-2/3$ \\\hline    
 $D_R$ &   $3({\bar 3}, 1)$  &  $I_{a c^{\ast}} $ &  
$-1$ & $0$ & $-1$ & $0$ & $1/3$ \\\hline    
$L$ &   $3(1, 2)$  &  $I_{bd} $ &  
$0$ & $-1$ & $0$ & $1$ & $-1/2$  \\\hline    
$N_R$ &   $3(1, 1)$  &  $I_{cd} $ &  
$0$ & $0$ & $1$ & $-1$ & $-1/2$  \\\hline    
$E_R$ &   $3(1, 1)$  &  $I_{cd^{\star}}$ &  
$0$ & $0$ & $-1$ & $-1$ & $0$  \\\hline    
\end{tabular}
\end{center}
\caption{\small Low energy fermionic spectrum of the
four stack
string scale 
$SU(3)_C \otimes
SU(2)_L \otimes U(1)_a \otimes U(1)_b \otimes U(1)_c 
\otimes U(1)_d $, D5-brane model together with its
$U(1)$ charges.
\label{spectrum7}}
\end{table}

\begin{table}[htb] \footnotesize
\renewcommand{\arraystretch}{0.8}
\begin{center}
\begin{tabular}{|c|c|c|c|c|c|c|c|c|}
\hline
Matter Fields & & Intersection & $Q_a$ & $Q_b$ & $Q_c$ & $Q_d$ & $Q_e$& Y
\\\hline
 $Q_L$ &  $(3, 2)$ & $I_{ab}$ & $1$ & $-1$ & $0$ & $0$ & $0$& $1/6$ \\\hline
 $q_L$  &  $2(3, 2)$ & $I_{a b^{\ast}}$ &  
$1$ & $1$ & $0$ & $0$  & $0$ & $1/6$  \\\hline
 $U_R$ & $3({\bar 3}, 1)$ & $I_{ac}$ & 
$-1$ & $0$ & $1$ & $0$ & $0$ & $-2/3$ \\\hline    
 $D_R$ &   $3({\bar 3}, 1)$  &  $I_{a c^{\ast}} $ &  
$-1$ & $0$ & $-1$ & $0$ & $0$ & $1/3$ \\\hline    
$L$ &   $2(1, 2)$  &  $I_{bd} $ &  
$0$ & $-1$ & $0$ & $1$ & $0$ & $-1/2$  \\\hline    
$l_L$ &   $(1, 2)$  &  $I_{b e} $ &  
$0$ & $-1$ & $0$ & $0$ & $1$ & $-1/2$  \\\hline    
$N_R$ &   $2(1, 1)$  &  $I_{cd}$ &  
$0$ & $0$ & $1$ & $-1$ & $0$ & $0$  \\\hline    
$E_R$ &   $2(1, 1)$  &  $I_{c d^{\ast}} $ &  
$0$ & $0$ & $-1$ & $-1$ & $0$ & $1$   \\\hline
  $\nu_R$ &   $(1, 1)$  &  $I_{c e} $ &  
$0$ & $0$ & $1$ & $0$ & $-1$ & $0$ \\\hline
$e_R$ &   $(1, 1)$  &  $I_{c e^{\ast}} $ &  
$0$ & $0$ & $-1$ & $0$ & $-1$  & $1$ \\\hline    
\hline
\end{tabular}
\end{center}
\caption{\small Low energy fermionic spectrum of the five stack 
string scale 
$SU(3)_C \otimes
SU(2)_L \otimes U(1)_a \otimes U(1)_b \otimes U(1)_c 
\otimes U(1)_d \otimes U(1)_e $, D5-brane model together with its
$U(1)$ charges.
\label{spectrum8}}
\end{table}

\begin{table}[htb] \footnotesize
\renewcommand{\arraystretch}{0.8}
\begin{center}
\begin{tabular}{|c|c|c|c|c|c|c|c|c|c|}
\hline
Matter Fields & & Intersection & $Q_a$ & $Q_b$ & $Q_c$
& $Q_d$ & $Q_e$ & $Q_f$ & Y \\\hline
 $Q_L$ &  $(3, 2)$ & $I_{ab}$ &
 $1$ & $-1$ & $0$ & $0$ & $0$ & $0$ & $1/6$ \\\hline
 $q_L$  & $2(3, 2)$ & $I_{a b^{\ast}}$ &  
$1$ & $1$ & $0$ & $0$  & $0$ & $0$ & $1/6$  \\\hline
 $U_R$ & $3({\bar 3}, 1)$ & $I_{ac}$ & 
$-1$ & $0$ & $1$ & $0$ & $0$ & $0$ & $-2/3$ \\\hline    
 $D_R$ &   $3({\bar 3}, 1)$  &  $I_{a c^{\ast}} $ &  
$-1$ & $0$ & $-1$ & $0$ & $0$ & $0$ & $1/3$ \\\hline    
$L^1$ &   $(1, 2)$  &  $I_{bd} $ &  
$0$ & $-1$ & $0$ & $1$ & $0$ & $0$ & $-1/2$  \\\hline
$L^2$ &   $(1, 2)$  &  $I_{be} $ &  
$0$ & $-1$ & $0$ & $0$ & $1$ & $0$ & $-1/2$  \\\hline
$L^3$ &   $(1, 2)$  &  $I_{bf} $ &  
$0$ & $-1$ & $0$ & $0$ & $0$ & $1$ &$-1/2$  \\\hline
$N_R^1$ &   $(1, 1)$  &  $I_{cd}$ &  
$0$ & $0$ & $1$ & $-1$ & $0$ & $0$ & $0$ \\\hline    
$E_R^1$ &   $(1, 1)$  &  $I_{c d^{\ast}} $ &
$0$ & $0$ & $-1$ & $-1$ & $0$ & $0$ & $1$  \\\hline
$N_R^2$ &   $(1, 1)$  &  $I_{ce}$ &  
$0$ & $0$ & $1$ & $0$ & $-1$ & $0$ & $0$ \\\hline    
$E_R^2$ &   $(1, 1)$  &  $I_{c e^{\ast}} $ &
$0$ & $0$ & $-1$ & $0$ & $-1$ & $0$ & $1$ \\\hline
$N_R^3$ &   $(1, 1)$  &  $I_{cf}$ &  
$0$ & $0$ & $1$ & $0$ & $0$ & $-1$ & $0$ \\\hline    
$E_R^3$ &   $(1, 1)$  &  $I_{c f^{\ast}} $ &
$0$ & $0$ & $-1$ & $0$ & $0$ & $-1$ & $1$ \\\hline
\end{tabular}
\end{center}
\caption{
\small Low energy fermionic spectrum of the six stack
string scale 
$SU(3)_C \otimes
SU(2)_L \otimes U(1)_a \otimes U(1)_b \otimes U(1)_c 
\otimes U(1)_d \otimes U(1)_e \otimes U(1)_f$, D5-brane model together with its
$U(1)$ charges.
\label{spectrum81}}
\end{table}

a) We note that there are a number of constraints that the chiral
spectra take into account.
First of all, as a result
of tadpole cancellation conditions, there is an equal number
of fundamental and
anti-fundamental representations. The models necessarily
include right handed neutrinos $\nu_R$ in the SM spectrum.
Thus from now on,
when we speak about obtaining the SM at low energies,
we will mean the SM with three generations of $\nu_R$'s.
\newline
b) The models accommodate various known low energy
gauged symmetries.
The latter can be expressed 
in terms of the $U(1)$ symmetries of the models. For example the
baryon number, B, is expressed as $B = 3 Q_a$ in all
models.
The study of 
Green-Schwarz mechanism will show us   
that baryon number is an unbroken gauged   
symmetry and as a result the corresponding gauge boson gets massive leaving
at low energies the baryon number as a global symmetry.
Thus proton is stable. 
\newline  
b) The mixed anomalies $A_{ij}$ of the $U(1)$'s 
with the non-abelian gauge groups $SU(N_a)$ of the theory
cancel through a generalized GS mechanism \cite{allo3, iru, sa},
involving
 close string modes couplings to worldsheet
 gauge fields \cite{ibanez1}.
 Two combinations of the $U(1)$'s are anomalous-
 model independent and become massive, their
 orthogonal 
 non-anomalous combination survives, combining to a
 single $U(1)$
 that remains massless, the latter to be identified with the
 hypercharge
generator.
\newline
c) The hypercharge operator in the model is defined as a linear combination
of the $U(1)_a$, $U(1)_c$, $U(1)_d$, $U(1)_e$, $U(1)_f$
gauge groups for the four-, five-, six-stack models respectively:
\beq
Y = \frac{1}{6}U(1)_{a}- \frac{1}{2} U(1)_c -
\frac{1}{2} U(1)_d\;.
\label{hyper1230}
\eeq 
\beq
Y = \frac{1}{6}U(1)_{a}- \frac{1}{2} U(1)_c -
\frac{1}{2} U(1)_d -
 \frac{1}{2} U(1)_e \;.
\label{hyper1231}
\eeq 
\beq
Y = \frac{1}{6}U(1)_{a}- \frac{1}{2} U(1)_c -
\frac{1}{2} U(1)_d -
 \frac{1}{2} U(1)_e - \frac{1}{2} U(1)_f \;.
\label{hyper1232}
\eeq 
\newline

d) Scalars, massless or tachyonic, appear naturally in the present
 classes of models. This is to 
be contrasted
with the D6 branes wrapping on $T^6$ orientifolded 
constructions \cite{inter1, inter2}, where one has to 
make some open 
string sectors supersymmetric in order to guarantee the presence of 
massless scalars in the models. The latter case was most obvious 
in the construction
of $N=0$ models, with just the SM at low energy from 
 five -\cite{kokos2}, six stacks \cite{kokos3} of
 D6-branes, as well
from constructions with a 
Pati-Salam $G_{422}$ D6 four stack GUT, on the
orientifolded $T^6$
backgrounds \cite{kokos4} at
the string scale.
The presence of scalars is necessary in order to break
the additional massless, beyond the hypercharge, 
anomaly free $U(1)$'s that survive
 the
Green-Schwarz mechanism.

e) A variety of SM solutions will be presented that are
directly related to the embedding of the SM
configurations in tables (1), (2), (3) in quiver
diagrams.
In general, when D-branes are localized on an
orbifold singularity ${\bf C}^n$,
the description of the local physics at the singularity
is obtained by keeping states that are invariant under the
 the combined geometric
and gauge action on the Chan-Paton index by
the action of the discrete group $\Gamma$ acting
on the singularity \cite{quiv1, quiv2, quiv3}. \newline
Each quiver records the action of the orbifold group
on the branes and their orientifold images. We chosen for
simplicity the orbifold group to be abelian, e.g. a $Z_3$,
however more general $Z_N$ choices 
are possible. \newline 
We should note that what is important for having the SM at low energy 
constructions is not the particular $Z_N$ quiver we examine, but rather
the four, five, six- stack configuration that 
localizes
the chiral three generation SM 
fermion content, as seen in tables (1), (2), (3). 
The structures of tables (1), (2), (3) have been used before in the D6
orientifolded models of \cite{luis1, kokos2, kokos3} respectively.

In a quiver,
each arrow between two nodes represents
a chiral fermion transforming
in the bifundamental representation of the two linked nodes.
Also the direction of the arrows denotes
the chirality of the representation and we choose left handed
fermions to correspond to arrows directed clockwise.
\newline 
At a four stack level, we found four different $Z_3$ quiver 
diagrams, that
give rise
to exactly the SM at low energies. They are
denoted as belonging to
the Q1-, Q2- Q3-, Q4-type and they are examined in
the next section. Also we provide alternative RR tadpole
cancellation conditions, to the ones appearing
in \cite{ibanez1}, for the four stack quivers of
a1-, a2-, a3-, a4-type.
Also, a number of five stack quivers are being
examined that are
characterized as belonging to the A1, A2, $\overline{A1}$, $\overline{A2}$. 
Moreover the six-stack quivers 
that are examined are characterized
as belonging to the C1, C2 type.
Care should be taken when treating the intersection numbers
of a SM embedding.
The sign of the intersection number that denotes the
chirality of
the relevant fermion, depends on the direction of the
arrows between the
nodes. For example, 
the following
intersection numbers hold in the four stack $a4$-quiver,
the five stack $A1$-quiver or the six stack C2-quiver
respectively :
\beqa
I_{ab} =\ -1, \ I_{ab^{\star}} = +2, \ I_{ac} = -3,
I_{cd^{\star}} = +3\nonumber\\
I_{bd} =\ -3, \ I_{cd^{\star}} = +3, \ I_{cd} = -3.
\label{interq1}
\eeqa
\beqa
I_{ab} =\ +1, \ I_{ab^{\star}} = -2,
\ I_{ac} = -3, \nonumber\\
 I_{ac^{\star}} = +3,
I_{bd} = +2, I_{cd}= -2,  I_{cd^{\star}} = +2,\nonumber\\
 I_{be}= +1, \ 
I_{ce}= -1, \ I_{ce^{\star}}= +1
\label{dour}
\eeqa
 \beqa
I_{ab} =\ +1,\ I_{ab^{\star}} = -2,\ I_{ac} = +3,
\ I_{ac^{\star}} = -3,
\nonumber\\
I_{bd} =\ +1,\ I_{be} = +1,\ I_{bf} = +1,\nonumber\\
I_{cd} =\ +1,\ I_{cd^{\star}} = -1,\ I_{ce} = +1, \nonumber\\
I_{ce^{\star}} = -1,\ I_{cf} = 1, \ I_{cf^{\star}}= -1.
\label{totrte}
\eeqa

%%%%%%%%%%%%%%%%%%%%%%%%%%%%%%%%%%%
%%%%%%%%%%%%%%%%%%%%%%%%%%%%%%%%%%%%%%%%%%%%%%%%%%%%%%
%
%
%
\section{Exact SM vacua from Four-Stack Quivers}

In this section, we discuss the appearance of exactly
the SM at low energy
from four stacks of D5 branes.
The orbifold structure of the SM configurations will
be depicted in terms
of quiver diagrams. In particular we examine $Z_3$ quivers.

Initially, 
we describe \footnote{,what was described in the
introduction as,}
the `reflected' quivers seen in figure (1).  
These four stack quivers give rise
to four different classes of SM vacua at low energy. For 
each of them we show that they give equivalent SM vacua 
to its `image' ai-quivers seen in figure (2). 
For the`image' ai-quivers, 
we discuss alternative RR tadpole solutions,
to the one's that have appeared in \cite{ibanez1}.

%\eps%%%%%%%%%% Figure here%%%%%%%%%%%%%%%%%%%%%%%%%%%%%%%%%%%%%%%%%
\begin{figure}
\begin{center}
\centering
\epsfysize=8cm
\leavevmode
%\hspace*{0in}\vspace*{.2in}
%\epsffile{anstan.eps}
\epsfbox{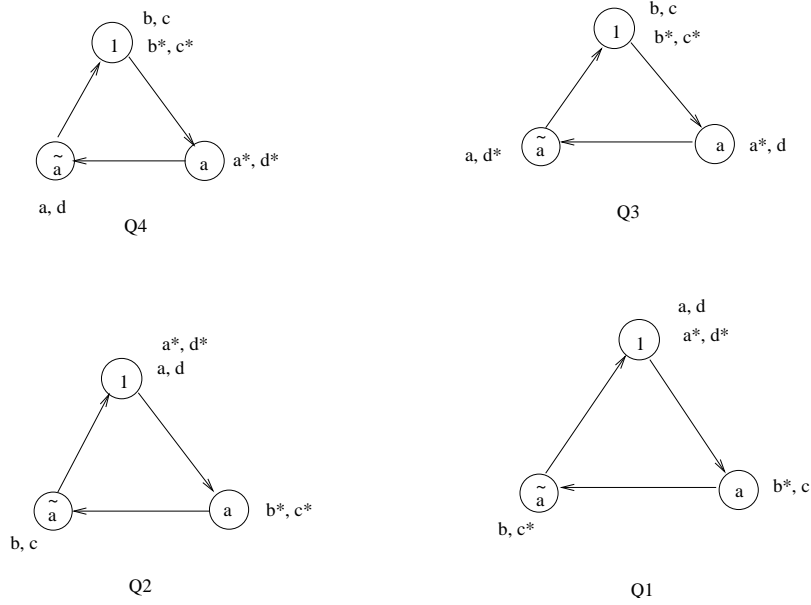}
\end{center}
\caption[]{\small
Assignment of SM embedding in  
configurations of four stacks of D5 branes depicted by the `reflected'
$Z_3$ quiver diagrams. At low energy we get only the SM.
Note that ${\tilde \alpha} = \alpha^{-1}$. These configurations give
equivalent vacua, with exactly the SM at low energy, to those vacua coming
from the `image' quivers of figure (2), under the correspondence
$a1 \iff Q1$,  $a2 \iff Q2$, 
$a3 \iff Q3$, $a4 \iff Q4$.
 }
\end{figure}
%%%%%%%%%%%%%%%%%end of figure %%%%%%%%%%%%%%%%%%%%%%%%%%%%%%%%%%%

%\eps%%%%%%%%%% Figure here%%%%%%%%%%%%%%%%%%%%%%%%%%%%%%%%%%%%%%%%%
\begin{figure}
\begin{center}
\centering
\epsfysize=8cm
\leavevmode
%\hspace*{0in}\vspace*{.2in}
\epsfbox{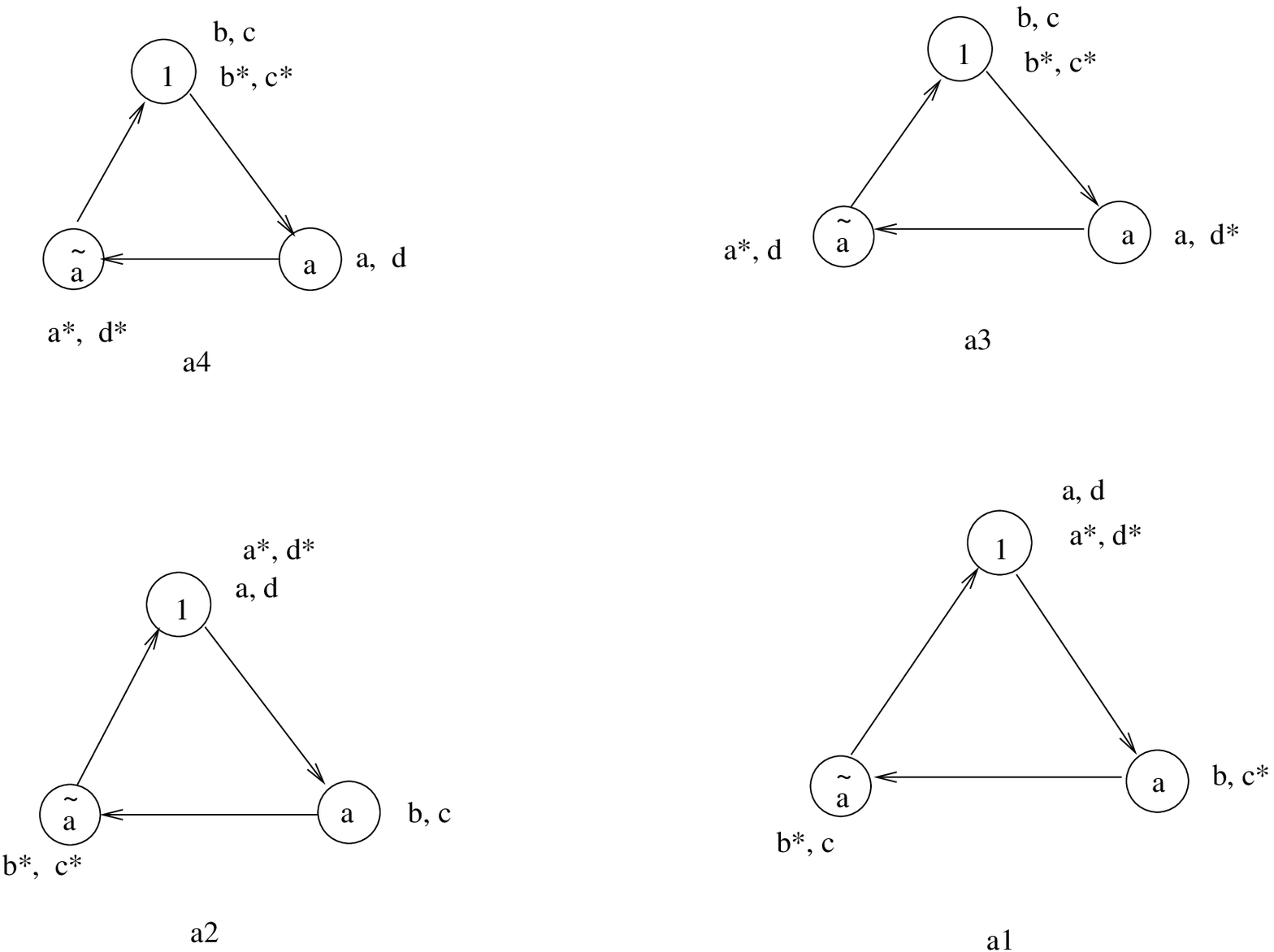}
\end{center}
\caption[]{\small
Assignment of SM embedding in  
configurations of four stacks of D5 branes depicted by
$Z_3$ quiver diagrams. At low energy we get only the SM.
Note that ${\tilde \alpha} = \alpha^{-1}$.
 }
\end{figure}
%%%%%%%%%%%%%%%%%end of figure %%%%%%%%%%%%%%%%%%%%%%%%%%%%%%%%%%%

\subsection{Exact SM vacua from the `reflected' Q1-Quiver}

In this subsection we examine the derivation of exactly
the SM at
low energies from the embedding of the four stack SM structure of table (1)
in a $Z_3$ quiver of $Q1$-type.

\begin{table}[htb]\footnotesize
\renewcommand{\arraystretch}{1.8}
\begin{center}
\begin{tabular}{||c||c|c|c||}
\hline
\hline
$N_i$ & $(n_i^1, m_i^1)$ & $(n_i^2, m_i^2)$ & $(n_i^3, m_i^3)$\\
\hline\hline
 $N_a=3$ & $(n_a^1, \epsilon {\tilde \ep}\b^1)$  &
$(3,  \frac{1}{2}{\tilde \epsilon} \epsilon )$ & $1_3$  \\
\hline
$N_b=2$  & $(1/\b_1, 0)$ & $(1, \frac{1}{2}{\epsilon}{\tilde \ep})$ & 
$\alpha^2 {\bf 1}_2$ \\
\hline
$N_c=1$ & $(1/\b_1, 0)$ &   $(0, {\epsilon{\tilde \ep}})$  & 
$\alpha$ \\    
\hline
$N_d=1$ & $(n_d^1, 3 \epsilon \b^1)$ &  $({\tilde \ep},  - \frac{1 
}{2}\epsilon)$  
  & $1$  \\\hline
$N_h$ & $(\epsilon_h/ \b^1, 0)$ &  
$(2, 0 )$  
  & $1_{N_h}$ 
\\\hline
\end{tabular}
\end{center}
\caption{\small General tadpole solutions for the four-stack 
$Q1$-type quiver of intersecting
D5-branes, giving rise to exactly the
standard model gauge group and observable chiral spectrum
at low energies.
The solutions depend 
on two integer parameters, 
$n_a^1$, $n_d^1$,
the NS-background $\beta^1= \ 1-b_i$, which is associated to 
the presence of the 
NS B-field by $b_i =0,\ 1/2$. 
 and
the phase parameters $\epsilon = {\tilde \epsilon}=\pm 1$, as well as the 
CP phase $\alpha$. 
\label{spectrum1oiel}}          
\end{table}

The solutions satisfying simultaneously the
intersection constraints and the
cancellation of the RR twisted crosscap tadpole cancellation
constraints
are given in parametric form in table (\ref{spectrum1oiel}). 
The multiparameter RR
tadpole solutions appearing in table (\ref{spectrum1oiel}) 
represent deformations of 
the D5-brane branes, of table (1), intersecting at angles,
 within the same homology class of the
factorizable two-cycles. 
The  
solutions of table (\ref{spectrum1oiel}) satisfy all tadpole 
equations, in (\ref{tadpoleO5b}), but the 
first. 
The latter reads :
\beq
9 n_a^1-\ \frac{1}{\b^1} +\ {\tilde \ep} \  n_d^1 + \
\frac{2\epsilon_h N_h}{\b^1} =\ -8.
\label{rampo}
\eeq

Note that we had added the presence of extra $N_h$ branes. 
Their contribution to the RR tadpole conditions is best 
described by placing them in the three-factorizable cycle 
\beq 
N_h \ (\epsilon_h/\b_1, 0)\ (2, 0)1_{N_h} \ .
\label{sda12el}
\eeq
The presence of an arbitrary number
of $N_h$ D5-branes, which give an extra $U(N_h)$ gauge group,
don't make any contribute to the rest of the tadpoles and
intersection constraints. Thus in terms of the
low energy theory their presence has no effect and
only the SM appears.

Most of the gauge theory anomalies of the four $U(1)$'s of the models
cancel as a consequence of the tadpole cancellation conditions.
As was shown in \cite{ibanez1}, the cubic
non-abelian anomaly
cancels as a consequence of the tadpole
cancellation conditions.
The mixed $U(1)-SU(N)^2$ anomaly partially cancels, as
a consequence of the tadpole cancellation conditions while a
non-cancelled piece remains, namely
\beq
\ca_{U(1)_{a,i}-SU(N_b^j)^2} = {-2 N_a^i \over N} 
\sum_{k=1}^N e^{2\pi i\frac{k\cdot i}{N}} 
c_k^2 \left( I_{ab} \ e^{-2\pi i\frac{k\cdot j}{N}}
+ \ I_{ab^*}  \ e^{2\pi i\frac{k\cdot j}{N}} \right).
\label{mixtaO5b}
\eeq
The latter cancels \cite{allo3} by the
exchange of four-dimensional
fields,
which come from the dimensional reduction of the
RR twisted forms living on the singularity \cite{ibanez1}. 
This is clearly seen in the T-dual picture of
fractional D7-branes wrapping the first two tori
in the presence of
non-trivial $F$ and $B$-fluxes. 
By integrating the 
couplings of the twisted RR p-forms \footnote{Appearing
in the worldvolume of each D7-brane.}
of even p, $A_o$,
$A_2$, $A_4$, $A_6$, over the compact dimensions
${\bf (T^2)_1 \times (T^2)_2}$, we can defining the
fields \cite{ibanez1}
\beq
\begin{array}{lcl}
B_0^{(k)} = A_0^{(k)}, 
& & B_2^{(k)} = \int_{\bf (T^2)_1 \times (T^2)_2} A_6^{(k)}, \\
C_0^{(k)} = \int_{\bf (T^2)_1 \times (T^2)_2} A_4^{(k)}, 
& & C_2^{(k)} =  A_2^{(k)}, \\
D_0^{(k)} = \int_{\bf (T^2)_2} A_2^{(k)},
& & D_2^{(k)} = \int_{\bf (T^2)_1} A_4^{(k)}, \\
E_0^{(k)} = \int_{\bf (T^2)_1} A_2^{(k)}, 
& & E_2^{(k)} = \int_{\bf (T^2)_2} A_4^{(k)}.
\end{array}
\label{pformas4}
\eeq
derive the four dimensional couplings as
\beqa & &
\begin{array}{c}
c_k N_a^i\, n^1_a n^2_a \int_{M_4} 
{\rm Tr} \left(\g_{k,a}-\g_{k,a^*}\right)\lam_i \
B_2^{(k)}\wedge {\rm Tr} F_{a,i}, \\
c_k N_a^i\, m^1_a m^2_a \int_{M_4} 
{\rm Tr} \left(\g_{k,a}-\g_{k,a^*}\right)\lam_i \
C_2^{(k)}\wedge {\rm Tr} F_{a,i}, \\
c_k N_a^i\, n^1_a m^2_a \int_{M_4} 
{\rm Tr} \left(\g_{k,a}+\g_{k,a^*}\right)\lam_i \
D_2^{(k)}\wedge {\rm Tr} F_{a,i}, \\
c_k N_a^i\, m^1_a n^2_a \int_{M_4} 
{\rm Tr} \left(\g_{k,a}+\g_{k,a^*}\right)\lam_i \
E_2^{(k)}\wedge {\rm Tr} F_{a,i},
\end{array} 
\label{acoplosdualO5} \\ & &
\begin{array}{c}
c_k m^1_a m^2_a \int_{M_4} 
{\rm Tr} \left(\g_{k,a}^{-1}+\g_{k,a^*}^{-1}\right)\lam_j^2 \
B_0^{(k)} \wedge {\rm Tr} \left(F_{a,j}\wedge F_{a,j}\right), \\
c_k n^1_a n^2_a \int_{M_4} 
{\rm Tr} \left(\g_{k,a}^{-1}+\g_{k,a^*}^{-1}\right)\lam_j^2 \
C_0^{(k)} \wedge {\rm Tr} \left(F_{a,j}\wedge F_{a,j}\right), \\
c_k m^1_a n^2_a \int_{M_4} 
{\rm Tr} \left(\g_{k,a}^{-1}-\g_{k,a^*}^{-1}\right)\lam_j^2 \
D_0^{(k)} \wedge {\rm Tr} \left(F_{a,j}\wedge F_{a,j}\right), \\
c_k n^1_a m^2_a \int_{M_4} 
{\rm Tr} \left(\g_{k,a}^{-1}-\g_{k,a^*}^{-1}\right)\lam_j^2 \
E_0^{(k)} \wedge {\rm Tr} \left(F_{a,j}\wedge F_{a,j}\right),
\end{array}
\label{acoplosdual2O5}
\eeqa
where $\lam$ is the Chan-Paton wavefunction for
the gauge
boson state, and $N_a^i$ factor arises from
normalization of
the $U(1)_{a,i}$ generator. 
 Since $B_2^{(k)}$, $C_2^{(k)}$, $D_2^{(k)}$  and $B_0^{(k)}$,
 $C_0^{(k)}$, $D_0^{(k)}$ are
four-dimensional Hodge duals respectively, 
the sum over the GS diagrams contributes a counterterm
that provides us with the structure required 
to cancel the residual mixed anomaly (\ref{mixtaO5b}).

In order to compute the twisted RR couplings for
a specific model, that means
taking into account the $U(1)$ anomaly constraints of
(\ref{acoplosdualO5}, \ref{acoplosdual2O5}), we should take into consideration
the number of twisted sectors available to us.
For the $Z_3$ orientifold quivers that we examine in this work
there is only one independent twisted
sector.
Thus for the Q1-quiver the constraints from $U(1)$
anomaly cancellation
appear into three terms
\beqa
B_2^{(1)} \wedge c_1 \left( \frac{ 2 
(\a^2 - \a ) }{\b^1 }
\right)F^b,&\nonumber\\       
D_2^{(1)} \wedge c_1 [ \ep {\tilde  \ep}]
\left( 3 n_a^1 F^a  
-\frac{1}{\b^1}F^b -\frac{1}{\b^1}F^c -{\tilde  \ep} n_d^1 F^d \right),&\nonumber\\
E_2^{(1)}  \wedge c_1 [6 \ep {\tilde  \ep}\b^1]
[3 F^a + F^d  ].&
\label{rr1a}
\eeqa

From (\ref{rr1a}) we conclude 
that there are two non-anomalous $U(1)$'s that become
massive
through their
couplings to the RR fields. They are the model
independent fields,
 $U(1)_b$ and the combination $3U(1)_a +
 U(1)_d$,
which become massive through their couplings to the 
RR 2-form fields $B_2^{(1)}, E_2^{(1)}$ respectively. In addition, there is 
a model dependent anomalous and massive $U(1)$ field
coupled to $D_2^{(1)}$ field.
That means that the 
 two non-anomalous free combinations are 
$U(1)_c$ and $U(1)_a - 3U(1)_d$.
Thus the combination of the $U(1)$'s, having no couplings to
twisted RR fields, remains light to low energies is
\beq  
Q^l =\ (Q_a -3 Q_d) +\
3  \b^1 ( n_a^1 + {\tilde \epsilon} n_d^1 ) Q_c   .
\label{hyper1}
\eeq
The subclass of tadpole solutions of (\ref{hyper1}) having
the SM hypercharge
assignment at low energies is exactly the one which is
proportional to
(\ref{hyper1230}). It
satisfies the condition, 
\beq
 \b_1 \ ( n_a^1 +\ {\tilde \epsilon} n_d^1) =\ -1.
\label{mashyper1}
\eeq
Summarizing, as long as (\ref{mashyper1}) holds
the $U(1)$ (\ref{hyper1}) is the hypercharge generator
of the SM.
Thus at low energies
we get the
chiral fermion content of the SM that gets localized according
to the open string sectors
appearing in table (1).

An alternative RR tadpole solution for the Q1-quiver
class  of SM's can be seen in table (\ref{spectrum1oiell}).
\begin{table}[htb]\footnotesize
\renewcommand{\arraystretch}{1.8}
\begin{center}
\begin{tabular}{||c||c|c|c||}
\hline
\hline
$N_i$ & $(n_i^1, m_i^1)$ & $(n_i^2, m_i^2)$ & $(n_i^3, m_i^3)$\\
\hline\hline
 $N_a=3$ & $(n_a^1, -\epsilon {\tilde \ep}\b^1)$  &
$(3,  -\frac{1}{2}{\tilde \epsilon} \epsilon )$ & $1_3$  \\
\hline
$N_b=2$  & $(1/\b_1, 0)$ & $(1, -\frac{1}{2}{\epsilon}{\tilde \ep})$ & 
$\alpha^2 {\bf 1}_2$ \\
\hline
$N_c=1$ & $(1/\b_1, 0)$ &   $(0, -{\epsilon{\tilde \ep}})$  & 
$\alpha$ \\    
\hline
$N_d=1$ & $(n_d^1, 3 \epsilon \b^1)$ &  $(-{\tilde \ep},  -
\frac{1}{2}\epsilon)$  
  & $1$  \\\hline
$N_h$ & $(\epsilon_h/ \b^1, 0)$ &  
$(2, 0 )$  
  & $1_{N_h}$ 
\\\hline
\end{tabular}
\end{center}
\caption{\small Alternative general tadpole
solutions for the four-stack
$Q1$-type quiver of intersecting
D5-branes, giving rise to exactly the
standard model gauge group and observable chiral spectrum
at low energies.
The solutions depend 
on two integer parameters, 
$n_a^1$, $n_d^1$,
the NS-background $\beta^1$ and
the phase parameters $\epsilon = {\tilde \epsilon}=\pm 1$, as well as the 
CP phase $\alpha$. 
\label{spectrum1oiell}}          
\end{table}
In this case the $U(1)$ couplings to twisted RR fields read
\beqa
B_2^{(1)} \wedge c_1 [2  \frac{({\a^2} -{\a})}{\b^1}]
F^b ,&\nonumber\\           
D_2^{(1)}\wedge c_1 [\ep {\tilde \epsilon}][-3 n_a^1 F^a +
\frac{1}{\b^1} F^b
+
\frac{1}{\b^1} F^c
- {\tilde \epsilon}  n_d^1 F^d ],   &\nonumber\\
E_2^{(1)}  \wedge c_1 [-6 \ep {\tilde \epsilon} \b^1 ][3 F^a + F^d ].&
\label{rrkl2asells1}
\eeqa
With the choice of tadpole solutions
of table (\ref{spectrum1oiell}) all tadpole equations in
(\ref{tadpoleO5b}) are satisfied but
the first, the latter giving
\beq                          
9 n_a^1 -\frac{1}{\b^1} - {\tilde \ep}
+ \frac{2 \ep_h N_h}{\b^1} = -8.
\label{latt567}
\eeq
The combination of the $U(1)$'s
\beq
 Q^l = (Q_a -3 Q_d)
+ 3 \b^1 (n_a^1 - {\tilde \epsilon}  n_d^1) Q_c.
\label{hyper2456ee}
\eeq
if the hypercharge condition 
\beq
\b^1 (n_a^1 - {\tilde \epsilon}n_d^1) = \ -1
\label{cond233ee}
\eeq
is fulfilled is the hypercharge.

%%%%%%%%%%%%%%%%%%%%%%%%%%%%%%%%%%%%%%%%%%%%%%%%%%%%%%
%%%%%%%%%%%%%%%%%%%%%%%%%%%%%%%%%
%%%%%%%%%%%%%%%%%%%%%

\subsection{Scalar sector of the `reflected' Q1-quiver}

In this class of models, the scalars appear from the
$aa^{\star}$, $ad^{\star}$, $ad$ sectors.
Lets us make a particular choice of the tadpole parameters
on these
SM's taking into account the constraints
(\ref{rampo}), (\ref{mashyper1}).
We choose
\beq
n_a^1 = -1, \ n_d^1 = -1, \ \b^1 = 1/2, \ \ep_h = 1, \
N_h = 2, \ \ep = 1, \ {\tilde \ep }= 1
\label{choiseof1}
\eeq

\begin{table}
[htb]\footnotesize
\renewcommand{\arraystretch}{1.5}
\begin{center}
\begin{tabular}{||c||c||c||}            
\hline                                   
\hline
Sector & Representation & $\a^{\prime} \cdot mass^2$ \\
\hline\hline
 $a {a^{\star}}$ & $2(3, 1)_{1/3} + (6, 1)_{-1/3}    $  &
$\pm \frac{1}{\pi}
\left(tan^{-1}\left(\frac{U^1}{2}\right)
- tan^{-1}\left(\frac{U^2}{6}\right)\right)
\stackrel{U^1 = \frac{U^2}{3}}{\rightarrow} 0$ \\\hline
$a d$  & $4(3,1)_{2/3}$ & 
$\pm \frac{1}{2\pi} [ tan^{-1}\left(\frac{3U^1}{2}\right)
- tan^{-1}
\left(\frac{U^1}{2}\right) -
tan^{-1}\left(\frac{U^2}{6}\right)-$ \\
& & $tan^{-1}\left(\frac{U^2}{2}\right)$ 
$\stackrel{U^1 = \frac{U^2}{3}}{\rightarrow}
\pm \frac{1}{\pi} tan^{-1}\left( \frac{U^2}{6}\right)$\\
\hline
$ad^{\star}$ & $2(3,1)_{-1/3}$ & 
$\pm \frac{1}{2 \pi}
[ - \frac{1}{\pi}
 tan^{-1}\left(\frac{U^1}{2}\right) - 
tan^{-1}\left(\frac{3U^1}{2}\right) +
 tan^{-1}\left(\frac{U^2}{2}\right)
- $\\
  & & $tan^{-1}\left(\frac{U^2}{6}\right)]$ 
$\stackrel{U^1 = \frac{U^2}{3}}{\rightarrow}
\pm \frac{1}{\pi} tan^{-1}\left( \frac{U^2}{6}\right)$
  \\\hline
\end{tabular}
\end{center}
\caption{\small 
Lightest scalar excitations for the four stack
Q1-quiver SM's.
\label{fortable}}          
\end{table}

There are two issues
that might concern us at this point. The first one
has to do with the existence of tachyons.
By choosing $U^1 = U^2/3$ the mass of the colour
$aa^{\star}$-sector goes to zero while the scalars from the
$ad$, $ad^{\star}$-sectors can get tachyonic values.
However, as has been emphasized in \cite{ibanez1}, the last problem
may be avoided as
scalars
may receive loop corrections from one gauge boson exchange in the form
that may stronger for colour scalars.
These corrections could drive their
masses to positive values (See also subsection (5.3)).
The lighter scalar excitations for this class of SM's can be seen
in table (\ref{fortable}).

%%%%%%%%%%%%%%%%%%%%%%%%%%%%%%%%%%%%%%%%%%%%%%%%%
%%%%%%%%%%%%%%%%%%%%%%%%%%%%%%%%%%%%%%%%%%%%%%%%%%
%
\subsection{Exact SM vacua from the `reflected' Q2-Quiver}

This is another class of models which may have exactly the SM
at low energy as the scalars appearing are coloured and thus we expect
loop corrections to lift the tachyonic directions.  
In this subsection we examine the derivation of exactly the SM at 
low energies from the embedding of the four stack SM structure of table (1)
in a $Z_3$ quiver of $Q2$-type.

\begin{table}[htb]\footnotesize
\renewcommand{\arraystretch}{1.8}
\begin{center}
\begin{tabular}{||c||c|c|c||}
\hline
\hline
$N_i$ & $(n_i^1, m_i^1)$ & $(n_i^2, m_i^2)$ & $(n_i^3, m_i^3)$\\
\hline\hline
 $N_a=3$ & $(n_a^1, \epsilon {\tilde \ep}\b^1)$  &
$(3,  \frac{1}{2}{\tilde \epsilon} \epsilon )$ & $1_3$  \\
\hline
$N_b=2$  & $(1/\b_1, 0)$ & $(1, \frac{1}{2}{\epsilon}{\tilde \ep})$ & 
$\alpha^2 {\bf 1}_2$ \\
\hline
$N_c=1$ & $(1/\b_1, 0)$ &   $(0, -{\epsilon{\tilde \ep}})$  & 
$\alpha^2$ \\    
\hline
$N_d=1$ & $(n_d^1, 3 \epsilon \b^1)$ &  $({\tilde \ep},  - \frac{1 
}{2}\epsilon)$  
  & $1$  \\\hline
$N_h$ & $(\epsilon_h/ \b^1, 0)$ &  
$(2, 0 )$  
  & $1_{N_h}$ 
\\\hline
\end{tabular}
\end{center}
\caption{\small General tadpole solutions for the
four-stack
$Q2$-type quiver of intersecting
D5-branes, giving rise to exactly the
standard model gauge group and observable chiral spectrum
at low energies.
The solutions depend 
on two integer parameters, 
$n_a^1$, $n_d^1$,
the NS-background $\beta^1$ and
the phase parameters $\epsilon = {\tilde \epsilon}=\pm 1$, as well as the 
CP phase $\alpha$. 
\label{spectrum1oielab1}}          
\end{table}

The solutions satisfying simultaneously the
intersection constraints and the
cancellation of the RR twisted crosscap tadpole cancellation
constraints
are given in parametric form in table (\ref{spectrum1oielab1}). 
These solutions represent
the most general solution of the RR tadpoles.
The 
solutions of table (\ref{spectrum1oielab1}) satisfy all RR tadpole 
cancellation conditions in (\ref{tadpoleO5b}) but the 
first. 
The latter reads :
\beq
9 n_a^1-\frac{1}{\b^1} + n_d^1{\tilde \ep} +
\frac{2\epsilon_h N_h}{\b^1} =-8.
\label{kato}
\eeq

In this case the $U(1)$ couplings to twisted RR fields read
\beqa
B_2^{(1)} \wedge c_1 [2  \frac{({\a^2} -{\a})}{\b^1}]
F^b ,&\nonumber\\           
D_2^{(1)}\wedge c_1 [\ep {\tilde \epsilon}][3 n_a^1 F^a -
\frac{1}{\b^1} F^b
+
\frac{1}{\b^1} F^c
- {\tilde \epsilon}  n_d^1 F^d ],   &\nonumber\\
E_2^{(1)}  \wedge c_1 [6\ep {\tilde \epsilon} \b^1 ][3 F^a + F^d ].&
\label{roo1}
\eeqa
The combination of the $U(1)$'s 
\beq
 Q^l = (Q_a -3 Q_d)
- 3 \b^1 (n_a^1 + {\tilde \epsilon}  n_d^1) Q_c.
\label{roo2}
\eeq
represents the hypercharge if the condition
\beq
\b^1 (n_a^1 + {\tilde \epsilon}n_d^1) = \ 1
\label{roo3}
\eeq
is satisfied.

An alternative RR tadpole solution for the Q2-quiver
class  of SM's can be seen in table (\ref{s2}).

\begin{table}[htb]\footnotesize
\renewcommand{\arraystretch}{1.8}
\begin{center}
\begin{tabular}{||c||c|c|c||}
\hline
\hline
$N_i$ & $(n_i^1, m_i^1)$ & $(n_i^2, m_i^2)$ & $(n_i^3, m_i^3)$\\
\hline\hline
 $N_a=3$ & $(n_a^1, -\epsilon {\tilde \ep}\b^1)$  &
$(3,  -\frac{1}{2}{\tilde \epsilon} \epsilon )$ & $1_3$  \\
\hline
$N_b=2$  & $(1/\b_1, 0)$ & $(1, -\frac{1}{2}{\epsilon}{\tilde \ep})$ & 
$\alpha^2 {\bf 1}_2$ \\
\hline
$N_c=1$ & $(1/\b_1, 0)$ &   $(0, -{\epsilon{\tilde \ep}})$  & 
$\alpha^2$ \\    
\hline
$N_d=1$ & $(n_d^1, 3 \epsilon \b^1)$ &  $(-{\tilde \ep},  -
\frac{1}{2}\epsilon)$  
  & $1$  \\\hline
$N_h$ & $(\epsilon_h/ \b^1, 0)$ &  
$(2, 0 )$  
  & $1_{N_h}$ 
\\\hline
\end{tabular}
\end{center}
\caption{\small Alternative general tadpole
solutions for the four-stack
$Q2$-type quiver of intersecting
D5-branes, giving rise to exactly the
standard model gauge group and observable chiral spectrum
at low energies.
The solutions depend 
on two integer parameters, 
$n_a^1$, $n_d^1$,
the NS-background $\beta^1$ and
the phase parameters $\epsilon = {\tilde \epsilon}=\pm 1$, as well as the 
CP phase $\alpha$. 
\label{s2}}          
\end{table}

In this case the $U(1)$ couplings to twisted RR fields read
\beqa
B_2^{(1)} \wedge c_1 [2  \frac{({\a^2} -{\a})}{\b^1}]
F^b ,&\nonumber\\           
D_2^{(1)}\wedge c_1 [\ep {\tilde \epsilon}][-3 n_a^1 F^a +
\frac{1}{\b^1} F^b
+
\frac{1}{\b^1} F^c
- {\tilde \epsilon}  n_d^1 F^d ],   &\nonumber\\
E_2^{(1)}  \wedge c_1 [-6 \ep {\epsilon} \b^1 ][3 F^a + F^d ].&
\label{s1}
\eeqa
The 
solutions of table (\ref{s2}) satisfy all RR tadpole 
cancellation conditions in (\ref{tadpoleO5b}) but the 
first. 
The latter being
\beq
9 n_a^1-\frac{1}{\b^1} + n_d^1{\tilde \ep} +
\frac{2\epsilon_h N_h}{\b^1} =-8.
\eeq
The $U(1)$  which remains massless reads
\beq
 Q^l = (Q_a -3 Q_d)
+ 3 \b^1 (n_a^1 - {\tilde \epsilon}  n_d^1) Q_c.
\label{roota1}
\eeq
The latter U(1) is the SM hypercharge if
\beq
\b^1 (n_a^1 - {\tilde \epsilon}n_d^1) = \ -1.
\label{roota2}
\eeq

\subsection{Scalar sector of the `reflected' Q2-quiver}

We will make the choice of RR tadpole parameters
by taking into account the constraints  
(\ref{kato}), (\ref{roo3}).
We choose
\beq
n_a^1 = 1, \ n_d^1 = 1, \ \b^1 = 1/2, \ \ep_h = -1, \
N_h = 4, \ \ep = 1, \ {\tilde \ep }= 1
\label{choiseof123}
\eeq

\begin{table}
[htb]\footnotesize
\renewcommand{\arraystretch}{1.5}
\begin{center}
\begin{tabular}{||c||c||c||}            
\hline                                   
\hline
Sector & Representation & $\a^{\prime} \cdot mass^2$ \\
\hline\hline
 $a {a^{\star}}$ & $2(3, 1)_{1/3} + (6, 1)_{-1/3}    $  &
$ 0$ \\\hline
$a d$  & $2(3,1)_{-2/3}$ & 
$0$\\
\hline
$ad^{\star}$ & $2(3,1)_{1/3}$ & 
$\pm \frac{1}{ \pi}
 tan^{-1}\left(\frac{U^2}{2}\right) $
  \\\hline
\end{tabular}
\end{center}
\caption{\small 
Lightest scalar excitations for the four stack
Q2-quiver SM's. The limit $U^1 =(U^2)/3$ is taken.
\label{fortablee1}}          
\end{table}

By choosing $U^1 = U^2/3$ the mass of the colour
$aa^{\star}$, $ad$ sector singlets goes to zero
while the scalars from the $ad^{\star}$-sectors have
a tachyonic direction. The latter should be lifted by loop
corrections as it has been emphasized before(see also (5.3)).
Thus at low energies only the SM should remain.

%%%%%%%%%%%%%%%%%%%%%%%%%%%%%%%%%%%%%%%%%%%%%%%%%
%%%%%%%%%%%%%%%%%%%%%%%%%%%%%%%%%%%%%%%%%%%%%%%%%%
%

\subsection{SM vacua from the `reflected' Q3-Quiver}

In this subsection we examine the derivation of
exactly the SM at
low energies from the embedding of the four stack SM structure of table (1)
in a $Z_3$ quiver of $Q3$-type.

\begin{table}[htb]\footnotesize
\renewcommand{\arraystretch}{1.8}
\begin{center}
\begin{tabular}{||c||c|c|c||}
\hline
\hline
$N_i$ & $(n_i^1, m_i^1)$ & $(n_i^2, m_i^2)$ & $(n_i^3, m_i^3)$\\
\hline\hline
 $N_a=3$ & $(1/\b_1, 0)$  &
$(1,  \frac{1}{2}{\tilde \epsilon} \epsilon )$ & $\alpha^2 1_3$  \\
\hline
$N_b=2$  & $(n_b^1, {\tilde \epsilon} \epsilon \b^1 )$ & $(1, 
\frac{3}{2}{\epsilon}{\tilde \ep})$ & 
${\bf 1}_2$ \\
\hline
$N_c=1$ & $(n_c^1, 3 \ep \b^1)$ &   $(0, -\epsilon)$  & 
$1$ \\    
\hline
$N_d=1$ & $(1/\b_1, 0)$ &  $(-1,  \frac{3 }{2}\epsilon {\tilde \ep})$  
  & $\alpha$  \\\hline
$N_h$ & $(\epsilon_h/ \b^1, 0)$ &  
$(2, 0 )$  
  & $1_{N_h}$ 
\\\hline
\end{tabular}
\end{center}
\caption{\small General tadpole solutions for the
four-stack
$Q3$-type quiver of intersecting
D5-branes, giving rise to models, with exactly the
standard model gauge group and observable chiral spectrum
at low energies.
The solutions depend 
on two integer parameters, 
$n_b^1$, $n_c^1$,
the NS-background $\beta^1$ and
the phase parameters $\epsilon = {\tilde \epsilon}=\pm 1$, as well as the 
CP phase $\alpha$. 
\label{spo3}}          
\end{table}
The solutions satisfying simultaneously the
intersection constraints and the
cancellation of the RR twisted crosscap tadpole cancellation
constraints
are given in parametric form in table (\ref{spo3}). 
The  
solutions of table (\ref{spo3}) satisfy all tadpole 
equations in ({\ref{tadpoleO5b}) but the 
first. 
The latter reads :
\beq
n_b^1 =  -4 +\frac{1}{2 \b^1}(1-  2\epsilon_h N_h).
\label{po1}
\eeq

Also in this case the $U(1)$ couplings to twisted RR fields read
\beqa
B_2^{(1)} \wedge c_1 [ \frac{({\a^2} -{\a})}{\b^1}]
[3 F^a + F^d] ,&\nonumber\\           
D_2^{(1)}\wedge c_1 [\ep {\tilde \epsilon}][-\frac{3}{2\b^1} F^a +
6 n_b^1 F^b
- 2 n_c^1 {\tilde \epsilon} F^c
- \frac{3}{2\b^1} F^d ],   &\nonumber\\
E_2^{(1)}  \wedge c_1 [4\ep {\tilde \epsilon}\b^1 ] F^b.&
\label{po2}
\eeqa

The low energy $U(1)$ combination
\beq
 Q^l = (Q_a -3 Q_d)
+  \frac{3 { \epsilon}}{2 \b^1 n_c^1} Q_c.
\label{po3}
\eeq
is the hypercharge if 
\beq
n_c^1 = -\frac{\epsilon}{2 \b^1} \ . 
\label{po4}
\eeq
From (\ref{po1}), (\ref{po4}) we conclude that $\b^1 = 1/2$.

An alternative RR tadpole cancellation to the one appearing in 
table (\ref{spo3}) appears in table (\ref{spo4}).

\begin{table}[htb]\footnotesize
\renewcommand{\arraystretch}{1.8}
\begin{center}
\begin{tabular}{||c||c|c|c||}
\hline
\hline
$N_i$ & $(n_i^1, m_i^1)$ & $(n_i^2, m_i^2)$ & $(n_i^3, m_i^3)$\\
\hline\hline
 $N_a=3$ & $(1/\b_1, 0)$  &
$(1,  -\frac{1}{2}{\tilde \epsilon} \epsilon )$ & $\alpha^2 1_3$  \\
\hline
$N_b=2$  & $(n_b^1, -{\tilde \epsilon} \epsilon \b^1 )$ & $(1, 
-\frac{3}{2}{\epsilon}{\tilde \ep})$ & 
${\bf 1}_2$ \\
\hline
$N_c=1$ & $(n_c^1, 3 \ep \b^1)$ &   $(0, -\epsilon)$  & 
$1$ \\    
\hline
$N_d=1$ & $(1/\b_1, 0)$ &  $(-1,   -\frac{3 }{2}\epsilon {\tilde \ep})$  
  & $\alpha$  \\\hline
$N_h$ & $(\epsilon_h/ \b^1, 0)$ &  
$(2, 0 )$  
  & $1_{N_h}$ 
\\\hline
\end{tabular}
\end{center}
\caption{\small Alternative general tadpole solutions for the
four-stack
$Q3$-type quiver of intersecting
D5-branes, giving rise to models with exactly the
standard model gauge group and observable chiral spectrum
at low energies.
The solutions depend 
on two integer parameters, 
$n_b^1$, $n_c^1$,
the NS-background $\beta^1$ and
the phase parameters $\epsilon = {\tilde \epsilon}=\pm 1$, as well as the 
CP phase $\alpha$. 
\label{spo4}}          
\end{table}

In this case, all tadpole conditions but the first,
in ({\ref{tadpoleO5b}), are satisfied, thus
\beq
n_b^1 =  -4 +\frac{1}{2 \b^1}(1-  2\epsilon_h N_h)
\label{sat1}
\eeq
The $U(1)$ couplings to twisted RR fields read:
\beqa
B_2^{(1)} \wedge c_1 [ \frac{({\a^2} -{\a})}{\b^1}]
[3 F^a + F^d] ,&\nonumber\\           
D_2^{(1)}\wedge c_1 [\ep {\tilde \epsilon}][\frac{3}{2\b^1} F^a -
6 n_b^1 F^b
- 2 n_c^1 {\tilde \epsilon} F^c
+ \frac{3}{2\b^1} F^d ],   &\nonumber\\
E_2^{(1)}  \wedge c_1 [-4\ep {\tilde \epsilon}\b^1 ] F^b.&
\label{sat2}
\eeqa
In this case, 
the combination of the $U(1)$'s which
is the SM hypercharge is
\beq
 Q^l = (Q_a -3 Q_d)
-  \frac{3 {\tilde \epsilon}}{2 \b^1 n_c^1} Q_c.
\label{sat3}
\eeq
if the hypercharge condition in satisfied, namely
\beq
n_c^1 = \frac{\tilde \epsilon}{2 \b^1}
\label{sat4}
\eeq
From (\ref{sat1}), (\ref{sat4}) we conclude that $\b^1 = 1/2$.

%
%%%%%%%%%%%%%%%%%%%%%%%%%%%%%%%%%%%%%%%%%%%%%%%%%%%%
%%%%%%%%%%%%%%%%%%%%%%%%%%%%%%%%%%%%%%%%%%%%%%%%%
%
%

\subsection{Scalar sector of the `reflected' Q3-quiver}

We will make the choice of RR tadpole parameters
by taking into account the constraints  
(\ref{kato}), (\ref{roo3}).
We choose
\beq
n_b^1 = 1, \ n_c^1 = 1, \ \b^1 = 1/2, \ \ep_h = -1, \
N_h = 4, \ \ep = -1, \ {\tilde \ep }= -1
\label{chonai}
\eeq

\begin{table}
[htb]\footnotesize
\renewcommand{\arraystretch}{1.5}
\begin{center}
\begin{tabular}{||c||c||c||}            
\hline                                   
\hline
Sector & Representation & $\a^{\prime} \cdot mass^2$ \\
\hline\hline
 $b {b^{\star}}$ & $3(1, 1)_{0}$  &
$   \pm \frac{1}{\pi}
[tan^{-1}(\frac{U^1}{2})
- tan^{-1}(\frac{U^2}{2})] $ \\\hline
$bc$  & $2(2,1)_{1/2}$ & $\pm \frac{1}{2\pi}
[ tan^{-1}(\frac{U^1}{2}) + tan^{-1}(\frac{3U^1}{2}) -
tan^{-1}(\frac{3U^2}{2}) + tan^{-1}(\frac{U^2}{0})]$  \\
\hline
$bc^{\star}$ & $(2,1)_{-1/2}$ & 
$   \pm \frac{1}{\pi}
[tan^{-1}(\frac{U^1}{2})  - tan^{-1}(\frac{3U^1}{2}) - 
tan^{-1}(\frac{3U^2}{2}) +  tan^{-1}(\frac{U^2}{0}) $
  \\\hline
\end{tabular}
\end{center}
\caption{\small 
Lightest scalar excitations for the four stack
Q3-quiver SM's. 
\label{fortableeee}}          
\end{table}

If we choose $U^1 = U^2$ then the mass of the
$aa^{\star}$ scalars goes to zero. However
the scalars from the $bc$, $bc^{\star}$ may tend to zero only when
the $U^1$ modulus assumes simultaneously the limit $U^1 \rightarrow \infty$, a decompactification limit.  
However, is is possible to lift these tachyonic 
directions by loop 
corrections  \cite{ibar, ibanez1}, that are not expected to be so strong as for colour scalars. Thus the SM may remain at low energies.
An alternative way to show that there always be directions in the moduli 
space that will give the SM at low energy will be shown in section 7 using
brane recombination.

%
%%%%%%%%%%%%%%%%%%%%%%%%%%%%%%%%%%%%%%%%%%%%%%%%%%%%
%%%%%%%%%%%%%%%%%%%%%%%%%%%%%%%%%%%%%%%%%%%%%%%%%
%
%
\subsection{SM vacua from the `reflected' Q4-Quiver}

In this subsection we examine the derivation of exactly the SM at 
low energies from the embedding of the four stack SM structure of table (1)
in a $Z_3$ quiver of $Q4$-type.

\begin{table}[htb]\footnotesize
\renewcommand{\arraystretch}{1.8}
\begin{center}
\begin{tabular}{||c||c|c|c||}
\hline
\hline
$N_i$ & $(n_i^1, m_i^1)$ & $(n_i^2, m_i^2)$ & $(n_i^3, m_i^3)$\\
\hline\hline
 $N_a=3$ & $(1/\b_1, 0)$  &
$(1,  \frac{1}{2}{\tilde \epsilon} \epsilon )$ & $\alpha^2 1_3$  \\
\hline
$N_b=2$  & $(n_b^1, {\tilde \epsilon} \epsilon \b^1 )$ & $(1, 
\frac{3}{2}{\epsilon}{\tilde \ep})$ & 
${\bf 1}_2$ \\
\hline
$N_c=1$ & $(n_c^1, 3 \ep \b^1)$ &   $(0, -\epsilon)$  & 
$1$ \\    
\hline
$N_d=1$ & $(1/\b_1, 0)$ &  $(1,  - \frac{3 }{2}\epsilon {\tilde \ep})$  
  & $\alpha^2$  \\\hline
$N_h$ & $(\epsilon_h/ \b^1, 0)$ &  
$(2, 0 )$  
  & $1_{N_h}$ 
\\\hline
\end{tabular}
\end{center}
\caption{\small General tadpole solutions for the
four-stack
$Q4$-type quiver of intersecting
D5-branes, giving rise to models with exactly the
standard model gauge group and observable chiral spectrum
at low energies.
The solutions depend 
on two integer parameters, 
$n_b^1$, $n_c^1$,
the NS-background $\beta^1$ and
the phase parameters $\epsilon = {\tilde \epsilon}=\pm 1$, as well as the 
CP phase $\alpha$. 
\label{llab1}}          
\end{table}

The solutions satisfying simultaneously the
intersection constraints and the
cancellation of the RR twisted crosscap tadpole cancellation
constraints
are given in parametric form in table (\ref{llab1}). 
These solutions represent
the most general solution of the twisted RR tadpoles (\ref{tadpoleO5b}).
The
solutions of table (\ref{llab1}) satisfy all tadpole 
equations, in ({\ref{tadpoleO5b}), but the 
first. 
The latter reads :
\beq
n_b^1 = -4 + \frac{1}{\b^1}(1-\epsilon_h N_h).
\label{comp} 
\eeq
Further, the $U(1)$ couplings to twisted RR fields read
\beqa
B_2^{(1)} \wedge c_1 [\frac{({\a^2} -{\a})}{\b^1}]
[3F^a + F^d] ,&\nonumber\\           
D_2^{(1)}\wedge c_1 [\ep {\tilde \epsilon}][-\frac{3}{2\b^1}F^a +
6 n_b^1 F^b
- 2 n_c^1 {\tilde \ep} F^c
+  \frac{3}{2\b^1}  F^d ],   &\nonumber\\
E_2^{(1)}  \wedge c_1 [4 \ep {\tilde \epsilon} \b^1 ]F^b .&
\label{roo11}
\eeqa
The `light' $U(1)$ combination
\beq
 Q^l = (Q_a -3 Q_d)
- \frac{3 {\epsilon}}{\b^1 n_c^1 } Q_c.
\label{roo22}
\eeq
is identified as the SM hypercharge if the condition 
\beq
n_c^1 = \frac{{\epsilon}}{ \b^1}
\label{roo3451}
\eeq
is satisfied.
From (\ref{comp}), (\ref{roo3451}) we conclude that $\b^1 = 1$.

\begin{table}[htb]\footnotesize
\renewcommand{\arraystretch}{1.8}
\begin{center}
\begin{tabular}{||c||c|c|c||}
\hline
\hline
$N_i$ & $(n_i^1, m_i^1)$ & $(n_i^2, m_i^2)$ & $(n_i^3, m_i^3)$\\
\hline\hline
 $N_a=3$ & $(1/\b_1, 0)$  &
$(1,  -\frac{1}{2}{\tilde \epsilon} \epsilon )$ & $\alpha^2 1_3$  \\
\hline
$N_b=2$  & $(n_b^1, -{\tilde \epsilon} \epsilon \b^1 )$ & $(1, 
-\frac{3}{2}{\epsilon}{\tilde \ep})$ & 
${\bf 1}_2$ \\
\hline
$N_c=1$ & $(n_c^1, 3 \ep \b^1)$ &   $(0, -\epsilon)$  & 
$1$ \\    
\hline
$N_d=1$ & $(1/\b_1, 0)$ &  $(1,   \frac{3 }{2}\epsilon {\tilde \ep})$  
  & $\alpha^2$  \\\hline
$N_h$ & $(\epsilon_h/ \b^1, 0)$ &  
$(2, 0 )$  
  & $1_{N_h}$ 
\\\hline
\end{tabular}
\end{center}
\caption{\small Alternative general tadpole solutions for the
four-stack
$Q4$-type quiver of intersecting
D5-branes, giving rise to SM-like models with exactly the
standard model gauge group and observable chiral spectrum
at low energies.
The solutions depend 
on two integer parameters, 
$n_b^1$, $n_c^1$,
the NS-background $\beta^1$ and
the phase parameters $\epsilon = {\tilde \epsilon}=\pm 1$, as well as the 
CP phase $\alpha$. 
\label{alter1}}          
\end{table}

An alternative RR tadpole cancellation to the one appearing in 
table (\ref{llab1}) appears in table (\ref{alter1}).
The twisted RR 
solutions of table (\ref{alter1}) satisfy all tadpole 
equations, in ({\ref{tadpoleO5b}), but the 
first. 
The latter reads :
\beq
n_b^1 = -4 + \frac{1}{\b^1}(1-\epsilon_h N_h).
\label{pao1} 
\eeq
In this case the $U(1)$ couplings to twisted RR fields read
\beqa
B_2^{(1)} \wedge c_1 [\frac{({\a^2} -{\a})}{\b^1}]
[3F^a + F^d] ,&\nonumber\\           
D_2^{(1)}\wedge c_1 [\ep {\tilde \epsilon}][\frac{3}{2\b^1}F^a -
6 n_b^1 F^b
- 2 n_c^1 {\tilde \ep} F^c
-  \frac{3}{2\b^1}  F^d ],   &\nonumber\\
E_2^{(1)}  \wedge c_1 [-4 \ep {\tilde \epsilon} \b^1 ]F^b .&
\label{pao2}
\eeqa
In this alternative solution
the $U(1)$ combination
\beq
 Q^l = (Q_a -3 Q_d)
+ \frac{3 {\tilde \epsilon}}{ n_c^1 \b^1 } Q_c.
\label{pao3}
\eeq
is identified as the hypercharge if
\beq
n_c^1 = \frac{{\tilde \epsilon}}{ \b^1}.
\label{pao4}
\eeq
From (\ref{pao1}), (\ref{pao4}) we conclude that $\b^1 = 1$.  
The Q4-models may give the SM at low energy as the scalar tachyonic directions
may always be lifted.

%%%%%%%%%%%%%%%%%%%%%%%%%%%%%%%%%%%%%%%%%%%%%%%%%%%%%%%%%%%%
%%%%%%%%%%%%%%%%%%%%%%%%%%%%%%%
%%%%%%%%%%%%%%%%%%%
%%%%%%%%

\subsection{Exact SM vacua from the a1-Quiver}

In this subsection we examine the derivation of exactly the SM at 
low energies from the embedding of the four stack SM structure of table (1)
in a $Z_3$ quiver of $a1$-type. 
The latter quiver can be seen in figure (2).
For an alternative RR tadpole solution of the $a1$-quiver see \cite{ibanez1}.

 \begin{table}[htb]\footnotesize
\renewcommand{\arraystretch}{1.8}
\begin{center}
\begin{tabular}{||c||c|c|c||}
\hline
\hline
$N_i$ & $(n_i^1, m_i^1)$ & $(n_i^2, m_i^2)$ & $(n_i^3, m_i^3)$\\
\hline\hline
 $N_a=3$ & $(n_a^1, {\tilde \epsilon} \epsilon \b^1 )$  &
$(3,  -\frac{1}{2}{\tilde \epsilon} \epsilon )$ & $ 1_3$  \\
\hline
$N_b=2$  & $(1/\b_1, 0)$ & $(1, 
-\frac{1}{2}{\epsilon}{\tilde \ep})$ & 
$\alpha {\bf 1}_2$ \\
\hline
$N_c=1$ & $(1/\b_1, 0)$ &   $(0, -\epsilon {\tilde \ep})$  & 
$\alpha^2$ \\    
\hline
$N_d=1$ & $(n_d^1, 3 \ep \b^1)$ &  $({\tilde \ep},   \frac{1}{2}\epsilon)$  
  & $1$  \\\hline
$N_h$ & $(\epsilon_h/ \b^1, 0)$ &  
$(2, 0 )$  
  & $1_{N_h}$ 
\\\hline
\end{tabular}
\end{center}
\caption{\small General tadpole solutions for the
four-stack
$a1$-type quiver of intersecting
D5-branes, giving rise to exactly the
standard model gauge group and observable chiral spectrum
at low energies.
%The solutions depend 
%on two integer parameters, 
%$n_a^1$, $n_d^1$,
%the NS-background $\beta^1$ and
%the phase parameters $\epsilon = {\tilde \epsilon}=\pm 1$, as well as the 
%CP phase $\alpha$. 
\label{dada1}}          
\end{table}

The solutions satisfying simultaneously the
intersection constraints and the
cancellation of the RR twisted crosscap tadpole cancellation
constraints
are given in parametric form in table (\ref{dada1}). 
These solutions represent
the most general solution of the twisted RR tadpoles (\ref{tadpoleO5b}).
They depend 
on two 
integer parameters $n_a^1$, $n_d^1$, 
the phase parameters
$\epsilon = \pm 1$, ${\tilde \epsilon} = \pm 1$, and the NS-background
parameter
$\beta_i =\ 1-b_i$, which is associated to the presence of the 
NS B-field by $b_i =0,\ 1/2$.

The  
solutions of table (\ref{dada1}) satisfy all tadpole 
equations, in ({\ref{tadpoleO5b}), but the 
first. 
The latter reads :
\beq
9 n_a^1 - \frac{1}{\b^1} + {\tilde \ep}n_d^1 + \frac{2 \epsilon_h N_h}{\b^1}=-8
\label{dasa1} 
\eeq

In this case the $U(1)$ couplings to twisted RR fields read
\beqa
B_2^{(1)} \wedge c_1 [-2\frac{({\a^2} -{\a})}{\b^1}]F^b ,&\nonumber\\           
D_2^{(1)}\wedge c_1 [\ep {\tilde \epsilon}][-3 n_a^1F^a +
\frac{1}{\b^1} F^b + \frac{1}{\b^1} F^c
+  {\tilde \epsilon} n_d^1 F^d ],   &\nonumber\\
E_2^{(1)}  \wedge c_1 [6 \ep {\tilde \epsilon} \b^1 ]( 3 F^a + F^d) .&
\label{dasa3}
\eeqa

From (\ref{dasa3}) it is easily identified the hypercharge
as the $U(1)$ combination
\beq
 Q^l = (Q_a -3 Q_d)
+ 3 \b^1 (n_a^1 + {\tilde \epsilon} n_d^1 ) Q_c.
\label{dasa4}
\eeq
which simultaneously satisfies
\beq
\b^1 (n_a^1 + {\tilde \epsilon} n_d^1 ) =\ -1
\label{dasa5}
\eeq

An alternative solution to the RR tadpole solutions for
 the a1-quiver, we found in
table (\ref{dada1}), can be seen in table (\ref{da2}).

\begin{table}[htb]\footnotesize
\renewcommand{\arraystretch}{1.8}
\begin{center}
\begin{tabular}{||c||c|c|c||}
\hline
\hline
$N_i$ & $(n_i^1, m_i^1)$ & $(n_i^2, m_i^2)$ & $(n_i^3, m_i^3)$\\
\hline\hline
 $N_a=3$ & $(n_a^1, -{\tilde \epsilon} \epsilon \b^1 )$  &
$(3,  \frac{1}{2}{\tilde \epsilon} \epsilon )$ & $ 1_3$  \\
\hline
$N_b=2$  & $(1/\b_1, 0)$ & $(1, 
\frac{1}{2}{\epsilon}{\tilde \ep})$ & 
$\alpha {\bf 1}_2$ \\
\hline
$N_c=1$ & $(1/\b_1, 0)$ &   $(0, \epsilon {\tilde \ep})$  & 
$\alpha^2$ \\    
\hline
$N_d=1$ & $(n_d^1, 3 \ep \b^1)$ &  $(-{\tilde \ep},   \frac{1}{2}\epsilon)$  
  & $1$  \\\hline
$N_h$ & $(\epsilon_h/ \b^1, 0)$ &  
$(2, 0 )$  
  & $1_{N_h}$ 
\\\hline
\end{tabular}
\end{center}
\caption{\small Alternative general twisted RR tadpole solutions for the
four-stack
$a1$-type quiver of intersecting
D5-branes, giving rise to exactly the
standard model gauge group and observable chiral spectrum
at low energies.
The solutions depend 
on two integer parameters, 
$n_a^1$, $n_d^1$,
the NS-background $\beta^1$ and
the phase parameters $\epsilon = {\tilde \epsilon}=\pm 1$, as well as the 
CP phase $\alpha$. 
\label{da2}}          
\end{table}

The  
solutions of table (\ref{da2}) satisfy all tadpole 
equations, in ({\ref{tadpoleO5b}), but the 
first. 
The latter reads :
\beq
9 n_a^1 - \frac{1}{\b^1} - {\tilde \ep}n_d^1 + \frac{2 \epsilon_h N_h}{\b^1}=-8
\label{ta1} 
\eeq

In this case the $U(1)$ couplings to twisted RR fields read
\beqa
B_2^{(1)} \wedge c_1 [2\frac{({\a} -{\a^2})}{\b^1}]F^b ,&\nonumber\\           
D_2^{(1)}\wedge c_1 [\ep {\tilde \epsilon}][3 n_a^1F^a -
\frac{1}{\b^1} F^b - \frac{1}{\b^1} F^c
+  {\tilde \epsilon} n_d^1 F^d ],   &\nonumber\\
E_2^{(1)}  \wedge c_1 [-6 \ep {\tilde \epsilon} \b^1 ]( 3 F^a + F^d) .&
\label{ta2}
\eeqa
Similarly the hypercharge gets identified as the
expression
\beq
 Q^l = (Q_a -3 Q_d)
+ 3 \b^1 (n_a^1 - {\tilde \epsilon} n_d^1 ) Q_c.
\label{ta3}
\eeq
which satisfies the additional condition
\beq
\b^1 (n_a^1 - {\tilde \epsilon} n_d^1 ) =\ -1
\label{ta4}
\eeq

%%%%%%%%%%%%%%%%%%%%%%%%%%%%%%%%%%%%%%%%%%%%%%%%%%%%
%%%%%%%%%%%%%%%%%%%%%%%%%%%%%%%%%%%%
%%%%%%%%%%%%%%%%%%%%%

\subsection{Exact SM vacua from the a2-Quiver}
In this subsection we examine the derivation of exactly the SM at 
low energies from the embedding of the four stack SM structure of table (1)
in a $Z_3$ quiver of $a2$-type. 
The latter quiver can be seen in figure (2).
For an alternative RR tadpole solution of the $a2$-quiver
see \cite{ibanez1}.

 \begin{table}[htb]\footnotesize
\renewcommand{\arraystretch}{1.8}
\begin{center}
\begin{tabular}{||c||c|c|c||}
\hline
\hline
$N_i$ & $(n_i^1, m_i^1)$ & $(n_i^2, m_i^2)$ & $(n_i^3, m_i^3)$\\
\hline\hline
 $N_a=3$ & $(n_a^1, {\tilde \epsilon} \epsilon \b^1 )$  &
$(3,  -\frac{1}{2}{\tilde \epsilon} \epsilon )$ & $ 1_3$  \\
\hline
$N_b=2$  & $(1/\b_1, 0)$ & $(1, 
-\frac{1}{2}{\epsilon}{\tilde \ep})$ & 
$\alpha {\bf 1}_2$ \\
\hline
$N_c=1$ & $(1/\b_1, 0)$ &   $(0, \epsilon {\tilde \ep})$  & 
$\alpha$ \\    
\hline
$N_d=1$ & $(n_d^1, 3 \ep \b^1)$ &  $({\tilde \ep},   \frac{1}{2}\epsilon)$  
  & $1$  \\\hline
$N_h$ & $(\epsilon_h/ \b^1, 0)$ &  
$(2, 0 )$  
  & $1_{N_h}$ 
\\\hline
\end{tabular}
\end{center}
\caption{\small General tadpole solutions for the
four-stack
$a2$-type quiver of intersecting
D5-branes, giving rise to exactly the
standard model gauge group and observable chiral spectrum
at low energies.
%The solutions depend
%on two integer parameters, 
%$n_a^1$, $n_d^1$,
%the NS-background $\beta^1$ and
%the phase parameters $\epsilon = {\tilde \epsilon}=\pm 1$, as well as the 
%CP phase $\alpha$. 
\label{lasa1}}          
\end{table}

The solutions satisfying simultaneously the
intersection constraints and the
cancellation of the RR twisted crosscap tadpole cancellation
constraints
are given in parametric form in table (\ref{lasa1}). 
These solutions represent
the most general solution of the twisted RR
tadpoles (\ref{tadpoleO5b}).
They depend
on two 
integer parameters $n_a^1$, $n_d^1$, 
the phase parameters
$\epsilon = \pm 1$, ${\tilde \epsilon} = \pm 1$, and the
NS-background
parameter
$\beta_i =\ 1-b_i$, which is associated to the presence of the 
NS B-field by $b_i =0,\ 1/2$. 
The 
solutions of table (\ref{lasa1}) satisfy all tadpole 
equations in ({\ref{tadpoleO5b}), but the 
first. 
The latter reads :
\beq
9 n_a^1 - \frac{1}{\b^1} + {\tilde \ep}n_d^1 + \frac{2 \epsilon_h N_h}{\b^1}=-8
\label{dasaa1} 
\eeq

In this case the $U(1)$ couplings to twisted RR fields read
\beqa
B_2^{(1)} \wedge c_1 [-2\frac{({\a^2} -{\a})}{\b^1}]F^b ,&\nonumber\\           
D_2^{(1)}\wedge c_1 [\ep {\tilde \epsilon}][-3 n_a^1F^a +
\frac{1}{\b^1} F^b - \frac{1}{\b^1} F^c
+  {\tilde \epsilon} n_d^1 F^d ],   &\nonumber\\
E_2^{(1)}  \wedge c_1 [6 \ep {\tilde \epsilon} \b^1 ]( 3 F^a + F^d) .&
\label{dasaa3}
\eeqa

The combination of the $U(1)$'s 
\beq
 Q^l = (Q_a -3 Q_d)
- 3 \b^1 (n_a^1 + {\tilde \epsilon} n_d^1 ) Q_c
\label{dasaa4}
\eeq
may be the SM hypercharge if the condition
\beq
\b^1 (n_a^1 + {\tilde \epsilon} n_d^1 ) =\ 1
\label{dasaa5}
\eeq

An alternative solution to the RR tadpole
solutions for the a2-quiver,
we found in table (\ref{lasa1}), can be seen in table (\ref{lasa3}).

\begin{table}[htb]\footnotesize
\renewcommand{\arraystretch}{1.8}
\begin{center}
\begin{tabular}{||c||c|c|c||}
\hline
\hline
$N_i$ & $(n_i^1, m_i^1)$ & $(n_i^2, m_i^2)$ & $(n_i^3, m_i^3)$\\
\hline\hline
 $N_a=3$ & $(n_a^1, -{\tilde \epsilon} \epsilon \b^1 )$  &
$(3,  \frac{1}{2}{\tilde \epsilon} \epsilon )$ & $ 1_3$  \\
\hline
$N_b=2$  & $(1/\b_1, 0)$ & $(1, 
\frac{1}{2}{\epsilon}{\tilde \ep})$ & 
$\alpha {\bf 1}_2$ \\
\hline
$N_c=1$ & $(1/\b_1, 0)$ &   $(0, -\epsilon {\tilde \ep})$  & 
$\alpha$ \\    
\hline
$N_d=1$ & $(n_d^1, 3 \ep \b^1)$ &  $(-{\tilde \ep},   \frac{1}{2}\epsilon)$  
  & $1$  \\\hline
$N_h$ & $(\epsilon_h/ \b^1, 0)$ &  
$(2, 0 )$  
  & $1_{N_h}$ 
\\\hline
\end{tabular}
\end{center}
\caption{\small Alternative general twisted RR tadpole solutions for the
four-stack
$a2$-type quiver of intersecting
D5-branes, giving rise to exactly the
standard model gauge group and observable chiral spectrum
at low energies.
The solutions depend 
on two integer parameters, 
$n_a^1$, $n_d^1$,
the NS-background $\beta^1$ and
the phase parameters $\epsilon = {\tilde \epsilon}=\pm 1$, as well as the 
CP phase $\alpha$. 
\label{lasa3}}          
\end{table}

In this case all tadpole equations but the first,
in ({\ref{tadpoleO5b}), are
satisfied. 
The latter reads :
\beq
9 n_a^1 - \frac{1}{\b^1} - {\tilde \ep}n_d^1 + \frac{2 \epsilon_h N_h}{\b^1}=-8
\label{lap1} 
\eeq

In this case the $U(1)$ couplings to twisted RR fields read
\beqa
B_2^{(1)} \wedge c_1 [2\frac{({\a} -{\a^2})}{\b^1}]F^b ,&\nonumber\\           
D_2^{(1)}\wedge c_1 [\ep {\tilde \epsilon}][3 n_a^1F^a -
\frac{1}{\b^1} F^b + \frac{1}{\b^1} F^c
+  {\tilde \epsilon} n_d^1 F^d ],   &\nonumber\\
E_2^{(1)}  \wedge c_1 [-6 \ep {\tilde \epsilon} \b^1 ]( 3 F^a + F^d) .&
\label{lap2}
\eeqa

Also in this case, the U(1)
\beq
 Q^l = (Q_a -3 Q_d)
- 3 \b^1 (n_a^1 - {\tilde \epsilon} n_d^1 ) Q_c
\label{lap3}
\eeq
is being identified as the hypercharge if
\beq
\b^1 (n_a^1 - {\tilde \epsilon} n_d^1 ) =\ 1
\label{lap4}
\eeq
is satisfied.

%%%%%%%%%%%%%%%%%%%%%%%%%%%%%%%%%%%%%%%%%%
%%%%%%%%%%%%%%%%%%%%%%%%
%%%%%%%%%%%%%%%%%%%%%%%%%%%%%%%%%%

\subsection{SM vacua from the a3-Quiver}
%
%                            i
%
In this subsection we examine the derivation of exactly the SM at 
low energies from the embedding of the four stack
SM structure of table (1)
in a $Z_3$ quiver of $a3$-type. 
The latter quiver can be seen in figure (2). 
For an alternative RR tadpole solution
of the $a3$-quiver see \cite{ibanez1}.

\begin{table}[htb]\footnotesize
\renewcommand{\arraystretch}{1.8}
\begin{center}
\begin{tabular}{||c||c|c|c||}
\hline
\hline
$N_i$ & $(n_i^1, m_i^1)$ & $(n_i^2, m_i^2)$ & $(n_i^3, m_i^3)$\\
\hline\hline
 $N_a=3$ & $(1/\b^1, 0)$  &
$(1,  -\frac{1}{2}{\tilde \epsilon} \epsilon )$ & $\alpha 1_3$  \\
\hline
$N_b=2$  & $(n_b^1, {\tilde \epsilon} \epsilon \b^1 )$ & $(1, 
-\frac{3}{2}{\epsilon}{\tilde \ep})$ & 
${\bf 1}_2$ \\
\hline
$N_c=1$ & $(n_c^1, 3 \ep \b^1)$ &   $(0, \epsilon )$  & 
$1$ \\    
\hline
$N_d=1$ & $(1/\b^1, 0)$ &  $(-1,   -\frac{3}{2}\epsilon{\tilde \epsilon} )$  
  & $\alpha^2$  \\\hline
$N_h$ & $(\epsilon_h/ \b^1, 0)$ &  
$(2, 0 )$  
  & $1_{N_h}$ 
\\\hline
\end{tabular}
\end{center}
\caption{\small General twisted RR tadpole solutions for the
four-stack
$a3$-type quiver of intersecting
D5-branes, giving rise to SM-like models with exactly the
standard model gauge group and observable chiral spectrum
at low energies.
%The solutions depend 
%on two integer parameters, 
%$n_b^1$, $n_c^1$,
%the NS-background $\beta^1$ and
%the phase parameters $\epsilon = {\tilde \epsilon}=\pm 1$, as well as the 
%CP phase $\alpha$. 
\label{sapo1}}          
\end{table}

The solutions satisfying simultaneously the
intersection constraints and the
cancellation of the RR twisted crosscap tadpole cancellation
constraints
are given in parametric form in table (\ref{sapo1}). 
These solutions represent
the most general solution of the twisted RR tadpoles (\ref{tadpoleO5b}).
They depend 
on two 
integer parameters $n_b^1$, $n_c^1$, 
the phase parameters
$\epsilon = \pm 1$, ${\tilde \epsilon} = \pm 1$, and the NS-background
parameter
$\beta_i =\ 1-b_i$, which is associated to the presence of the 
NS B-field by $b_i =0,\ 1/2$.

The 
RR tadpole solutions of table (\ref{sapo1}) satisfy all tadpole 
equations, in ({\ref{tadpoleO5b}),  but the 
first. 
The latter reads :
\beq
 n_b^1 = -4 + \frac{1}{2\b^1}(1 - 2 \epsilon_h N_h) 
\label{sapo2} 
\eeq

In this case the $U(1)$ couplings to twisted RR fields read
\beqa
B_2^{(1)} \wedge c_1 [\frac{({\a} -{\a^2})}{\b^1}](3 F^a + F^d),
&\nonumber\\           
D_2^{(1)}\wedge c_1 [\ep {\tilde \epsilon}][\frac{3}{2\b^1} F^a - 6
n_b^1 F^b + 2 n_c^1 {\tilde \ep} F^c
+ \frac{3}{2\b^1}  F^d ],   &\nonumber\\
E_2^{(1)}  \wedge c_1 [4 \ep {\tilde \epsilon} \b^1 ]F^b .&
\label{sapo3}
\eeqa
The $U(1)$ combination given by
\beq
 Q^l = (Q_a -3 Q_d)
- \frac{3 {\tilde \epsilon}}{2\b^1 n_c^1}  Q_c.
\label{sapo4}
\eeq
is the SM hypercharge if 
\beq
n_c^1 =\ \frac{{\tilde \ep}}{2\b^1}
\label{sapo5}
\eeq
From (\ref{sapo2}) and (\ref{sapo5}) we conclude that $\b^1 = 1/2$.

An alternative solution to the RR tadpole solutions for
 the a3-quiver, we found in table (\ref{sapo1}),
 can be seen in table (\ref{sapo6}).

\begin{table}[htb]\footnotesize
\renewcommand{\arraystretch}{1.8}
\begin{center}
\begin{tabular}{||c||c|c|c||}
\hline
\hline
$N_i$ & $(n_i^1, m_i^1)$ & $(n_i^2, m_i^2)$ & $(n_i^3, m_i^3)$\\
\hline\hline
 $N_a=3$ & $(1/\b^1, 0)$  &
$(1,  \frac{1}{2}{\tilde \epsilon} \epsilon )$ & $\alpha 1_3$  \\
\hline
$N_b=2$  & $(n_b^1, -{\tilde \epsilon} \epsilon \b^1 )$ & $(1, 
\frac{3}{2}{\epsilon}{\tilde \ep})$ & 
${\bf 1}_2$ \\
\hline
$N_c=1$ & $(n_c^1, 3 \ep \b^1)$ &   $(0, \epsilon )$  & 
$1$ \\    
\hline
$N_d=1$ & $(1/\b^1, 0)$ &  $(-1,   \frac{3}{2}\epsilon{\tilde \epsilon} )$  
  & $\alpha^2$  \\\hline
$N_h$ & $(\epsilon_h/ \b^1, 0)$ &  
$(2, 0 )$  
  & $1_{N_h}$ 
\\\hline
\end{tabular}
\end{center}
\caption{\small Alternative general twisted RR tadpole solutions for the
four-stack
$a3$-type quiver of intersecting
D5-branes, giving rise to models with exactly the
standard model gauge group and observable chiral spectrum
at low energies.
The solutions depend 
on two integer parameters, 
$n_b^1$, $n_c^1$,
the NS-background $\beta^1$ and
the phase parameters $\epsilon = {\tilde \epsilon}=\pm 1$, as well as the 
CP phase $\alpha$. 
\label{sapo6}}          
\end{table}

The 
solutions of table (\ref{sapo6}) satisfy all tadpole 
equations, in ({\ref{tadpoleO5b}),  but the 
first. 
The latter reads :
\beq
 n_b^1 = -4 + \frac{1}{2\b^1}(1 - 2 \epsilon_h N_h) 
\label{kara2} 
\eeq

In this case the $U(1)$ couplings to twisted RR fields read
\beqa
B_2^{(1)} \wedge c_1 [\frac{({\a} -{\a^2})}{\b^1}](3 F^a + F^d),
&\nonumber\\           
D_2^{(1)}\wedge c_1 [\ep {\tilde \epsilon}][-\frac{3}{2\b^1} F^a
+ 6
n_b^1 F^b + 2 n_c^1 {\tilde \ep} F^c
- \frac{3}{2\b^1}  F^d ],   &\nonumber\\
E_2^{(1)}  \wedge c_1 [-4 \ep {\tilde \epsilon} \b^1 ]F^b .&
\label{kara3}
\eeqa
From (\ref{kara3}) we conclude that the U(1) combination
\beq
 Q^l = (Q_a -3 Q_d)
- \frac{3 {\tilde \epsilon}}{2\b^1 n_c^1}  Q_c
\label{kara4}
\eeq
is the hypercharge if
\beq
n_c^1 =\ \frac{{\tilde \ep}}{2\b^1}
\label{kara5}
\eeq
From (\ref{kara2}) and (\ref{kara5}) we conclude that $\b^1 = 1/2$. 
That means that the alternative solution of
table (\ref{sapo6}) gives at low energy a SM
vacuum with identical parameter as that of (\ref{sapo1}).
Thus there is no need for us to redefine the background to show the
equivalence between the two vacua.

%%%%%%%%%%%%%%%%%%%%%%%%%%%%%%%%%%%%%%%%%%
%%%%%%%%%%%%%%%%%%%%%%%%
%%%%%%%%%%%%%%%%%%%%%%%%%%%%%%%%%%

\subsection{SM vacua from the a4-Quiver}
In this subsection we examine the derivation of exactly
the SM at
low energies from the embedding of the four stack SM structure of table (1)
in a $Z_3$ quiver of $a4$-type. 
The latter quiver can be seen in figure (2). 
For an alternative RR tadpole solution of the $a4$-quiver see \cite{ibanez1}.

\begin{table}[htb]\footnotesize
\renewcommand{\arraystretch}{1.8}
\begin{center}
\begin{tabular}{||c||c|c|c||}
\hline
\hline
$N_i$ & $(n_i^1, m_i^1)$ & $(n_i^2, m_i^2)$ & $(n_i^3, m_i^3)$\\
\hline\hline
 $N_a=3$ & $(1/\b^1, 0)$  &
$(1,  -\frac{1}{2}{\tilde \epsilon} \epsilon )$
& $\alpha 1_3$  \\
\hline
$N_b=2$  & $(n_b^1, {\tilde \epsilon} \epsilon \b^1 )$
& $(1, -\frac{3}{2}{\epsilon}{\tilde \ep})$ & 
${\bf 1}_2$ \\
\hline
$N_c=1$ & $(n_c^1, 3 \ep \b^1)$ &   $(0, \epsilon )$  & 
$1$ \\    
\hline
$N_d=1$ & $(1/\b^1, 0)$ &  $(1,  \frac{3}{2}\epsilon{\tilde \epsilon} )$  
  & $\alpha$  \\\hline
$N_h$ & $(\epsilon_h/ \b^1, 0)$ &
$(2, 0 )$  
  & $1_{N_h}$ 
\\\hline
\end{tabular}
\end{center}
\caption{\small General twisted RR tadpole solutions for the
four-stack
$a4$-type quiver of intersecting
D5-branes, giving rise to SM models with exactly the
standard model gauge group and observable chiral spectrum
at low energies.
The solutions depend 
on two integer parameters, 
$n_b^1$, $n_c^1$,
the NS-background $\beta^1$ and
the phase parameters $\epsilon = {\tilde \epsilon}=\pm 1$, as well as the 
CP phase $\alpha$. 
\label{rapo1}}          
\end{table}

The solutions satisfying simultaneously the
intersection constraints and the
cancellation of the RR twisted crosscap tadpole cancellation
constraints
are given in parametric form in table (\ref{rapo1}). 
The 
solutions of table (\ref{rapo1}) satisfy all tadpole 
equations in ({\ref{tadpoleO5b}), but the 
first. 
The latter reads :
\beq
 n_b^1 = -4 + \frac{1}{\b^1}(1-  \epsilon_h N_h) 
\label{rapo2} 
\eeq

In this case the $U(1)$ couplings to twisted RR fields read
\beqa
B_2^{(1)} \wedge c_1 [\frac{({\a} -{\a^2})}{\b^1}](3 F^a + F^d),
&\nonumber\\           
D_2^{(1)}\wedge c_1 [\ep {\tilde \epsilon}][-\frac{3}{2\b^1} F^a + 6
n_b^1 F^b + 2 n_c^1 {\tilde \ep} F^c
+ \frac{3}{2\b^1}  F^d ],   &\nonumber\\
E_2^{(1)}  \wedge c_1 [4 \ep {\tilde \epsilon} \b^1 ]F^b .&
\label{rapo3}
\eeqa
From (ref{rapo3}) we conclude that
the $U(1)$ combination 
\beq
 Q^l = (Q_a -3 Q_d)
+ \frac{3 {\tilde \epsilon}}{\b^1 n_c^1}  Q_c.
\label{rapo4}
\eeq
represents the SM hypercharge if the hypercharge condition is
satisfied
\beq
n_c1 =\ -\frac{{\tilde \ep}}{\b^1}.
\label{rapo5}
\eeq
From (\ref{rapo2}) and (\ref{rapo5}) we conclude that $\b^1 = 1$.

An alternative solution to the RR tadpole solutions for
the a4-quiver can be seen in table (\ref{rapo6}).

\begin{table}[htb]\footnotesize
\renewcommand{\arraystretch}{1.8}
\begin{center}
\begin{tabular}{||c||c|c|c||}
\hline
\hline
$N_i$ & $(n_i^1, m_i^1)$ & $(n_i^2, m_i^2)$ & $(n_i^3, m_i^3)$\\
\hline\hline
 $N_a=3$ & $(1/\b^1, 0)$  &
$(1,  \frac{1}{2}{\tilde \epsilon} \epsilon )$ & $\alpha 1_3$  \\
\hline
$N_b=2$  & $(n_b^1, -{\tilde \epsilon} \epsilon \b^1 )$ & $(1, 
\frac{3}{2}{\epsilon}{\tilde \ep})$ & 
${\bf 1}_2$ \\
\hline
$N_c=1$ & $(n_c^1, 3 \ep \b^1)$ &   $(0, \epsilon )$  & 
$1$ \\    
\hline
$N_d=1$ & $(1/\b^1, 0)$ &  $(1,   -\frac{3}{2}\epsilon{\tilde \epsilon} )$  
  & $\alpha$  \\\hline
$N_h$ & $(\epsilon_h/ \b^1, 0)$ &  
$(2, 0 )$  
  & $1_{N_h}$ 
\\\hline
\end{tabular}
\end{center}
\caption{\small Alternative general twisted RR tadpole solutions for the
four-stack
$a4$-type quiver of intersecting
D5-branes, giving rise to to models with the
standard model gauge group and observable chiral spectrum
at low energies and also some tachyon scalars. The tachyon directions may 
always be lifted to positive values. 
The solutions depend 
on two integer parameters, 
$n_b^1$, $n_c^1$,
the NS-background $\beta^1$ and
the phase parameters $\epsilon = {\tilde \epsilon}=\pm 1$, as well as the 
CP phase $\alpha$. 
\label{rapo6}}          
\end{table}
The 
solutions of table (\ref{rapo6}) satisfy all tadpole 
equations in ({\ref{tadpoleO5b}), but the 
first. 
The latter reads :
\beq
 n_b^1 = -4 + \frac{1}{\b^1}(1-  \epsilon_h N_h) 
\label{rapo7} 
\eeq
 In this case the $U(1)$ couplings to twisted RR fields read
\beqa
B_2^{(1)} \wedge c_1 [\frac{({\a} -{\a^2})}{\b^1}](3 F^a + F^d),
&\nonumber\\           
D_2^{(1)}\wedge c_1 [\ep {\tilde \epsilon}][-\frac{3}{2\b^1} F^a + 6
n_b^1 F^b + 2 n_c^1 {\tilde \ep} F^c
+ \frac{3}{2\b^1}  F^d ],   &\nonumber\\
E_2^{(1)}  \wedge c_1 [4 \ep {\tilde \epsilon} \b^1 ]F^b .&
\label{rapo8}
\eeqa
The combination of the $U(1)$'s which remains light at low 
energies is identical to (\ref{rapo4}).
Also the hypercharge condition in this case is identical
to (\ref{rapo5}). In addition $\b^1 = 1$.

%%%%%%%%%%%%%%%%%%%%%%%%%%%%%%%%%%%%%%%%%%%%
%%%%%%%%%%%%%%%%%%%%%%%%%%%%%%%%%%
%%%%%%%%%%%%%%%%%%%%

\subsection{Effective SM Theories from the Qi and ai Quivers}

As we have already pointed out the theories described by the
$Qi$-, $ai$-quivers, $i=1,..,4$, and seen in figures (1), (2), 
are in one to one correspondence.
Lets us show that the theories described by
these quivers are equivalent. The correspondence is
\beq
a_1 \Leftrightarrow Q_1, \  a_2 \Leftrightarrow Q_2, \
a_3 \Leftrightarrow Q_3, \ a_4 \Leftrightarrow Q_4,
\label{equiv}
\eeq

Lets us discuss the
relation  $a_1 \Leftrightarrow Q_1$.
A direct comparison of the theories between the two quivers, e.g.
using the relations (\ref{sda12el}), (\ref{hyper1}), 
(\ref{mashyper1}) for the Q1-quiver and
(\ref{dasa1}), (\ref{dasa4}), (\ref{dasa5}) for the a1-quiver, 
shows that the first tadpole condition,
and the hypercharge of the initially different
classes of models are identical \footnote{Also identical
appear
to be the couplings to RR fields}. That means that the
low energy content of the the two quivers is identical as
matter as it concerns the fermion sector; the low energy 
effective theories are identical.
For the two classes of theories to be actually equivalent
we have to show that their scalar sector as well the couplings of the antisymmetric
fields to various two form RR fields are identical.
This is obvious, since the data
that we have to take into account to calculate the
scalar sector are based on the selection of a particular set of
wrapping numbers satisfying the first tadpole
condition and the hypercharge condition. The latter equations
are identical, thus the
 scalar sector is identical for
both theories. Also the various couplings to twisted RR fields are
the same. In the following, we compare the theories coming from their general
RR tadpole solutions and not their alternative solutions seen in tables
(8), (10), (12), (16), (18), (20).

Identical relations hold for the other pair
$a_2 \Leftrightarrow Q_2$.
\newline
For the pair
$a_3 \Leftrightarrow Q_3$ we need to transform
\beq 
{\ep}  \stackrel{Q3 \rightarrow a_3}{\rightarrow} {\tilde
\ep},
\label{repla1}
\eeq
in the $Q_3$ quiver relations to show that the two class of SM
theories are identical.
\newline
For the pair
$a_4 \Leftrightarrow Q_4$ we need to transform
\beq 
{\ep}  \stackrel{Q4 \rightarrow a_4}{\rightarrow} -{\tilde
\ep},
\label{repla2}
\eeq
in the $Q_4$ quiver relations to show that the two class of SM
theories are identical.
We note that the correspondence between the `reflected' quivers $Qi$
and their `images' $ai$ is most easily seen in the solutions
presented in this work. \newline
Alternatively, we could try
to prove the equivalence e.g. of the effective theories for the 
$Qi$-quivers presented in this work and those one's that follow from 
the RR tadpole solutions for the $ai$-quivers,
that appeared in \cite{ibanez1}.
In this case the transformations needed for the two theories to be 
proven equivalent are more complicated,
as can be seen by the field redefinition,
\beq
\b^1 \stackrel{Q2 \rightarrow 4.a_1}{\rightarrow} {\tilde \ep} \b^1 , \ 
{\tilde \ep} n_d^1 \stackrel{Q1 \rightarrow 4.a_1}{\rightarrow} n_d^1 \ .
\label{isa1}
\eeq
We note that we have used the RR tadpole solutions
of the $4.a_1$-quiver (the a2-quiver of the present work)
of \cite{ibanez1} and also relations (\ref{dasaa1}), (\ref{dasaa3}), (\ref{dasaa4}),  
 (\ref{dasaa5}), of the present work.

%%%%%%%%%%%%%%%%%%%%%%%%%%%%%%%%%%%%%%%%%%%%%%%             
%%%%%%%%%%%%%%%%%%%%%%%%%%%%%%%%%%%%
%%%%%%%%%%%%%%%%%%%%%%

\section{Standard Model Vacua from
Five-Stack Quivers }

%
%
%
%

%\eps%%%%%%%%%% Figure here%%%%%%%%%%%%%%%%%%%%%%%%%%%%%%%%%%%%%%%%%
\begin{figure}
\begin{center}
\centering
\epsfysize=8cm
\leavevmode
%\hspace*{0in}\vspace*{.2in}
\epsfbox{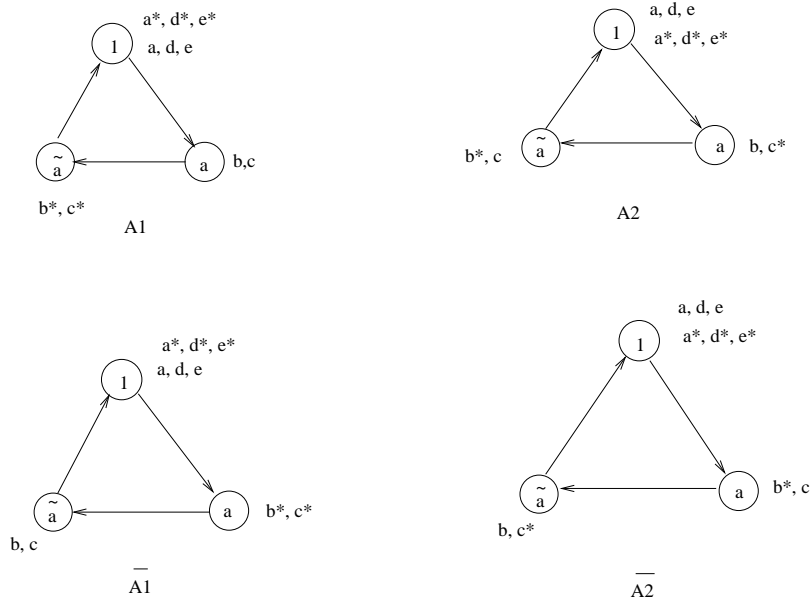}
\end{center}
\caption[]{\small
Assignment of SM embedding in  
configurations of five stacks of D5 branes depicted by
$Z_3$ quiver diagrams. At low energy we get the SM.
Note that ${\tilde \alpha} = \alpha^{-1}$.
 }
\end{figure}
%%%%%%%%%%%%%%%%%end of figure %%%%%%%%%%%%%%%%%%%%%%%%%%%%%%%%%%%

In this chapter, we will discuss the basic
elements of the
five stack SM quiver embedding classification, namely the classification
of the different quivers describing the SM embedding 
of chiral fermions of
different intersecting D5 branes that appear in table (2).
We discuss for simplicity the embedding in $Z_3$ quivers.
 For each class of models, we present the general
 class of solutions to the twisted RR tadpole
solutions, the two massless
$U(1)$'s, surviving the Green-Schwarz mechanism,
as well the
hypercharge condition on the parameters of the tadpole
solutions, that is needed for one of the massless $U(1)$'s to be
identified with the hypercharge
generator of the SM.
We will explicitly describe the 
scalars generically present in the models only for 
the $A_1$ quiver. The scalars
generically
present in these classes of models are expected
to get massive from the loop corrections (\ref{oneloop}).
Also, the extra, beyond the hypercharge combination, anomaly 
free $U(1)$'s get massive by the use of the some charged scalars 
getting a vev and/or their non-zero coupling to twisted RR fields.
Thus in the classes of theories coming from
five stack quivers,
the SM survives in general down to low energy. In general one expects 
to find always tachyonic scalars that might look problematic.
However, the tachyonic diretions might be lifted by loop corrections
(see section (5.3)) or
 as we will show in section
7, the present SM-like structure of the models is not permanent as using
brane recombination the models flow to the associated four stack
quiver \footnote{
which is obtained by naively deleting the e-brane from the five stack 
quiver}. In this procedure we have assumed that brane recombination 
e.g. $d + e \rightarrow {\tilde d}$ takes place .

The multiparameter
tadpole solutions appearing in tables (\ref{um1}),
 (\ref{spectrum1oi}),
(\ref{spectrum2}), (\ref{spectrum2oip2}),
 represent deformations of
the D5-brane branes, of table (2), intersecting at angles,
 within the same homology class of the
factorizable two-cycles.

\subsection{SM Vacua from the
Five-Stack A1-type quiver}

The first class of models, that give the SM
at low energy, are associated
with the A1-quiver and can be seen in figure (3).
% We put quotes in the SM-like title as in section 7 we will lift
%this impression (also for the rest of five and six stack quivers).

\begin{table}[htb]\footnotesize
\renewcommand{\arraystretch}{1.8}
\begin{center}
\begin{tabular}{||c||c|c|c||}
\hline
\hline
$N_i$ & $(n_i^1, m_i^1)$ & $(n_i^2, m_i^2)$ & $(n_i^3, m_i^3)$\\
\hline\hline
 $N_a=3$ & $(n_a^1, -\epsilon \b^1)$  &
$(3,  \frac{1}{2}{\tilde \epsilon} \epsilon )$ & $1_3$  \\
\hline
$N_b=2$  & $(1/\b_1, 0)$ & $({\tilde \epsilon}, \frac{1}{2}{\epsilon})$ & 
$\alpha 1_2$ \\
\hline
$N_c=1$ & $(1/\b_1, 0)$ &   $(0, -{\epsilon})$  & 
$\alpha$ \\    
\hline
$N_d=1$ & $(n_d^1, -2 \epsilon \b^1)$ &  $(1,  - \frac{1}{2}\epsilon 
{\tilde \epsilon})$  
  & $1$  \\\hline
$N_e = 1$ & $(n_e^1, -\epsilon \b^1)$ &  
$(1, - \frac{1}{2}\epsilon 
{\tilde \epsilon} )$  
  & $1$ 
\\\hline
$N_h$ & $(\epsilon_h/ \b^1, 0)$ &  
$(2, 0 )$  
  & $1_{N_h}$ 
\\\hline
\end{tabular}
\end{center}
\caption{\small General tadpole solutions for the
five-stack
A1-type quiver of intersecting
D5-branes, giving rise to SM at low energies.
The solutions depend 
on three integer parameters, 
$n_a^1$, $n_d^1$, $n_e^1$,
the NS-background $\beta^1$ and
the phase parameters $\epsilon = {\tilde \epsilon}=\pm 1$, as well as the 
CP phase $\alpha$. 
\label{um1}}          
\end{table}

The solutions satisfying simultaneously the
intersection constraints and the
cancellation of the RR crosscap tadpole constraints
are given in parametric form in table (\ref{um1}). 
These solutions represent
the most general solution of the twisted RR tadpoles.

The 
solutions of table (\ref{um1}) satisfy all
tadpole equations in ({\ref{tadpoleO5b}),
but the
first. The latter becomes :
\beq
9 n_a^1-\frac{\tilde \epsilon}{\b^1} + n_d^1 + n_e^1 +
\frac{2\epsilon_h N_h}{\b^1} =-8
\label{idio1}
\eeq
Note that we had added the presence of extra $N_h$ branes. 
Their contribution to the RR tadpole conditions is best 
described by placing them in the three-factorizable cycle 
\beq 
N_h \ (\epsilon_h/\b_1, 0) \ (2, 0)1_{N_h}
\label{sda12}
\eeq
The presence
of $N_h$ D5-branes, give an extra $U(N_h)$ gauge group,
which don't contribute to the rest of the tadpoles and
intersection constraints. Thus in terms of the
low energy theory their presence has no effect.
For the A1-quiver the constraints from $U(1)$
anomaly cancellation
appear into three terms
\beqa
B_2^{(1)} \wedge c_1 \left( \frac{ 2 {\tilde \epsilon}
(\a - \a^2 ) }{\b^1 }
\right)F^b,&\nonumber\\       
D_2^{(1)} \wedge c_1 \left( \ep {\bar \ep}
(3 n_a^1 F^a - n_d^1 F^d -n_e^1  F^e)
-\frac{\ep}{\b^1}(F^b-F^c) \right),&\nonumber\\
E_2^{(1)}  \wedge c_1 [-2 \ep \b^1]
[9 F^a + 2F^d + F^e ].&
\label{rr1}
\eeqa
A U(1), which is orthogonal to the U(1)'s coupled to the
RR fields in(\ref{rr1}), is given by
\beq
Q^l = (Q_a -3 Q_d -3 Q_e)
-3 {\tilde \epsilon} \b^1 ( n_a^1 + n_d^1 + n_e^1 ) Q_c.   
\label{hyper}
\eeq
The subclass of tadpole solutions of (\ref{hyper}) having
the SM hypercharge
assignment at low energies is exactly the one which is
proportional to
(\ref{hyper1231}). It
satisfies the condition, 
\beq
{\tilde \epsilon} \b_1 \ ( n_a^1 +\  n_d^1 
+\  n_e^1) = 1.
\label{mashyper}
\eeq
Summarizing, as long as (\ref{mashyper}) holds
(\ref{hyper}) is the hypercharge generator
of the SM. Thus at low energies
we get the
chiral fermion content of the SM that gets localized
according
to open string sectors
appearing in table (2).
We note that there is one extra anomaly free $U(1)$
beyond the
hypercharge
combination which is
\beq
Q^{(5)} =\ Q_d -2 Q_e + Q_c
\label{extra}
\eeq
The latter U(1) is orthogonal to the U(1)'s seen in
(\ref{rr1}), if
\beq
-n_d^1 + 2 n_e^1 + \frac{{\tilde \epsilon}}{\beta_1} = 0
\label{last}
\eeq
The following choice of wrapping numbers is consistent
with the constraints (\ref{idio1}), (\ref{mashyper}),
(\ref{last})
\beq
\beta_1 =1, \ n_d^1 =3, \ n_e^1 = 1,\ n_a^1 = -3, \
{\tilde \epsilon} = 1, \ \epsilon = 1, \ \epsilon_h = -1,
\ N_h = 8
\label{foloch}
\eeq

One comment is in order. 
The A1-quiver five stack diagram is similar to
the a2-four stack
quiver of table (2). Their only difference stems from
 the
addition of the 5-th stack of $e$-branes in the quiver node which transforms
with CP phase $\bf 1$. 
By taking the limit of vanishing $e$ brane,
at the same time taking the limit
where the contributions of the $e$-brane to the
hypercharge generator
vanish, e.g. $n_e^1 \rightarrow 0$, we recover
the corresponding four
stack hypercharge that is associated with the a2-quiver
in (\ref{dasa4}) only if the moduli tadpole parameters
of the five stack A1-quiver SM-like model change
as follows:
\beq
{\tilde \ep} \b^1 \rightarrow \b^1, \ n_d^1
\rightarrow {\tilde \ep} n_d^1, \
\ep_h \rightarrow \frac{\ep_h}{\tilde \ep}.
\label{change1}
\eeq
Thus quivers on different stacks appear to have the
same effective low energy theory, the SM, at low energy.
Actually, the tadpole constraint at the limit $n_e^1=0$
are the same for both theories (see (\ref{idio1}),
(\ref{dasaa1})).
However, it appears at this stage  that their scalar
sectors are not identical. It will not be the case.               
Later in section 7 we will see that all five stack
quivers are identical theories to their associated
four stack quivers thus they have only the SM at low energy, also
suggesting that string vacua in the present models
are continuously connected.

\subsection{Higgs sector of the Five-Stack A1-quiver}

A general comment is in order.
The A1 five-stack quiver describes the embedding of the
SM chiral fermions of table 2, in the A1-quiver.
However, all quivers described in this work, are
constructed be deforming around the QCD intersections
$I_{ab}$, $I_{ab^{\star}}$, $I_{ac}$, $I_{ac^{\star}}$.
Thus the Higgs sector which is coming from b,c, branes
is of common origin to the present D5-branes and also in
the corresponding five-stack D6 models of \cite{kokos2}.
This is easily understood as it is the SM embedding
of table 2 which is embedded in either the present
D5-brane models or the D6-brane models of \cite{kokos2}.
For the sake of
completeness we will repeat some aspects of
this mechanism.

The set of Higgs present in the models get localized
on the $bc$, $bc^{\star}$ intersections. The Higgs
available in the models can be seen in
\cite{kokos2}
Apriori, this set of Higgs ,
are part of the massive spectrum of fields
localized in the intersections $bc, bc^{\star}$.
However,
the Higgses $H_i$, $h_i$ become massless, and effectively tachyonic
triggering electroweak symmetry breaking, 
 by varying
the distance along the
first tori between the $b, c^{\star}$, $b, c$ branes
respectively.
Similar set of Higgs fields appear in the four 
\cite{luis1}, five \cite{kokos2}, six stack $D6$-SM's
\cite{kokos3} as
well the Pati-Salam four
stack D6-models \cite{kokos4}.

For the models presented in this work,
the number of complex scalar doublets
is equal to the non-zero
intersection number product between the $bc$, $bc^{\star}$
branes in the
second complex plane. Thus
%\beqa
$n_{H^{\pm}} =\  I_{bc^{\star}}\ =\ 1$,
%&
$n_{h^{\pm}} =\  I_{bc}\ =\ 1$  .
%\label{inter1}
%\eeqa
As it have been discussed before
in \cite{luis1, kokos2, kokos3, kokos4}
the presence of scalar doublets $H^{\pm}, h^{\pm}$,
can be seen as coming from a field theory mass matrix 
involving the fields $H_i$ and $h_i$ defined as
\beq
H^{\pm}={1\over2}(H_1^*\pm H_2); \ h^{\pm}={1\over2}(h_1^*\pm h_2)  \ .
\label{presca}
\eeq
and
making the effective potential which 
corresponds to the spectrum of Higgs scalars to be given by
\beqa
V_{Higgs}\ =\ m_H^2 (|H_1|^2+|H_2|^2)\
+\ m_h^2 (|h_1|^2+|h_2|^2)\  \nonumber\\
+\ m_B^2 H_1H_2\
+\ m_b^2 h_1h_2\
+\ h.c.,
\label{Higgspot}
\eeqa
where
\beqa
 {m_h}^2 \ =\ { M_s^2 { Z_1^{(bc)}}\over {4\pi ^2}}\ & ; & \ 
{m_H}^2 \ =\ {M_s^2 {Z_1^{(bc^*)}}\over {4\pi ^2}}\nonumber \\
m_b^2\ =\ \frac{M_s^2}{2\alpha '}|\vartheta_2^{(bc)}| \ & ;&
m_B^2\ =\ \frac{M_s^2}{2\alpha '}|\vartheta_2^{(bc^*)}|,
\label{masilla}
\eeqa
and
\beq
\vartheta_2^{(bc)} \ = \frac{\pi}{2} + tan^{-1}\left( \frac{U^2}{2}\right), \
\vartheta_2^{(bc^{\star})} \ = \frac{\pi}{2} - tan^{-1}\left( \frac{U^2}{2}\right),
\label{gonies}
\eeq
and we have 
made the
choice of parameters (\ref{foloch}).

We note that the $Z_1$ is a free parameter, a moduli,  
and thus can become very small 
in relation to the Planck scale.

\subsection{Scalar sectors of the Five-Stack A1-quiver}

Scalars appear in the theory in sectors 
that share the same CP phase. Thus for example in the
A1-quiver
that we will examine in detail, the lightest
scalars appear from the nine intersections
$a {a^{\star}}$,  $d {d^{\star}}$, $e {e^{\star}}$,  $a d$, $a {d^{\star}}$,
$ae$,  $a {e^{\star}}$, $de$, $d {e^{\star}}$.
They are listed in table (\ref{scalartable}). We note
that the subscript
in the scalar representations denotes the hypercharge.

Let us make the
choice (\ref{foloch}).
Within this choice, the angles that the branes $a$, $b$, $c$, $d$, $e$
 form with the
orientifold plane appear as follows :
  
\beqa
\vartheta^1_a =  -\pi +
tan^{-1}\left(\frac{U^1}{5}\right), & & \vartheta^1_d =
- tan^{-1}
\left( \frac{2U^1}{3}\right),  \\
\vartheta^2_a =  -tan^{-1}\left(\frac{U^2}{6} \right), & & 
\vartheta^2_d =  \tan^{-1}\left( \frac{U^2}{2}  \right)  ,\\
\vartheta^1_e =  -tan^{-1}\left(U^1 \right), & & 
\vartheta^2_e =  \tan^{-1}\left( \frac{U^2}{2}  \right)  .\\
\label{anglesorien}
\eeqa

\begin{table}
[htb]\footnotesize
\renewcommand{\arraystretch}{1.6}
\begin{center}
\begin{tabular}{||c||c||l||}
\hline
\hline
Sector & Representation & $\a^{\prime} \cdot mass^2$ \\
\hline\hline
 $a {a^{\star}}$ & $10(3, 1)_{1/3} + 8(6, 1)_{-1/3}    $  &
$\pm  \left(-1 + \frac{1}{\pi} tan^{-1}(
\frac{U^1}{3}) -
\frac{1}{\pi}tan^{-1}(\frac{U^2}{6})\right) $ \\
\hline
$d {d^{\star}}$  & $4(1, 1)_{-1}$ &
$\pm \frac{1}{\pi} \left( -
tan^{-1}(\frac{2U^1}{3}) +
tan^{-1}(\frac{U^2}{2})\right)$ \\
\hline
$e {e^{\star}}$  & No \ scalars \ present &  \\
\hline
$ad$ & $18(3,1)_{-2/3}$ &   $\pm \frac{1}{2}
[ \frac{1}{\pi}
 tan^{-1}(\frac{U^1}{3}) - \frac{1}{\pi}
tan^{-1}(\frac{U^2}{6}) +
 \frac{1}{\pi} tan^{-1}(\frac{U^2}{2})$ \\
&& $-\frac{1}{\pi} tan^{-1}(\frac{2 U^1}{3})]$
\\\hline
$a {d^{\star}}$ & $3(3,1)_{-1/3}$ &  
$  \pm \frac{1}{2\pi}[ -\pi + 
 tan^{-1}(\frac{U^2}{2}) -
tan^{-1}(\frac{2U^1}{3})
+ tan^{-1}(\frac{U^1}{3})$
\\
 && $- tan^{-1}(\frac{U^2}{6})]$
\\\hline
$ae^{\star}$ & $2(3,1)_{1/3}$ &
$  \pm \frac{1}{2\pi}[ -\pi +
 tan^{-1}(\frac{U^1}{3})
+tan^{-1}(\frac{U^2}{2}) -
tan^{-1}(\frac{U^2}{6}) $\\
&&$- tan^{-1}(U^1) ] $
\\\hline
$ae$ & $8(3,1)_{-2/3}$ &  
$ \pm \frac{1}{2\pi}
[-\pi - tan^{-1}(\frac{2U^2}{3})
+tan^{-1}(\frac{U^2}{2}) +
tan^{-1}(\frac{U^1}{3}) $ \\
&& $+tan^{-1}(U^1)]$ \\\hline
$d e$ & $(1,1)_{0}$ &  
$\pm \frac{1}{2\pi}\left(
-tan^{-1}(\frac{2U^1}{3}) +
tan^{-1}(U^1)  \right)
+ \frac{Z^2}{4\pi^2}$
\\\hline
$d {e^{\star}}$ & $5(1,1)_{0}$ &  
$ \pm \frac{1}{2\pi}\left( -tan^{-1}(U^1)
-tan^{-1}(\frac{2U^1}{3}) +
2 tan^{-1}(\frac{U^2}{2}) \right) $  
\\\hline
\end{tabular}
\end{center}
\caption{\small 
Lightest scalar excitations for the A1-quiver models. The subscript
denotes the hypercharge.
\label{scalartable}}          
\end{table}

As can be seen from table (\ref{scalartable}) there is a number of 
scalars present in the models.
These scalars may receive, as the models are non-SUSY,
important higher loop corrections which are stronger for coloured scalars.
As it has been emphasized before \cite{allo3, ibanez1}, the 
precise form of the one loop corrections takes the form
\beq
\Delta m^2 (\mu)= \sum_a \frac{4 C^a_F \a_a(M_s)}{4\pi}M_s^2 f_a \log(M_s/\mu) 
 \Delta M^2_{KK/W},
\label{oneloop}
\eeq
where $t=2 \log(M_s/\mu)$ and $b_a$ the one-loop $\beta$-function 
coefficients, $C^a_F$ the quadratic Casimir in the fundamental representation
and the sum over the index $a$ runs over
the different gauge representation
of the individual scalar fields.
Also, $\Delta M^2_{KK/W}$ receives contributions from Kaluza-Klein,
winding and string excitations lighter than the
string
scale \footnote{Where $f_a = \frac{2 + b_a
\frac{\alpha_a (M_s)}{4 \pi} t}{1 + b_a \frac{\alpha_a (M_s)}{4 \pi} t
}$, and $b_a$ the coefficients of the one-loop
$\beta$-functions.  }.
Thus all scalars in the models are expected to
receive non-zero
contributions to
their mass, the latter being driven to
 positive values
for these scalars. 

The complete absence of tachyons in the models is difficult
to be maintained just by fixing the complex structure parameter.
E.g. the choice
\beqa
\frac{2U^1}{3} = \frac{U^2}{2},
\label{tacyo}
\eeqa
sends to zero the mass of the $dd^{\star}$ scalars.
Also the mass of the de-sector singlet
scalars may become positive by varying the distance
between the branes. In addition, all scalars receive corrections
from (\ref{oneloop}).
 Thus the A1-models may receive important loop corrections which may lift
 their tachyonic directions. The latter argument will be further justified in
section 7, where the models will be shown to be equivalent to the four stack
$a1$ quivers under brane recombination.
A comment is in order.
From the $ee^{\star}$ sector there are no scalars present, as 
can be seen from the fact that the ``existing'' scalars
transform in the antisymmetric representation of the $U(1)$ group.

\subsection{ Vacua from
the Five-Stack ${\overline{A1}}$-type quiver}

In this section we examine the derivation of a class of
Standard models with exactly the SM
at low energies, from the embedding of the five stack
SM structure of table (2)
in the quiver of ${\overline{A1}}$-type.

\begin{table}[htb]\footnotesize
\renewcommand{\arraystretch}{1.8}
\begin{center}
\begin{tabular}{||c||c|c|c||}
\hline
\hline
$N_i$ & $(n_i^1, m_i^1)$ & $(n_i^2, m_i^2)$ & $(n_i^3, m_i^3)$\\
\hline\hline
 $N_a=3$ & $(n_a^1, \epsilon {\tilde \ep}\b^1)$  &
$(3,  \frac{1}{2}{\tilde \epsilon} \epsilon )$ & $1_3$  \\
\hline
$N_b=2$  & $(1/\b_1, 0)$ & $(1, \frac{1}{2}{\epsilon}{\tilde \ep})$ & 
$\alpha^2 {\bf 1}_2$ \\
\hline
$N_c=1$ & $(1/\b_1, 0)$ &   $(0, -{\epsilon{\tilde \ep}})$  & 
$\alpha^2$ \\    
\hline
$N_d=1$ & $(n_d^1, 2 \epsilon \b^1)$ &  $({\tilde \ep},  - \frac{1 
}{2}\epsilon)$  
  & $1$  \\\hline
$N_e = 1$ & $(n_e^1, \epsilon \b^1)$ &  
$({\tilde \ep}, -\frac{1 }{2}\epsilon )$  
  & $1$ 
\\\hline
$N_h$ & $(\epsilon_h/ \b^1, 0)$ &  
$(2, 0 )$  
  & $1_{N_h}$ 
\\\hline
\end{tabular}
\end{center}
\caption{\small General tadpole solutions for the five-stack 
${\overline{A1}}$-type quiver of intersecting
D5-branes, giving rise to exactly the
standard model to low energies.
The solutions depend 
on three integer parameters, 
$n_a^1$, $n_d^1$, $n_e^1$,
the NS-background $\beta^1$ and
the phase parameters $\epsilon = {\tilde \epsilon}=\pm 1$, as well as the 
CP phase $\alpha$. 
\label{spectrum1oi}}          
\end{table}

The solutions satisfying simultaneously the
intersection constraints and the
cancellation of the RR twisted crosscap tadpole cancellation
constraints
are given in parametric form in table (\ref{spectrum1oi}). 
These solutions represent
the most general solution of the RR tadpoles. 
The twisted RR 
solutions of table (\ref{spectrum1oi}) satisfy all tadpole 
equations in ({\ref{tadpoleO5b}), but the 
first. 
The latter reads :
\beq
9 n_a^1-\frac{1}{\b^1} + n_d^1{\tilde \ep} + n_e^1{\tilde \ep} +
\frac{2\epsilon_h N_h}{\b^1} =-8.
\label{katou}
\eeq
Note that we had added the presence of extra $N_h$ branes. 
Their contribution to the RR tadpole conditions is best 
described by placing them in the three-factorizable cycle 
\beq 
N_h \ (\epsilon_h/\b_1, 0)\ (2, 0)1_{N_h} \ .
\label{sda124}
\eeq
The presence of an arbitrary number
of $N_h$ D5-branes, which give an extra $U(N_h)$ gauge group,
don't make any contribution to the rest of the tadpoles and
intersection constraints. Thus in terms of the
low energy theory their presence has no effect.
The $U(1)$ couplings to twisted RR fields read:
\beqa
B_2^{(1)} \wedge c_1 [2  \frac{({\a^2} -{\a})}{\b^1}]
F^b ,&\nonumber\\           
D_2^{(1)}\wedge c_1 [\ep {\tilde \epsilon}][3 n_a^1 F^a -
\frac{1}{\b^1} F^b
+
\frac{1}{\b^1} F^c
- {\tilde \epsilon}  n_d^1 F^d    
- {\tilde \epsilon}  n_e^1 F^e ],   &\nonumber\\
E_2^{(1)}  \wedge c_1 [2 \ep {\tilde \epsilon} \b^1 ][9 F^a + 2 F^d + F^e ].&
\label{rrkl2as}
\eeqa

The combination of the $U(1)$'s which remains light at low 
energies is given by
\beq
 Q^l = (Q_a -3 Q_d -3 Q_e )
- 3 \b^1 (n_a^1 + {\tilde \epsilon}  n_d^1 +{\tilde \epsilon} n_e^1) Q_c.
\label{hyper2456}
\eeq
In addition, the hypercharge condition in this case is given uniquely by
\beq
\b^1 (n_a^1 + {\tilde \epsilon}n_d^1 + {\tilde \epsilon}n_e^1) = \ 1.
\label{cond233}
\eeq
The, other than hypercharge, anomaly free,
$U(1)$ is given by
 \beq
 Q^{(5)} =
(Q_d -2 Q_e + Q_c ) .
\label{hyper2asd1}
\eeq
Using the consition coming from the requirement
the fifth U(1) to survive massless the Green-SChwarz
mechanism, e.g. if it is orthogonal to the
model dependent U(1) coupled to the RR field $D_2^{(1)}$
give us $(1/\beta_1) -{\tilde \epsilon} n_d^1 +
2  {\tilde \epsilon} n_e^1 = 0$. Wrappings consistent
with this constraint and (\ref{katou}) are given by
\beq
n_d= 4, \ n_e = 3, \ {\tilde \ep} = -1, \ \beta_1 =1/2, \ \ep = 1, \
\ep_h = 1, \ N_h = 20, \
n_a^1 = \ 9.
\label{setlet1}
\eeq
If the scalar fields $(1,1)_0$, which are localized in
the $de$-sector with mass given by \footnote{where ${\bar Z}^{(2)}$ is 
the transverse distance
between the
branes $d$, $e$ in the 
second tori}
\beq
\alpha^{\prime} m_{de}^2 = \ \pm \frac{1}{2\pi}|
  tan^{-1}(\frac{U^1}{4}) -
tan^{-1}(\frac{U^1}{6})| + \frac{{\bar Z}^{(2)}}{4 \pi^2}, 
\label{reads1}
\eeq
 receive a vev then they may break the $ Q^{(5)}$ symmetry.
 All scalars that may ne present in the models may receive important loop
 corrections frm (\ref{oneloop}) which may lift their tachyonic directions.
Thus at                    
low energies only the SM survives.
\subsection{SM Vacua from the Five-Stack
A2-type quiver}

In this section we examine the derivation of a class of
theories with exactly the SM at low energies. They come from the
embedding of the five stack SM structure of table (2)
in the quiver of $A2$-type. 

\begin{table}[htb]\footnotesize
\renewcommand{\arraystretch}{1.8}
\begin{center}
\begin{tabular}{||c||c|c|c||}
\hline
\hline
$N_i$ & $(n_i^1, m_i^1)$ & $(n_i^2, m_i^2)$ & $(n_i^3, m_i^3)$\\
\hline\hline
 $N_a=3$ & $(n_a^1, -\epsilon \b^1)$  &
$(3,  \frac{1}{2}{\tilde \epsilon} \epsilon )$ & $1_3$  \\
\hline
$N_b=2$  & $(1/\b_1, 0)$ & $({\tilde \epsilon}, \frac{1}{2} {\epsilon})$ 
&$ \alpha 1_2$ \\
\hline
$N_c=1$ & $(1/\b_1, 0)$ &   $(0, {\epsilon})$  & 
$\alpha^2$ \\    
\hline
$N_d=1$ & $(n_d^1, -2 \epsilon \b^1)$ &  $(1,  - \frac{1}{2}\epsilon 
{\tilde \epsilon})$  
  & $1$  \\\hline
$N_e = 1$ & $(n_e^1, -\epsilon \b^1)$ &  
$(1, - \frac{1}{2}\epsilon 
{\tilde \epsilon} )$  
  & $1$ 
\\\hline
$N_h$ & $(\epsilon_h/ \b^1, 0)$ &  
$(2, 0 )$  
  & $1_{N_h}$ 
\\\hline
\end{tabular}
\end{center}
\caption{\small General tadpole solutions for the five-stack 
A2-type quiver of intersecting
D5-branes, giving rise to the SM
at low
energies.
The solutions depend 
on three integer parameters, 
$n_a^1$, $n_d^1$, $n_e^1$,
the NS-background $\beta^1$ and
the phase parameters $\epsilon = {\tilde \epsilon} =\ \pm 1$, as well as the
CP phase $\alpha$. 
\label{spectrum2}}          
\end{table}

With the choice of tadpole solutions of
table (\ref{spectrum2}) all tadpole
solutions in ({\ref{tadpoleO5b}), but the first, are satisfied,
the latter giving
\beq
9 n_a^1 -\frac{{\tilde \epsilon}}{\b^1} + n_d^1 + n_e^1
+ \frac{2\epsilon_h N_h}{\b^1}= -8.
\label{additio2}
\eeq
The $U(1)$ couplings to twisted RR fields read:
\beqa
B_2^{(1)} \wedge c_1 [- 2 {\tilde \epsilon} \frac{\a -{\a}^2}{\b^1}]
F^b ,&\nonumber\\           
D_2^{(1)}\wedge c_1 [  {\tilde \epsilon} {\epsilon}][
3 n_a^1 F^a -
\frac{{\tilde \epsilon}}{\b^1} F^b
-
\frac{{\tilde \epsilon}}{\b^1} F^c
-   n_d^1 F^d    
-  n_e^1 F^e ],   &\nonumber\\
E_2^{(1)}  \wedge c_1 (-2  {\epsilon} \b^1 )[9 F^a + 2 F^d + F^e ].&
\label{rrkl2}
\eeqa
The combination of the $U(1)$'s which remains light at low 
energies is given by
\beq
 Q^l = (Q_a -3 Q_d -3 Q_e )
+ 3 {\tilde \ep}\b^1 (n_a^1 + n_d^1 + n_e^1) Q_c   .
\label{hyper2}
\eeq
while the hypercharge condition in this case is given uniquely by
\beq
{\tilde \ep}\b^1 (n_a^1 + n_d^1 + n_e^1) = \ -1.
\label{cond2}
\eeq
The fifth anomaly free $U(1)$
given by
 \beq
 Q^{(5)} = (Q_d -2 Q_e + Q_c )
\label{hyper20}
\eeq          
survives massless the generalized Green-Schwarz
mechanism if $-n_d^1 + 2n_e^1 -{\tilde \epsilon}/\beta^1 =0$.
Wrapping numbers consistent with this constraint and
(\ref{additio2}) are given by
\beq
n_e^1 = 1, \ n_a^1 = -3, \ n_d^1 = 4, \ \b_1 =1/2, \
\ep =1, \ {\tilde \ep}=1,  \ep_h = 1, N_h =3.
\label{setwra1}
\eeq  
Then U(1) gauge boson associated with 
(\ref{hyper20}) gets broken with the help of
$de$-sector scalars
\beq
(1,1)_{0}
\label{data1}
\eeq
with
\beq
\alpha^{\prime} m_{de}^2 = \ \pm \frac{1}{2\pi}\left(
- \frac{1}{\pi}tan^{-1}(\frac{U^1}{4})
+  
tan^{-1}( \frac{U^1}{2})\right)
+ \frac{{\bar Z}^2}{4 \pi^2},
\label{reads11}
\eeq
where ${\bar Z}$ is the distance between the branes $d$, $e$ in the 
second tori.
The latter scalars receive a vev and
break the $Q^{(5)}$
symmetry leaving only the SM gauge symmetry at low energies.
Any tachyonic directions that may be present in the models may
receive important corrections
from the couplings (\ref{oneloop}).

%
%
%%%%
%%%%%%%%%%

\subsection{SM Vacua from the
Five-Stack ${\overline{A2}}$-type quiver}

\begin{table}[htb]\footnotesize
\renewcommand{\arraystretch}{1.8}
\begin{center}
\begin{tabular}{||c||c|c|c||}
\hline
\hline
$N_i$ & $(n_i^1, m_i^1)$ & $(n_i^2, m_i^2)$ & $(n_i^3, m_i^3)$\\
\hline\hline
 $N_a=3$ & $(n_a^1, \epsilon {\tilde \ep }\b^1)$  &
$(3,  \frac{1}{2}{\tilde \epsilon} \epsilon )$ & $1_3$  \\
\hline
$N_b=2$  & $(1/\b_1, 0)$ & $(1, \frac{1}{2} {\tilde \epsilon}{\epsilon})$ 
&$ \alpha^2 {\bf 1}_2$ \\
\hline
$N_c=1$ & $(1/\b_1, 0)$ &   $(0, {\epsilon} {\tilde \ep }  )$  & 
$\alpha$ \\    
\hline
$N_d=1$ & $(n_d^1, 2 \epsilon \b^1)$ &  $({\tilde \ep },  - \frac{1} 
{2}\ep  )$  
  & $1$  \\\hline
$N_e = 1$ & $(n_e^1, \epsilon \b^1)$ &  
$({\tilde \ep }, - \frac{1}{2} \ep )$  
  & $1$ 
\\\hline
$N_h$ & $(\epsilon_h/ \b^1, 0)$ &  
$(2, 0 )$  
  & $1_{N_h}$ 
\\\hline
\end{tabular}
\end{center}
\caption{\small General tadpole solutions for the five-stack 
$\overline{A2}$-type quiver of intersecting
D5-branes, giving rise to the SM at low
energies.
The solutions depend 
on three integer parameters, 
$n_a^1$, $n_d^1$, $n_e^1$,
the NS-background $\beta^1$ and
the phase parameters $\epsilon = {\tilde \epsilon}=\pm 1$, as well as the 
CP phase $\alpha$. 
\label{spectrum2oip2}}          
\end{table}

In this section we examine the derivation of
a class of models, with exactly the SM at
low energies, from the embedding of the five stack
SM structure of table (2)
in the quiver of ${\overline{A2}}$-type. 
With the choice of tadpole solutions of
table (\ref{spectrum2oip2})
all tadpole 
solutions in ({\ref{tadpoleO5b}), but the first, are
satisfied, the latter giving
\beq
9 n_a^1 -\frac{1}{\b^1} + {\tilde \ep}(n_d^1 + n_e^1) + 
\frac{2\epsilon_h N_h}{\b^1} = -8. 
\label{additio2oberli}
\eeq
The $U(1)$ couplings to twisted RR fields read
\beqa
B_2^{(1)} \wedge c_1 [-2 \frac{\a -{\a^2}}{\b^1}]
F^b ,&\nonumber\\           
D_2^{(1)}\wedge c_1 [  {\tilde \epsilon} {\epsilon} ][ 3 n_a^1 F^a -
\frac{1}{\b^1} F^b
-
\frac{1}{\b^1} F^c
- {\tilde \epsilon}   n_d^1 F^d    
- {\tilde \epsilon}  n_e^1 F^e ],   &\nonumber\\
E_2^{(1)}  \wedge c_1 [2 \ep {\epsilon} \b^1 ][9 F^a + 2 F^d + F^e ].&
\label{rrkl2over}
\eeqa
The combination of the $U(1)$'s which remains light at low 
energies is given by
\beq
 Q^l = (Q_a -3 Q_d -3 Q_e )
+ 3 \b^1 (n_a^1 + {\tilde \epsilon}n_d^1 + {\tilde \epsilon}n_e^1) Q_c   .
\label{hyper2over}
\eeq
The hypercharge condition in this case is given uniquely by
\beq
\b^1 (n_a^1 + {\tilde \epsilon}n_d^1 + {\tilde \epsilon} n_e^1) = \ -1
\label{cond2over}
\eeq
The other anomaly free $U(1)$, beyond hypercharge,
given by
 \beq
 Q^{(5)} = Q_d -2 Q_e + Q_c 
\label{hyper2over1}
\eeq          
survive massless the Green-Schwarz mechanism if
$-n_d^1 + 2 n_e^1 -({\tilde \epsilon}/\beta^1) = 0$.
A set of wrappings consistent with the last constraint and
(\ref{additio2oberli}) are given by
\beq
n_a^1 = 3, \ n_d= 4, \ n_e = 1, \ {\tilde \ep} = -1, \
\ep =1, \ \ep_h =1, \ N_h = 1. \
 \ \b^1 =1/2.
\label{setlet1new}
\eeq
The scalar fields breaking $ Q^{(5)}$ get localized in the
$de$-sector, e.g. $(1,1)_0$
with mass given by
\beq
\alpha^{\prime} m_{de}^2 = \ \pm \frac{1}{2\pi}|
tan^{-1}(\frac{ U^1}{4}) - tan^{-1}( \frac{U^1}{2})|
+ \frac{{\bar Z}^{(2)}}{4 \pi^2},
\label{reads1new}
\eeq
where ${\bar Z}^{(2)}$ is the transverse distance between the 
branes $d$, $e$ in the 
second tori.
The latter scalars receive a vev and break the $Q^{(5)}$
symmetry.
Tachyonic scalar directions may be lifted by their corrections from
the couplings (\ref{oneloop}).
Thus at low energy we get
the SM gauge group and chiral content.

\section{Standard Models from Six-Stack $Z_3$ Quivers}

In this section, we discuss the embedding of SM configurations 
of table (3) in six stacks of intersecting D5-branes at the string scale,
 that in
general, give at low energy the SM together with some extra 
scalar fields.
Part of these scalar fields is responsible
for breaking the two anomaly free free U(1)'s, beyond the the U(1)
corresponding to the SM, surviving
massless the generalized Green-Schwarz mechanism.
Nevertheless, all scalars may receive important loop corrections from
(\ref{oneloop}) which may lift all their tachyonic directions. The latter
will be further justidied in section 7 using brane recombination. 
The embedding of the six stack SM structure of 
table (3) in various quivers can be seen
in figures (4) and (6). In the present work, we will describe explicitly only
the C1, C2 quivers. Similar results hold for the other six stack quivers. 
However, in the next section we will show by using brane recombination that 
there is continuous flow of all the six stack quivers to their 
four stack quiver `counterparts'.

\subsection{SM Vacua from the Six-Stack C1-quiver}

%\eps%%%%%%%%%% Figure here%%%%%%%%%%%%%%%%%%%%%%%%%%%%%%%%%%%%%%%%%
\begin{figure}
\begin{center}
\centering
\epsfysize=8cm
\leavevmode
%\hspace*{0in}\vspace*{.2in}
\epsfbox{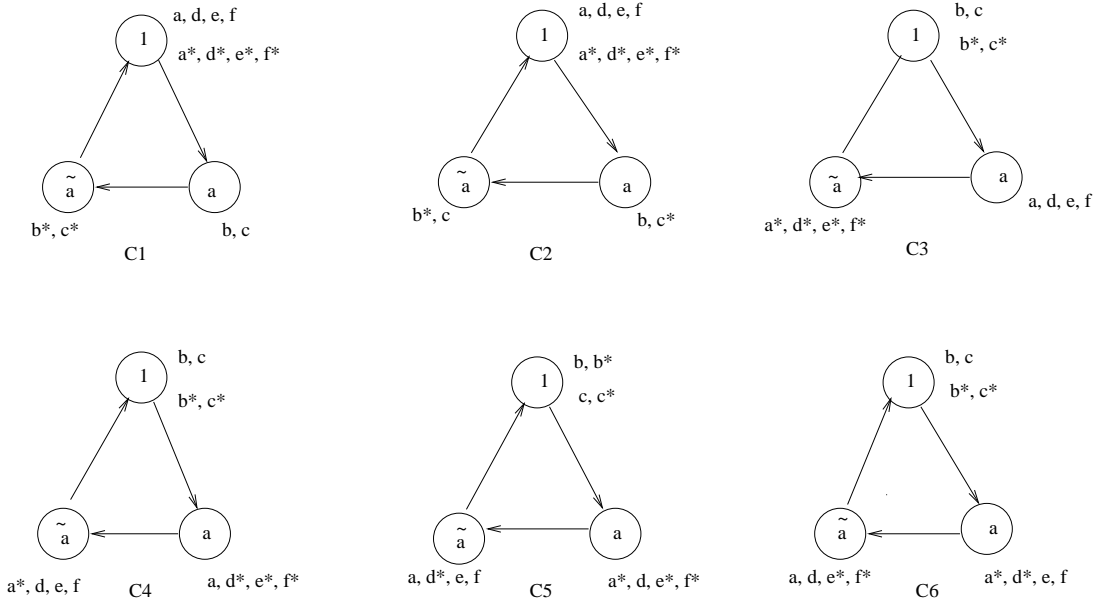}
\end{center}
\caption[]{\small
Assignment of SM embedding in  
configurations of six stacks of D5 branes depicted in
$Z_3$ quiver diagrams; six out of a total of ten quivers.
In all the cases we get 
the SM at low energy.
Note that ${\tilde \alpha} = \alpha^{-1}$.
 }
\end{figure}
%%%%%%%%%%%%%%%%%end of figure %%%%%%%%%%%%%%%%%%%%%%%%%%%%%%%%%%%

The C1-quiver of SM embedding  
seen in figure (3) is solved by the RR tadpole solutions
seen in Appendix II.
With the latter choice the tadpole
solutions in ({\ref{tadpoleO5b}), but the first,
are satisfied, the latter giving 
\beq
9 n_a^1 -\frac{{\tilde \epsilon}}{\b^1} + n_d^1 + n_e^1
+ n_f^1
+ \frac{2\epsilon_h N_h}{\b^1}= -8.
\label{six2}
\eeq
Also the $U(1)$ couplings to twisted RR fields read:
\beqa
B_2^{(1)} \wedge c_1 [2 {\tilde \epsilon}
\frac{\a -{\a}^2}{\b^1}]
F^b ,&\nonumber\\           
D_2^{(1)}\wedge c_1 [{\tilde \epsilon} {\epsilon}]
[ 3 n_a^1 F^a - \frac{{\tilde \epsilon}}{\b^1} F^b
+  \frac{{\tilde \epsilon}}{\b^1} F^c
-   n_d^1 F^d    
-   n_e^1 F^e -   n_f^1 F^f],   &\nonumber\\
E_2^{(1)}  \wedge c_1 (-2  {\epsilon} \b^1 )[9 F^a +
F^d + F^e + F^f].&
\label{six3}
\eeqa
There are three U(1)'s which survive massless the Green-Schwarz
mechanism. One of them is automatically 
orthogonal to all U(1)'s seen in (\ref{six3}). It is 
\beq
 Q^l = (Q_a -3 Q_d -3 Q_e -3 Q_f)
- 3 {\tilde \ep}\b^1 (n_a^1 + n_d^1 + n_e^1 + n_e^1) Q_c.
\label{six4}
\eeq
The latter U(1) represents the hypercharge (\ref{hyper1232})
if the hypercharge condition is satisfied
\beq
{\tilde \ep}\b^1 (n_a^1 + n_d^1 + n_e^1 + n_f^1) = \ 1.
\label{six5}
\eeq
The other $U(1)$'s which survive massless the Green-Schwarz
mechanism are given by
 \beqa
 Q^{(5)} &=&  2Q_d - Q_e - Q_f ,\\
 Q^{(6)} &=& Q_e - Q_f
\label{six6}
\eeqa          
The U(1)'s (\ref{six6}) remain massless, and
orthogonal to the U(1)'s (\ref{six3}) if
$n_d^1 = n_e^1 = n_f^1$.
The set of wrappings
\beq
n_a^1 = 4, n_d^1 = n_e^1 = n_f^1 = -1, \beta^1 =1,
{\tilde \epsilon} = 1, \epsilon_h =-1, N_h = 20
\eeq
is consistent with all the constraints present.

The scalars from the $de^{\star}$ sector 
\beq
I_{de^{\star}}(1,1)_0 \equiv 2(1,1)_0 
\label{actu11}
\eeq
break the extra U(1) $Q^{(5)}$. Also the scalars from the $df^{\star}$ sector
\beq
I_{df^{\star}}(1,1)_0 \equiv 2(1,1)_0 
\label{actu1101}
\eeq
break the extra U(1) $Q^{(6)}$.
The masses of the $de^{\star}$, $df^{\star}$ scalars may be given
respectively
by
\beq
\alpha^{\prime} m_{de^{\star}}^2 = \ \pm |1-
\frac{1}{\pi}tan^{-1}( U^1) - \frac{1}{\pi}tan^{-1}(\frac{U^2}{2})|,
\label{c1sca1}
\eeq
\beq
 m_{df^{\star}}^2 = \  m_{de^{\star}}^2
\label{c1sca2}
\eeq
Because all scalars receive important corrections to their mass from the
couplings (\ref{oneloop}) all tachyonic directions may be lifted.
Thus at low energies, only the SM survives.

%
%
%
%%%%%%%%%%%%%%%%%%%%%%%%%%

\subsection{SM Vacua from the Six-Stack C2-quiver}

The C2-quiver embedding of the SM chiral
fermions of table (3)
can be seen in figure (4). 
With the choice of tadpole solutions seen in 
appendix II all tadpole
solutions in ({\ref{tadpoleO5b}), but the first,
are satisfied, the latter giving
\beq
9 n_a^1 -\frac{1}{\b^1} + n_d^1 + n_e^1
+ n_f^1
+ \frac{2\epsilon_h N_h}{\b^1}= -8.
\label{six8}
\eeq
The $U(1)$ couplings to twisted RR fields follows:
\beqa
B_2^{(1)} \wedge c_1 [2
\frac{\a -{\a}^2}{\b^1}]
F^b ,&\nonumber\\           
D_2^{(1)}\wedge c_1 [{\tilde \epsilon} {\epsilon}]
[ 3 n_a^1 F^a - \frac{1}{\b^1} F^b
-
\frac{1}{\b^1} F^c
-   n_d^1 F^d    
-   n_e^1 F^e -   n_f^1 F^f],   &\nonumber\\
E_2^{(1)}  \wedge c_1 (-2 \ep  {\epsilon} \b^1 )[9 F^a +
F^d + F^e + F^f].&
\label{six9}
\eeqa
For the C2 quiver the SM hypercharge may be associated with the 
U(1) combination 
\beq
 Q^l = (Q_a -3 Q_d -3 Q_e -3 Q_f)
+ 3 \b^1 (n_a^1 + n_d^1 + n_e^1 + n_e^1) Q_c.
\label{six10}
\eeq
if the hypercharge condition is given uniquely by
\beq
\b^1 (n_a^1 + n_d^1 + n_e^1 + n_f^1) = \ -1.
\label{six11}
\eeq
The other $U(1)$'s which survive masslss the Green-SChwarz mechanims 
are given by
 \beqa
 Q^{(5)} &=& 
 (2Q_d - Q_e - Q_f),\\
 Q^{(6)} &=& Q_e - Q_f
\label{six111}
\eeqa
The existence of the latter U(1)'s give us the model
dependent condition $n_d^1 = n_e^1 = n_f^1$.
E.g making the choice $n_d^1 = n_e^1 = n_f^1 =1$,
$n_a^1 = -4$, $\epsilon =1$, $\beta^1 =1$,
$\epsilon_h  = -1$, $N_h =21$, the scalars
from $de^{\star}$-sector $2(1, 1)_{0}$ break $ Q^{(5)}$,
while the scalars from $df^{\star}$-sector
$2(1, 1)_{0}$ break $ Q^{(6)}$.
Thus at low energy only the SM remains as all tachyonic directions
may be lifted by the corrections (\ref{oneloop}).

%
%
%%%%%%%%%%%%%%%%%%%%%%%%%%%%%%%%%%%%%%%%%%%%%%%
%%%%%%%%%%%%%%%%%%%%%
%%%%%%%%

\section{Brane recombination flows of SM string vacua with different
number of stacks}

We have seen that when examining five and six stack vacua,
some additional features appeared in comparison to the
corresponding four stack `counterpart' 
quiver vaccum \footnote{By `counterpart' we mean the 
quiver obtained by `naively' deleting the extra branes from the
five, six, stack quivers so that they become four stack one's.}. 
Thus we find more scalars to be present in these models.
 We should emphasize however, that this is an artifact
of our procedure.
The appearance of the scalar spectrum in the models
we studied, as opposed to the chiral multiplets that is fixed
by the intersection numbers, depends on our choice of
wrapping numbers entering the RR tadpole cancellation
conditions. Their number for a general brane $j$ comes from
the sector which the brane transforms with a CP phase of unity.

In this section we will show that the appearance of fermion and scalars
in five and six stack quiver models is in one to one correspondence  with
their counterpart four stack quivers.
That is we show using
the mechanism of brane
recombination (BR), that there are directions in homology space
such that the five and six stack D5 vacua are equivalant to four stack 
vacua and thus have only the SM at low energy.

Brane recombination in the picture of the D5-brane models we are examining
may correspond exactly to the picture suggested by Sen, namely that the 
tachyon condensed at its minimum is exactly the final configuration of the 
recombined branes (which are stretched along a minimal volume cycle in its
homology class).
In the context of intersecting branes BR effects have been considered 
in \cite{allo3} (and further in \cite{allo8} ) where
it corresponds to a stringy version of the electroweak symmetry breaking 
mechanism. In this case it was the recombination of one of the U(1) of the  
`left''
b-brane with the U(1) `right'' c-brane that induced the Higgs mechanism. 
In our case the origin of BR effects are different.
In gereral the BR mechanism corresponds to recombining D5-branes wrapping 
different 
intersecting cycles. We note that the D5 branes that we will recombine
are not necessarily paralled across some two dimensional tori.
Thus we will be able to show that by recombining always the U(1) branes
that are not involved in the QCD intersection numbers (not the a, b, c, branes)
there is a continuous flow between the 
six, five and four stack models. 

In order to illuminate our points we will examine for simplicity
some examples in the five and in the six 
stack quivers.  We note that there are two general types of quivers occuring
in the present work. In the first class of 
quivers the coloured (or baryonic)a- brane is on the same node (transforming 
trivially under CP) with another
two d, e (resp. three, the d, e, f) U(1) branes (and/or images) in the
five (resp. six) stack quiver.
In the second class of 
quivers the left U(2) b, brane (and/or images) is on the 
node (transforming trivially under CP) with the right c brane (and/or 
images).

\subsection{Five stack D5-models flowing to four stack SM D5's}

Take for example one representative of the first type of quivers, e.g.
the five stack A1-quiver. Its intersection numbers before BR were 
given in                  
eqn. (\ref{dour}). Lets us assume that two of the branes 
recombine, e.g. the d with e brane into a new brane j. 
Thus instead of the original five stack model with a, b, c, d, e branes we 
are left with a four stack model made of a, b, c, j branes.
The new intersection numbers are computed easily by noticing that 
$I_{ab}$ is now an additive quantity, 
thus $I_{aj} = I_{ad} + I_{ae}= 0 + 0 = 0$, and e.g. 
$I_{bj} = I_{bd} + I_{be}= 2 + 1= 3 $. 
Thus after recombination into a single brane 
$d + e \rightarrow j$ we are left with the following intersection numbers 
\beqa
I_{bj}=3,   & I_{cj} = -3,        &  I_{c j^{\star}} = 3,\nonumber\\
I_{ab} = 1, & I_{ab^{\star}} = 2, &  I_{ac} = -3, \  I_{ac^{\star}} = 3  
\label{a1quiafte}
\eeqa
which we regognize to be the intersection numbers of the four stack 
$\alpha_1$ quiver (the quiver obtained by naively deleting the e-brane
from the A1-quiver!).  
Pictorially
\beq
A1 \stackrel{d + e  = j}{\Longrightarrow} a1
\eeq

Let us now examine a quiver of the second type appearing in the present work, one
representative of which is the B3-quiver seen in figure (\ref{ell1}).
Again we recombine the d, e branes \footnote{Even though they are transforming
non-trivially with a CP phase}.   
Its intersection numbers before BR are given by
\beqa 
 I_{ab} = -1, &  I_{ab^{\star}} = -2 \ , &  I_{ac} = 3, \ I_{ac^{\star}} = 3,
 \nonumber\\
I_{bd} = -2, &  I_{be} = -1 \ , &  I_{cd} = 2, \ I_{cd^{\star}} = 2, 
\nonumber\\
I_{ce} = 1, &  I_{ce^{\star}} = 1 &
\label{B3quibe}
\eeqa
After BR, $d + e \rightarrow j$ and we are left with the following 
intersection numbers, namely
\beqa 
 I_{ab} = -1, &  I_{ab^{\star}} = -2 \ , &  I_{ac} = 3, \ I_{ac^{\star}} = 3,
 \nonumber\\
I_{bj} = -3, &  I_{cj} = 3 \ , &  I_{ce^{\star}} = 3 
\label{B3quiafte}
\eeqa
which are the intersection numbers of the four stack 
$\alpha_4$ quiver (the quiver obtained by naively deleting the e-brane
from the B3-quiver).
Pictorially
\beq
B3 \stackrel{d + e = j}{\Longrightarrow} a4
\eeq

Thus something novel happens here, as the the original five 
stack quiver A1(resp. B3) flows into the $\alpha_1$ 
(resp. $\alpha_4$) quiver. The latter quiver has only 
coloured scalars and they are expected to receive strong loop 
corrections from the couplings (\ref{oneloop}). Therefore the A1-quiver (resp. B3) 
at low 
energy may have only the SM at low energy. 
What is BR is really telling us is that there are flat directions in the moduli space
of the A1-quiver (resp. B3) which can be used to break the extra U(1) 
symmetry carried by
the e-brane. Moreover, they confirm that we are speaking about two
equivalent theories.

In addition, they confirm the uniqueness of our
original choice of
constructing five \footnote{These results also hold for the five stack D6 SM's
of \cite{kokos2} which under d + e brane BR, flow to the four stack SM examples
of  \cite{luis1}, while the six stack SM's of \cite{kokos3} flows into 
the SM's of \cite{luis1} under d + e + f BR.} 
stack deformations out of the four stack vacua of \cite{ibanez1}
by deforming around its
QCD intersection number structure \footnote{Following the choice of 
universal five stack SM intersection number ansatz of
table (\ref{spectrum8}).}.
In the same way it can be shown e.g. that the B4 quiver under
$d^{\star}$ + $e^{\star}$  BR into a single brane $\tilde d$,
flows to the the four stack a3-quiver. Similar results hold
for the other five stack quivers.

\subsection{Six stack D5-models flowing to four stack D5 SM's}

Using BR in the branes d, e, f, the six stack quivers can be
shown to ``flow'' to their corresponding quiver SM that
is generated by naively deleting the presence of the branes
d, e, f, from the corresponding node and replacing it by the new
recombined brane.
To show the latter we will consider two representative examples
that are based on the C1 and C3 quivers.

The intersection numbers of the C1 quiver before BR are given by
\beqa 
 I_{ab} = 1, &  I_{ab^{\star}} = 2 \ , &  I_{ac} = -3, \
 I_{ac^{\star}} = 3,
 \nonumber\\
I_{bd} = 1, &  I_{be} = 1 \ , &  I_{bf} = 1, \ I_{cd} = -1\nonumber\\
I_{cd^{\star}} = 1, &  I_{ce} = -1 \ , &  I_{ce^{\star}} = 1, \
I_{cf} = -1,\nonumber\\
I_{cf^{\star}} = 1 &&
\label{C1quiafte}
\eeqa

After BR, $d + e + f \rightarrow l$, one can easily seen that
the new intersection numbers are those of (\ref{a1quiafte}).
Thus we have evident the chain transition
\beq
C1 \stackrel{d+e+f = l}{\Longrightarrow} a1
\eeq
Let us now turn our attention to the C3 quiver.
In this case, the intersection numbers before
BR are given by
\beqa 
 I_{ab} = -1, &  I_{ab^{\star}} = -2 \ , &  I_{ac} = 3, \
 I_{ac^{\star}} = 3,
 \nonumber\\
I_{bd} = -1, & I_{be} = -1 \ , &  I_{bf} = -1, \ I_{cd} =1\nonumber\\
I_{cd^{\star}} = 1, &  I_{ce} = 1 \ , &  I_{ce^{\star}} = 1, \
I_{cf} =1,\nonumber\\
I_{cf^{\star}} = 1 &&
\label{C3qubefo}
\eeqa
while after BR, $d + e + f \rightarrow l$, they become those of the
4-stack a4-quiver, the latter naively obtained by deleting the
e, f, branes from the B3-quiver dagram.
\beq
C3 \stackrel{d+e+f = l}{\Longrightarrow} a4
\eeq

\subsection{Six stack D5-models flowing to five stack SM D5's}

There are more chirality non-changing transitions. We note that
brane recombination as we proposed to be carried out does not
change the chiral content of the models but rather by changing
the positions of the chiral matter in the target space reduces also
the scalar content of the models present. A more appropriate name for the
transitions taking place
will be {\em gauge and scalar changing transitions} as they help us avoid the
presence of extra scalars generically present in the five and six stack
SM's.

By BR on the d, e branes under the chain
$e + f \rightarrow {\tilde e}$ it may be shown that the following
chirality non-changing transitions take place on the :
\beq
C1 \stackrel{e+f = {\tilde e}}{\Longrightarrow}{B3}^{\prime}
\eeq
\beq
C3 \stackrel{e+f = {\tilde e}}{\Longrightarrow}{B3}^{\prime}
\eeq
As ${B3}^{\prime}$ quiver we denote the intersection numbers of the
B3 quiver where the d, e branes are interchanged (this is a
symmetry of the B3 quiver,
${B3}^{\prime} \stackrel{d \longleftrightarrow e}{\equiv} B3 $).
Subequently adopting BR, $ d + e \rightarrow {\tilde d}$, 
the ${B3}^{\prime}$ quiver flows to the $a_1$ (resp. $a_4$) quivers.
The transitions described can be seen in figure (\ref{reco6}).

%%%%%%%%%% Figure here%%%%%%%%%%%%%%%%%%%%%%%%%%%%%%%%%%%%%%%%%
\begin{figure}
\begin{center}
\centering
\epsfysize=3.5in
\leavevmode
%\hspace*{0in}\vspace*{.2in}
\epsfbox{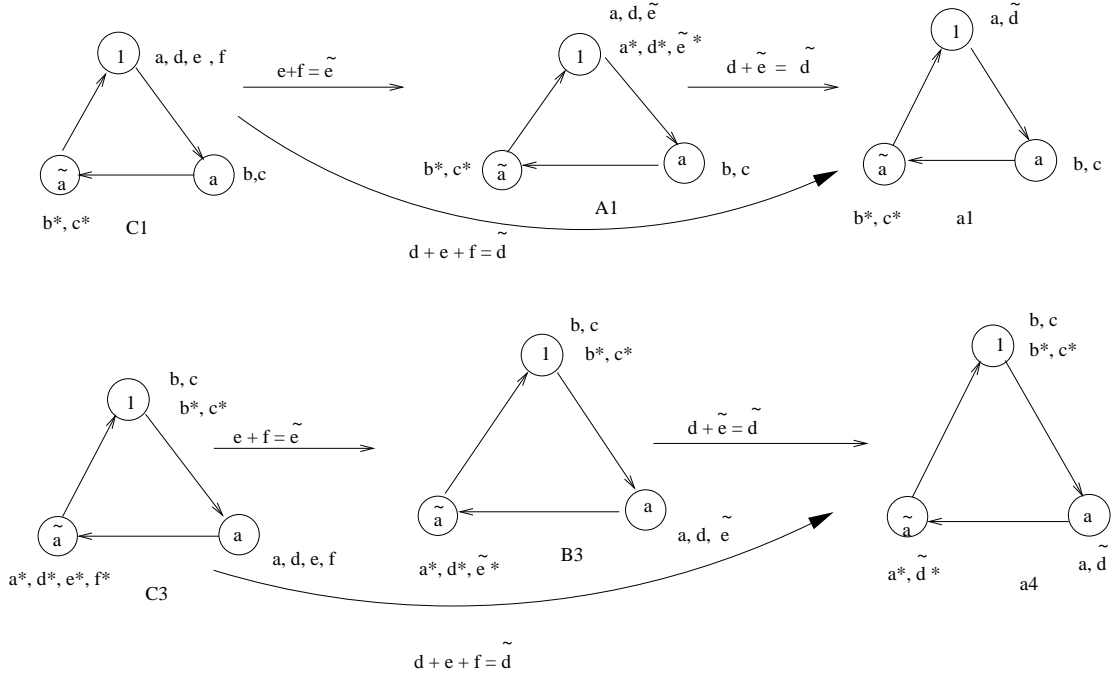}
\end{center}
\caption[]{\small
Recombination flow for some of the six-stack quivers. All models flow to four 
stack SM quivers after recombination.}
\label{reco6}
\end{figure}
%%%%%%%%%%%%%%%%%end of figure %%%%%%%%%%%%%%%%%%%%%%%%%%%%%%%%%%%

\section{Epilogue}

In this work, 
we have discussed the appearance of non-supersymmetric
compactifications with D5 branes intersesting at angles
that may have exactly the SM at low energies. Our main purpose was
to classify all the possible configurations that may give the SM
at low energy and may be of phenomenological interest.

The models of this work, are build on a
background of D5-branes, the latter being part of a
four, five and six-stack 
structure, with
intersecting D5-branes
on an orientifold of $T^4 \otimes \frac{C}{Z_N}$
\cite{ibanez1}.
The orbifold structure of the background is encoded
in different quiver diagrams (see figures 1-7).  The
quiver diagrams
reflect the geometric and orbifold action on the
brane/orientifold-image system placed on the quiver nodes.

We completed the discussion of four stack SM's
initiated in
\cite{ibanez1} by discussing the four, `reflected',
quivers Q1, Q2, Q3, Q4. 
 We pointed out that these quivers describe 
equivalent theories to the quiver
 examples presented in \cite{ibanez1}, thus they
describe vacua with exactly the SM at low energies.

We showed that reflected $Z_3$ quivers,
that are obtained from an `image' quiver by keeping
the quiver transformation of the nodes fixed, while at
the same time
interchanging the brane/orientifold content of the nodes
with CP
phases different from unity, describe equivalent theories.
Also, we presented some new tadpole solutions to the four
stack quivers of \cite{ibanez1}, the latter seen in figure
(2). The new solutions describe, at the level of low
energy
effective theories equivalent SM theories to those
of \cite{ibanez1}.

In addition, we discussed the Standard model
embedding in five and
six stack quivers, denoted as Ai, ${\bar A}$i;/Bj ($i=1,.., 2; j=1,..,7$),
 Ck ($k=1,..,10$) quivers
respectively \footnote{These particular SM embedding have 
been used before in
five and six-stack orientifolded $T^6$ backgrounds 
\cite{kokos2, kokos3} respectively and produced string vacua
with exactly the Standard Model at low energy.}.
In this case, the appearance of the SM at low energies is achieved easily
but also make their appearance singlet, doublet and colour scalars. In all 
cases singlet scalars may be used  to break the extra U(1)'s, while 
tachyonic directions that exist among scalars may be lifted by the loop
corrections (\ref{oneloop}).
Further justification to the latter was provided by using the mechanism of 
brane
recombination (BR).
BR showed us that there are gauge breaking transitions \footnote{that do not
change the chira content of the models} in the context
of intersecting D5-branes that allow a continuous flow between the six,
five and four quivers of the same kind, namely the one's that can be created by
adding leptonic U(1) e, f, branes, on the nodes of the four stack quiver having
a baryonic (a) brane.
This is quite interesting result, since gauge breaking transitions are known
to exist 
only in the context of M-theory to occur through small
instanton transitions \cite{ovru}.

All the models have a gauged baryon number as it happens in
their 
orientifolded $T^6$ tori counterparts \cite{kokos2}.
We note that  
the same property is being shared
in all the constructions involving modelling with D5-branes 
in the same backgrounds or in the SM vacua from intersecting 
D6-brane models
on the orientifolded $T^6$ tori \cite{luis1, kokos2, kokos3, kokos4}.
The models based on the present backgrounds 
have a stability defect as they have closed string tachyons.
Thus their full
stability is an open question. Similar issues have been examined in
the context of non-compact orbifolds \cite{adams}.

Crucial for the four-, five-, six- stack constructions
of D5-branes used in this work,
was the
configurations of tables (1), (2), (3) that
exhibit the particular localization of fermions
at the different
intersections.
The same configurations 
have been used before in the construction of four- five-
six-
stack SM's in orientifolded $T^6$
tori \cite{luis1, kokos2, kokos3}.

The most important difference of the present study with the relevant
four- five or six- stack SM's found in their 
counterparts SM's
 in orientifolded
 $T^6$ tori \cite{luis1, kokos2, kokos3, kokos4} is related to
the existence
of scalars in the models. In the models
of \cite{kokos2, kokos3, kokos4}
the only scalars present in the models where the
one's appearing after
imposing $N=1$ SUSY in a particular sector involving the
right handed
neutrino. The latter supersymmetric scalars
were responsible \footnote{the latter scalars are necessary 
if the corresponding $U(1)$ gauge bosons have a zero 
coupling to the RR fields.}
for the breaking of the extra massless $U(1)$'s, beyond the one associated
with the hypercharge, at low energies.
On the contrary in the present intersecting brane backgrounds there is a 
variety
of different scalars in the models, color, singlet, or doublet 
scalars. All scalars receive loop corrections that are stronger for the 
coloured scalars. Also singlet scalars are being used to break the 
extra U(1)'s, beyond hypercharge, that survive massless to low 
energies.
The kind of scalars present in the SM-like models
depends on our choise
of parameters entering the RR tadpole cancellation conditions.
Thus a different choice of parameters will provide us with 
different scalars.

In section 7, the novel
use of the brane recombination (BR) mechanism showed us that there
is
continous ``flow'' from six to five and four stack vacua depicted
by quiver diagrams. In fact, we were able to show that by starting with the
six stack vacua and recombining the d, e, f branes(resp. e, f)
the six stack (resp. five stack) quiver intersection numbers ``flow'' to 
the four
stack intersection numbers. Thus instead of solving 
the five and six stack quiver one could also solve the associated four 
stack model. 
As the intersection numbers $I_{ab}$ are practically involved in the RR tadpole
cancellation conditions through the relation
\beq
\sum_a N_a [\Pi_a ]= 0
\label{rcer1}
\eeq
which `flows' to
 \beq
\sum_a N_a [\Pi_b]\cdot [\Pi_a ]= \sum_a N_a I_{ab} = 0,
\label{rrc}
\eeq
the latter being the cancellation of the cubic non-abelian anomalies,
we have found \footnote{The recombination changed the intersection numbers,
such the RR charge (\ref{rcer1}) was conserved, and (\ref{rrc}) before and after
recombination was preserved.} that the six and five stack models are equivalent
in homology space to the associated four stack models. 
Thus we were able to show that classes of D5 brane quiver vacua 
which may have the SM at low energy, and thuis may be of phenomenological
interest, are continously connected \footnote{It can be easily seen,
using BR along the lines of section 7, that the six and five
D6 brane stack deformations of \cite{kokos2} \cite{kokos3} are
continously connected to the D6-models of \cite{luis1}.}. The mechanism 
responsible for BR flow is rather topological at our current level
of understanding BR. It will be interesting if a deeper
explanation may be found, perhaps in relation with a lifting of the 
present D5-models to their dual models that may result from an 
M-theory compactification. The latter perhaps could be justified in 
relation to the small instanton transition analysis of \cite{ovru}. 

The Standard Models discussed in this work, have a
natural low scale \footnote{ Thus they may avoid
the standard gauge hierarchy problem of Higgs scalars.}
of order of the TeV \cite{ibanez1, anto},
 as the volume of the 
two dimensional manifold, transverse to the D5 branes,
over which the singularity
structure of $C/Z_N$ can be embedded can be made large enough,
thus allowing for a low string scale $M_s << M_p$.

\section{Acknowledgments}
I am grateful to D. Cremades, L. Ib\'a\~nez, F. Marchesano and  
A. Uranga for useful discussions.

%\newpage

%%%%%%%%%% Figure here%%%%%%%%%%%%%%%%%%%%%%%%%%%%%%%%%%%%%%%%%
\begin{figure}
\begin{center}
\centering
\epsfysize=7cm
\leavevmode
%\hspace*{0in}\vspace*{.2in}
%\epsffile{anstan.eps}
\epsfbox{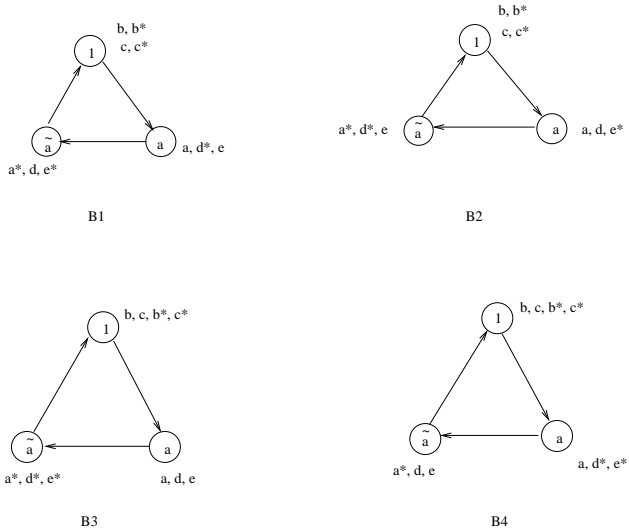}
\end{center}
\caption[]{\small
Configurations of D5 branes with $Z_3$ quiver diagrams, which 
give at low energy the SM.}
\label{ell1}
\end{figure}
%%%%%%%%%%%%%%%%%end of figure %%%%%%%%%%%%%%%%%%%%%%%%%%%%%%%%%%%

%
%
%

%%%%%%%%%% Figure here%%%%%%%%%%%%%%%%%%%%%%%%%%%%%%%%%%%%%%%%%
\begin{figure}
\begin{center}
\centering
\epsfysize=8cm
\leavevmode
%\hspace*{0in}\vspace*{.2in}
%\epsffile{anstan.eps}
\epsfbox{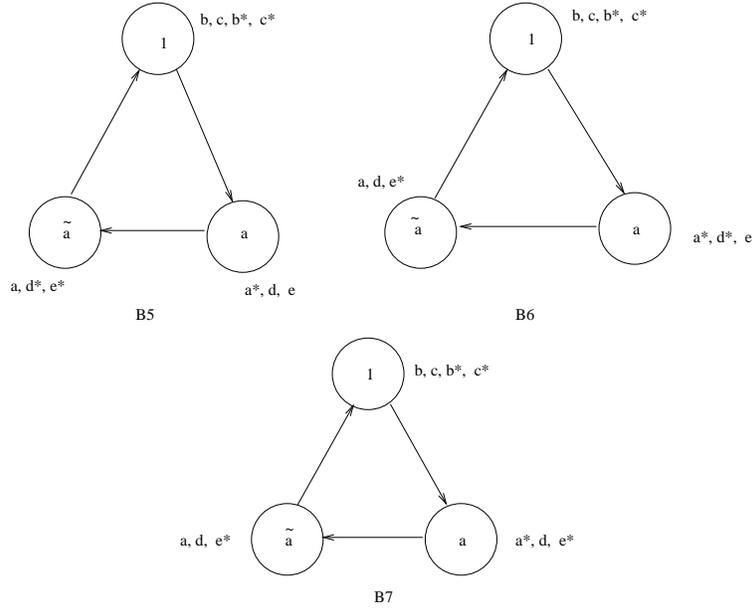}
\end{center}
\caption[]{\small
Configurations of five stacks of intersecting
D5 branes placed in $Z_3$ quiver diagrams and giving rise
to the SM at low energy. }
\end{figure}
%%%%%%%%%%%%%%%%%end of figure %%%%%%%%%%%%%%%%%%%%%%%%%%%%%%%%%%%

%
%
%
%

%
%
%
%
%
%
%
%

%\eps%%%%%%%%%% Figure here%%%%%%%%%%%%%%%%%%%%%%%%%%%%%%%%%%%%%%%%%
\begin{figure}
\begin{center}
\centering
\epsfysize=8cm
\leavevmode
%\hspace*{0in}\vspace*{.2in}
%\epsffile{anstan.eps}
\epsfbox{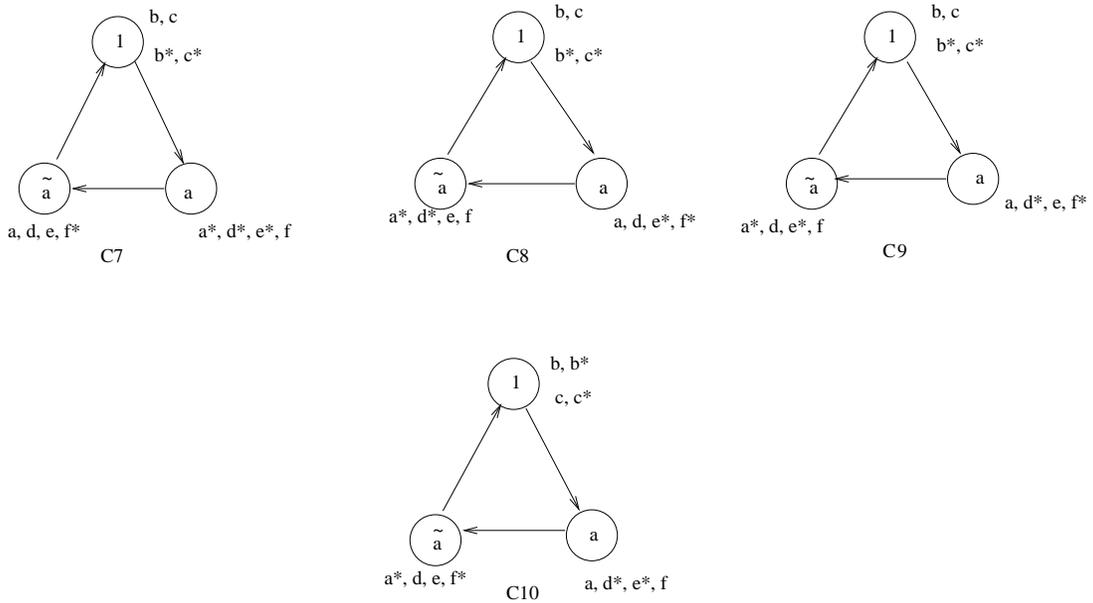}
\end{center}
\caption[]{\small
Assignment of SM embedding in  
configurations of six stacks of D5 branes depicted in
$Z_3$ quiver diagrams; four out of a total of ten quivers,
giving rise to the SM at low energy.
Note that ${\tilde \alpha} = \alpha^{-1}$.
 }
\end{figure}
%%%%%%%%%%%%%%%%%end of figure %%%%%%%%%%%%%%%%%%%%%%%%%%%%%%%%%%%

%\newpage

\section{Appendix A}

%{RR tadpole solutions for five-stack quivers}

The solutions of the five stack D5 quivers depend
on the integer parameters, $n_a^1$, $n_b^1$, $n_c^1$,
$n_d^1$, $n_e^1$, the NS-background $\beta^1$ and
the phase parameters $\epsilon = {\tilde \epsilon}
= \epsilon_h =\pm 1$,
 also on the CP phase $\alpha$.

%\begin{table}
%[htb]\footnotesize
%\renewcommand{\arraystretch}{1.8}
\begin{center}
\begin{tabular}{||c||c|c|c||}
\hline
\hline
\multicolumn{4}{c}{RR tadpole solutions for B1-quiver}\\\hline
\hline
\hline
$N_i$ & $(n_i^1, m_i^1)$ & $(n_i^2, m_i^2)$ & $(n_i^3, m_i^3)$\\
\hline\hline
 $N_a=3$ & $(1/\b^1, 0)$  &
$(3,  \frac{1}{2}{\tilde \epsilon} \epsilon )$ & $\alpha 1_3$  \\
\hline
$N_b=2$  & $(n_b^1, -{\tilde \epsilon} \epsilon \b^1)$ & 
$(1, \frac{1}{2} {\tilde \epsilon}{\epsilon})$ &$ 1_2$ \\
\hline
$N_c=1$ & $(n_c^1,  \epsilon \b^1 )$ &   
$(0, {\epsilon} )$  & 
$1$ \\    
\hline
$N_d=1$ & $(1/\b^1, 0)$ &  $(-2,   \epsilon 
{\tilde \epsilon})$  
  & $\alpha^2$  \\\hline
$N_e = 1$ & $(1/ \b^1, 0)$ &  
$(1, -\frac{1}{2}\epsilon 
{\tilde \epsilon} )$  
  & $\alpha$ 
\\\hline
$N_h$ & $(\epsilon_h/ \b^1, 0)$ &  
$(2, 0 )$  
  & $1_{N_h}$ 
\\\hline
\end{tabular}
\end{center}
%\caption{\small General twisted RR tadpole
%cancellation solutions for the 
%B1-type quiver of intersecting
%five-stacks of D5-branes.
%The solutions depend 
%on two integer parameters, 
%$n_b^1$, $n_c^1$,
%the NS-background $\beta^1$ and
%the phase parameters $\epsilon = {\tilde \epsilon}=\pm 1$, as well as the 
%CP phase $\alpha$. At low energies we get the SM.
%\label{spectrum5}}          
%\end{table}

%\begin{table}
%[htb]\footnotesize
%\renewcommand{\arraystretch}{2}
\begin{center}
\begin{tabular}{||c||c|c|c||}
\hline
\hline
\multicolumn{4}{c}{RR tadpole solutions for B2-quiver}\\\hline
\hline
\hline
$N_i$ & $(n_i^1, m_i^1)$ & $(n_i^2, m_i^2)$ & $(n_i^3, m_i^3)$\\
\hline\hline
 $N_a=3$ & $(1/\b^1, 0)$  &
$(3,  \frac{1}{2}{\tilde \epsilon} \epsilon )$ & $\alpha 1_3$  \\
\hline
$N_b=2$  & $(n_b^1, -{\tilde \epsilon} \epsilon \b^1)$ & 
$(1, \frac{1}{2} {\tilde \epsilon}{\epsilon})$ &$ 1_2$ \\
\hline
$N_c=1$ & $(n_c^1,  \epsilon \b^1 )$ &   
$(0, {\epsilon} )$  & 
$1$ \\    
\hline
$N_d=1$ & $(1/\b^1, 0)$ &  $(2,   -\epsilon 
{\tilde \epsilon})$  
  & $\alpha$  \\\hline
$N_e = 1$ & $(1/ \b^1, 0)$ &  
$(-1, \frac{1}{2}\epsilon 
{\tilde \epsilon} )$  
  & $\alpha^2$ 
\\\hline
$N_h$ & $(\epsilon_h/ \b^1, 0)$ &  
$(2, 0 )$  
  & $1_{N_h}$ 
\\\hline
\end{tabular}
\end{center}
%\caption{\small General tadpole solutions for the five-stack 
%B2-type quiver of intersecting
%D5-branes.
%At low energy we get the SM.
%and an extra anomaly free
%$U(1)$ symmetry
%under which the SM fields get charged.
%The solutions depend 
%on two integer parameters, 
%$n_b^1$, $n_c^1$, 
%the NS-background $\beta^1$ and
%the phase parameters $\epsilon = {\tilde \epsilon}=\pm 1$, as well as the 
%CP phase $\alpha$. 
%\label{spectrum6}}          
%\end{table}

%\begin{table}
%[htb]\footnotesize
%\renewcommand{\arraystretch}{2}
\begin{center}
\begin{tabular}{||c||c|c|c||}
\hline
\hline
\multicolumn{4}{c}{RR tadpole solutions for B3-quiver}\\\hline
\hline
\hline
$N_i$ & $(n_i^1, m_i^1)$ & $(n_i^2, m_i^2)$ & $(n_i^3, m_i^3)$\\
\hline\hline
 $N_a=3$ & $(1/\b^1, 0)$  &
$(3,  \frac{1}{2}{\tilde \epsilon} \epsilon )$ & $\alpha 1_3$  \\
\hline
$N_b=2$  & $(n_b^1, -{\tilde \epsilon} \epsilon \b^1)$ & 
$(1, \frac{1}{2} {\tilde \epsilon}{\epsilon})$ &$ 1_2$ \\
\hline
$N_c=1$ & $(n_c^1, -{\tilde \epsilon} \epsilon \b^1 )$ &   
$(0, -{\epsilon}{\tilde \epsilon} )$  & 
$1$ \\    
\hline
$N_d=1$ & $(1/\b^1, 0)$ &  $(2,   -\epsilon 
{\tilde \epsilon})$  
  & $\alpha$  \\\hline
$N_e = 1$ & $(1/ \b^1, 0)$ &  
$(1, -\frac{1}{2}\epsilon 
{\tilde \epsilon} )$  
  & $\alpha$ 
\\\hline
$N_h$ & $(\epsilon_h/ \b^1, 0)$ &  
$(2, 0 )$  
  & $1_{N_h}$ 
\\\hline
\end{tabular}
\end{center}
%\caption{\small General tadpole solutions for the five-stack 
%B3-type quiver of intersecting
%D5-branes.
%At low energy we get the SM.
%and an extra $U(1)$ symmetry
%under which the SM fields get charged.
%The solutions depend 
%on two integer parameters, 
%$n_b^1$, $n_c^1$, 
%the NS-background $\beta^1$ and
%the phase parameters $\epsilon = {\tilde \epsilon}=\pm 1$, as well as the 
%CP phase $\alpha$. 
%\label{spectrum3}}          
%\end{table}

%\begin{table}
%[htb]\footnotesize
%\renewcommand{\arraystretch}{2}
\begin{center}
\begin{tabular}{||c||c|c|c||}
\hline
\hline
\multicolumn{4}{c}{RR tadpole solutions for B4-quiver}\\\hline
\hline
\hline
$N_i$ & $(n_i^1, m_i^1)$ & $(n_i^2, m_i^2)$ & $(n_i^3, m_i^3)$\\
\hline\hline
 $N_a=3$ & $(1/\b^1, 0)$  &
$(3,  \frac{1}{2}{\tilde \epsilon} \epsilon )$ & $\alpha 1_3$  \\
\hline
$N_b=2$  & $(n_b^1, -{\tilde \epsilon} \epsilon \b^1)$ & 
$(1, \frac{1}{2} {\tilde \epsilon}{\epsilon})$ &$ 1_2$ \\
\hline
$N_c=1$ & $(n_c^1,  \epsilon \b^1 )$ &   
$(0, {\epsilon} )$  & 
$1$ \\    
\hline
$N_d=1$ & $(1/\b^1, 0)$ &  $(-2,   \epsilon 
{\tilde \epsilon})$  
  & $\alpha^2$  \\\hline
$N_e = 1$ & $(1/ \b^1, 0)$ &  
$(-1, \frac{1}{2}\epsilon 
{\tilde \epsilon} )$  
  & $\alpha^2$ 
\\\hline
$N_h$ & $(\epsilon_h/ \b^1, 0)$ &  
$(2, 0 )$  
  & $1_{N_h}$ 
\\\hline
\end{tabular}
\end{center}
%\caption{\small General tadpole solutions for the five-stack 
%B4-type quiver of intersecting
%D5-branes.
%At low energy we get the SM.
% and an extra $U(1)$ symmetry
%under which the SM fields get charged.
%The solutions depend 
%on two integer parameters, 
%$n_b^1$, $n_c^1$, 
%the NS-background $\beta^1$ and
%the phase parameters $\epsilon = {\tilde \epsilon}=\pm 1$, as well as the 
%CP phase $\alpha$. 
%\label{spectrum4}}          
%\end{table}

%\begin{table}
%[htb]\footnotesize
%\renewcommand{\arraystretch}{2}
\begin{center}
\begin{tabular}{||c||c|c|c||}
\hline
\hline
\multicolumn{4}{c}{RR tadpole solutions for B5-quiver}\\\hline
\hline
\hline
$N_i$ & $(n_i^1, m_i^1)$ & $(n_i^2, m_i^2)$ & $(n_i^3, m_i^3)$\\
\hline\hline
 $N_a=3$ & $(1/\b^1, 0)$  &
$(3,  \frac{1}{2}{\tilde \epsilon} \epsilon )$ & $\alpha^2 1_3$  \\
\hline
$N_b=2$  & $(n_b^1, {\tilde \epsilon} \epsilon \b^1)$ & 
$(1, \frac{1}{2} {\tilde \epsilon}{\epsilon})$ &$ 1_2$ \\
\hline
$N_c=1$ & $(n_c^1,  \epsilon {\tilde \epsilon} \b^1 )$ &   
$(0, -{\epsilon}{\tilde \epsilon}  )$  & 
$1$ \\    
\hline
$N_d=1$ & $(1/\b^1, 0)$ &  $(-2,   \epsilon 
{\tilde \epsilon})$  
  & $\alpha$  \\\hline
$N_e = 1$ & $(1/ \b^1, 0)$ &  
$(-1, \frac{1}{2}\epsilon 
{\tilde \epsilon} )$  
  & $\alpha$ 
\\\hline
$N_h$ & $(\epsilon_h/ \b^1, 0)$ &  
$(2, 0 )$  
  & $1_{N_h}$ 
\\\hline
\end{tabular}
\end{center}
%\caption{\small General tadpole solutions for the five-stack 
%B5-type quiver of intersecting
%D5-branes.
%At low energy we get the SM.
%% and an extra $U(1)$ symmetry
%under which the SM fields get charged.
%The solutions depend 
%on two integer parameters, 
%$n_b^1$, $n_c^1$, 
%the NS-background $\beta^1$ and
%the phase parameters $\epsilon = {\tilde \epsilon}=\pm 1$, as well as the 
%CP phase $\alpha$. 
%\label{spectrum4b5}}          
%\end{table}

%\begin{table}
%[htb]\footnotesize
%\renewcommand{\arraystretch}{2}
\begin{center}
\begin{tabular}{||c||c|c|c||}
\hline
\hline
\multicolumn{4}{c}{RR tadpole solutions for B6-quiver}\\\hline
\hline
\hline
$N_i$ & $(n_i^1, m_i^1)$ & $(n_i^2, m_i^2)$ & $(n_i^3, m_i^3)$\\
\hline\hline
 $N_a=3$ & $(1/\b^1, 0)$  &
$(3,  \frac{1}{2}{\tilde \epsilon} \epsilon )$ & $\alpha^2 1_3$  \\
\hline
$N_b=2$  & $(n_b^1, {\tilde \epsilon} \epsilon \b^1)$ & 
$(1, \frac{1}{2} {\tilde \epsilon}{\epsilon})$ &$ 1_2$ \\
\hline
$N_c=1$ & $(n_c^1,  \epsilon {\tilde \epsilon} \b^1 )$ &   
$(0, -{\epsilon}{\tilde \epsilon}  )$  & 
$1$ \\    
\hline
$N_d=1$ & $(1/\b^1, 0)$ &  $(2,   -\epsilon 
{\tilde \epsilon})$  
  & $\alpha^2$  \\\hline
$N_e = 1$ & $(1/ \b^1, 0)$ &  
$(-1, \frac{1}{2}\epsilon 
{\tilde \epsilon} )$  
  & $\alpha$ 
\\\hline
$N_h$ & $(\epsilon_h/ \b^1, 0)$ &  
$(2, 0 )$  
  & $1_{N_h}$ 
\\\hline
\end{tabular}
\end{center}
%\caption{\small General tadpole solutions for the five-stack 
%B6-type quiver of intersecting
%D5-branes.
%At low energy we get the SM.
% and an extra $U(1)$ symmetry
%under which the SM fields get charged.
%The solutions depend 
%on two integer parameters, 
%$n_b^1$, $n_c^1$, 
%the NS-background $\beta^1$ and
%the phase parameters $\epsilon = {\tilde \epsilon}=\pm 1$, as well as the 
%CP phase $\alpha$. 
%\label{spectrum4b6}}          
%\end{table}

%\begin{table}
%[htb]\footnotesize
%\renewcommand{\arraystretch}{2}
\begin{center}
\begin{tabular}{||c||c|c|c||}
\hline
\hline
\multicolumn{4}{c}{RR tadpole solutions for B7-quiver}\\\hline
\hline
\hline
$N_i$ & $(n_i^1, m_i^1)$ & $(n_i^2, m_i^2)$ & $(n_i^3, m_i^3)$\\
\hline\hline
 $N_a=3$ & $(1/\b^1, 0)$  &
$(3,  \frac{1}{2}{\tilde \epsilon} \epsilon )$ & $\alpha^2 1_3$  \\
\hline
$N_b=2$  & $(n_b^1, -{\tilde \epsilon} \epsilon \b^1)$ & 
$(-1, -\frac{1}{2} {\tilde \epsilon}{\epsilon})$ &$ 1_2$ \\
\hline
$N_c=1$ & $(n_c^1,  -\epsilon {\tilde \epsilon} \b^1 )$ &   
$(0, {\epsilon}{\tilde \epsilon}  )$  & 
$1$ \\    
\hline
$N_d=1$ & $(1/\b^1, 0)$ &  $(-2,   \epsilon 
{\tilde \epsilon})$  
  & $\alpha$  \\\hline
$N_e = 1$ & $(1/ \b^1, 0)$ &  
$(1, -\frac{1}{2}\epsilon 
{\tilde \epsilon} )$  
  & $\alpha^2$ 
\\\hline
$N_h$ & $(\epsilon_h/ \b^1, 0)$ &  
$(2, 0 )$  
  & $1_{N_h}$ 
\\\hline
\end{tabular}
\end{center}
%\caption{\small General tadpole solutions for the five-stack 
%B7-type quiver of intersecting
%D5-branes.
%At low energy, we get an exotic structure involving the SM left 
%handed quarks and with the rest of SM fermion representations have exotic 
%hypercharges.
%The solutions depend 
%on two integer parameters, 
%$n_b^1$, $n_c^1$,
%the NS-background $\beta^1$ and
%the phase parameters $\epsilon = {\tilde \epsilon}=\pm 1$, as well as the 
%CP phase $\alpha$. 
%\label{spectrum4b7}}          
%\end{table}

%\newpage

\section{Appendix B}

%{RR tadpole solutions for six-stack quivers}

In this Appendix, we will provide the explicit RR tadpole solutions
corresponding to
the classification of embedding the six stack SM
configurations of
table (3), using
intersecting D5 models, in $Z_3$ quivers.
We list all tadpole solutions appearing in the C1,..., C10 quivers.

%\begin{table}[htb]\footnotesize
%\renewcommand{\arraystretch}{1}
%\begin{center}
%\begin{tabular}{||c||}\hline
%RR tadpole solutions for C1-quiver\hline
%\end{tabular}
\begin{center}
\begin{tabular}{||c||c|c|c||}\hline
\hline
\multicolumn{4}{c}{RR tadpole solutions for C1-quiver}\\\hline
$N_i$ & $(n_i^1, m_i^1)$ & $(n_i^2, m_i^2)$ &
$(n_i^3, m_i^3)$\\
\hline
\hline
 $N_a=3$ & $(n_a^1, -\epsilon \b^1)$  &
$(3,  \frac{1}{2}{\tilde \epsilon} \epsilon )$ & $1_3$  \\
\hline
$N_b=2$  & $(1/\b_1, 0)$ & $({\tilde \epsilon}, \frac{1}{2}{\epsilon})$ & 
$\alpha 1_2$ \\
\hline
$N_c=1$ & $(1/\b_1, 0)$ &   $(0, -{\epsilon})$  & 
$\alpha$ \\    
\hline
$N_d=1$ & $(n_d^1, - \epsilon \b^1)$ &  $(1,  - \frac{1}{2}\epsilon 
{\tilde \epsilon})$  
  & $1$  \\
\hline
$N_e = 1$ & $(n_e^1, -\epsilon \b^1)$ &  
$(1, - \frac{1}{2}\epsilon 
{\tilde \epsilon} )$  
  & $1$ 
\\
\hline
$N_f = 1$ & $(n_f^1, -\epsilon \b^1)$ &  
$(1, - \frac{1}{2}\epsilon 
{\tilde \epsilon} )$  
  & $1$ 
\\
\hline
$N_h$ & $(\epsilon_h/ \b^1, 0)$ &  
$(2, 0 )$  
  & $1_{N_h}$ 
\\
\hline
\end{tabular}
\end{center}
%\caption{\small General tadpole solutions for the six-stack 
%C1-type quiver 
%\label{asix1}}           
%\end{center}
%\end{table}

%%%%%%%%%%%%%%%%%%%%%%%%%%%%
%%%%%%%%%%%%%%%%%%%%
%%%%%%%%%%%%%

%\begin{table}[htb]\footnotesize
%\renewcommand{\arraystretch}{1}
\begin{center}
\begin{tabular}{||c||c|c|c||}
\hline\hline
\multicolumn{4}{c}{\small{RR tadpoles for 6-stack 
C2 quiver}}\\\hline
$N_i$ & $(n_i^1, m_i^1)$ & $(n_i^2, m_i^2)$ &
$(n_i^3, m_i^3)$\\
\hline\hline
 $N_a=3$ & $(n_a^1, -\epsilon {\tilde \epsilon} \b^1)$  &
$(3,  \frac{1}{2}{\tilde \epsilon} \epsilon )$ & $1_3$  \\
\hline
$N_b=2$  & $(1/\b_1, 0)$ & $(1, \frac{1}{2}{\epsilon}{\tilde \ep})$ & 
$\alpha 1_2$ \\
\hline
$N_c=1$ & $(1/\b_1, 0)$ &   $(0, {\epsilon}{\tilde \epsilon} )$  & 
$\alpha^2$ \\    
\hline
$N_d=1$ & $(n_d^1, - \epsilon {\tilde \epsilon} \b^1)$
&  $(1,  - \frac{1}{2}\epsilon{\tilde \epsilon})$  
  & $1$  \\
\hline
$N_e = 1$ & $(n_e^1, -\epsilon{\tilde \epsilon}  \b^1)$ &  
$(1, - \frac{1}{2}\epsilon 
{\tilde \epsilon} )$  
  & $1$ 
\\
\hline
$N_f = 1$ & $(n_f^1, -\epsilon {\tilde \epsilon} \b^1)$ &  
$(1, - \frac{1}{2}\epsilon 
{\tilde \epsilon} )$  
  & $1$ 
\\
\hline
$N_h$ & $(\epsilon_h/ \b^1, 0)$ &  
$(2, 0 )$  
  & $1_{N_h}$ 
\\
\hline
\end{tabular}
\end{center}
%\caption{\small General tadpole solutions for the six-stack 
%C2-type quiver 
%\label{asix2}}           
%\end{center}
%\end{table}

%%%%%%%%%%%%%%%%%%%%%%%%%%%%
%%%%%%%%%%%%%%%%%%
%%%%%%%%%%%%

%\begin{table}[htb]\footnotesize
%\renewcommand{\arraystretch}{1}
\begin{center}
\begin{tabular}{||c||c|c|c||}\hline
\hline
\multicolumn{4}{c}{\small{RR tadpoles for 6-stack 
C3 quiver}}\\\hline\hline
$N_i$ & $(n_i^1, m_i^1)$ & $(n_i^2, m_i^2)$ &
$(n_i^3, m_i^3)$\\
\hline
\hline
 $N_a=3$ & $(1/\b^1, 0)$  &
$(3,  \frac{1}{2}{\tilde \epsilon} \epsilon )$ & $\alpha 1_3$  \\
\hline
$N_b=2$  & $(n_b^1, -{\tilde \epsilon} \epsilon \b^1)$
& $(1, \frac{1}{2}{\epsilon}{\tilde \epsilon})$ &
$ 1_2$ \\
\hline
$N_c=1$ & $(n_c^1, -{\tilde \epsilon} \epsilon \b^1)$
&   $(0, -{\epsilon}{\tilde \epsilon} )$  &
$1$ \\    
\hline
$N_d=1$ & $(1/\b^1, 0)$
&  $(1,  - \frac{1}{2}\epsilon{\tilde \epsilon})$  
  & $\alpha$  \\
\hline
$N_e = 1$ & $(1/\b^1, 0)$ &  
$(1, - \frac{1}{2}\epsilon 
{\tilde \epsilon} )$  
  & $\alpha$
  \\
\hline
$N_f = 1$ & $(1/\b^1, 0)$ &  
$(1, - \frac{1}{2}\epsilon 
{\tilde \epsilon} )$  
  & $\alpha$ 
\\
\hline
$N_h$ & $(\epsilon_h/ \b^1, 0)$ &  
$(2, 0 )$  
  & $1_{N_h}$ 
\\
\hline
\end{tabular}
\end{center}
%\caption{\small General tadpole solutions for the six-stack 
%C3-type quiver 
%\label{asix3}}           
%\end{center}
%\end{table}

%%%%%%%%%%%%%%%%%%%%%%%%%%%%
%%%%%%%%%%%%%%%%%%%
%%%%%%%%%%%%%

%\begin{table}[htb]\footnotesize
%\renewcommand{\arraystretch}{1}
\begin{center}
\begin{tabular}{||c||c|c|c||}\hline\hline
\multicolumn{4}{c}{\small{RR tadpoles for 6-stack 
C4 quiver}}\\\hline
$N_i$ & $(n_i^1, m_i^1)$ & $(n_i^2, m_i^2)$ &
$(n_i^3, m_i^3)$\\
\hline
\hline
 $N_a=3$ & $(1/\b^1, 0)$  &
$(3,  \frac{1}{2}{\tilde \epsilon} \epsilon )$ & $\alpha 1_3$  \\
\hline
$N_b=2$  & $(n_b^1, -{\tilde \epsilon} \epsilon \b^1)$
& $(1, \frac{1}{2}{\epsilon}{\tilde \epsilon})$ &
$ 1_2$ \\
\hline
$N_c=1$ & $(n_c^1, -{\tilde \epsilon} \epsilon \b^1)$
&   $(0, -{\epsilon}{\tilde \epsilon} )$  &
$1$ \\    
\hline
$N_d=1$ & $(1/\b^1, 0)$
&  $(-1,   \frac{1}{2}\epsilon{\tilde \epsilon})$  
  & $\alpha^2$  \\
\hline
$N_e = 1$ & $(1/\b^1, 0)$ &  
$(-1,  \frac{1}{2}\epsilon 
{\tilde \epsilon} )$  
  & $\alpha^2$ 
\\
\hline
$N_f = 1$ & $(1/\b^1, 0)$ &  
$(-1,  \frac{1}{2}\epsilon 
{\tilde \epsilon} )$  
  & $\alpha^2$ 
\\
\hline
$N_h$ & $(\epsilon_h/ \b^1, 0)$ &  
$(2, 0 )$  
  & $1_{N_h}$ 
\\
\hline
\end{tabular}
\end{center}
%\caption{\small General tadpole solutions for the six-stack 
%C4-type quiver 
%\label{asix4}}           
%\end{center}
%\end{table}

%\begin{table}[htb]\footnotesize
%\renewcommand{\arraystretch}{1}
\begin{center}
\begin{tabular}{||c||c|c|c||}\hline\hline
\multicolumn{4}{c}{\small{RR tadpoles for 6-stack 
C5 quiver}}\\\hline
\hline
\hline
$N_i$ & $(n_i^1, m_i^1)$ & $(n_i^2, m_i^2)$ &
$(n_i^3, m_i^3)$\\\hline\hline
 $N_a=3$ & $(1/\b^1, 0)$  &
$(3,  \frac{1}{2}{\tilde \epsilon} \epsilon )$ & $\alpha^2 1_3$  \\
\hline
$N_b=2$  & $(n_b^1, -{\tilde \epsilon} \epsilon \b^1)$
& $(-1, -\frac{1}{2}{\epsilon}{\tilde \epsilon})$ &
$ 1_2$ \\
\hline
$N_c=1$ & $(n_c^1, -{\tilde \epsilon} \epsilon \b^1)$
&   $(0, {\epsilon}{\tilde \epsilon} )$  &
$1$ \\    
\hline
$N_d=1$ & $(1/\b^1, 0)$
&  $(-1,   \frac{1}{2}\epsilon{\tilde \epsilon})$  
  & $\alpha$  \\\hline
$N_e = 1$ & $(1/\b^1, 0)$ &  
$(1, - \frac{1}{2}\epsilon 
{\tilde \epsilon} )$  
  & $\alpha^2$ 
\\\hline
$N_f = 1$ & $(1/\b^1, 0)$ &  
$(1, - \frac{1}{2}\epsilon 
{\tilde \epsilon} )$  
  & $\alpha^2$ 
\\\hline
$N_h$ & $(\epsilon_h/ \b^1, 0)$ &  
$(2, 0 )$  
  & $1_{N_h}$ 
\\\hline
\end{tabular}
\end{center}
%\caption{\small General tadpole solutions for the six-stack 
%C5-type quiver of intersecting
%D5-branes, giving rise to the
%SM at low energies.
%The SM  chiral fields get charged under the $U(1)$'s.
%The solutions depend 
%on two integer parameters, 
%$n_b^1$, $n_c^1$, 
%the NS-background $\beta^1$ and
%the phase parameters $\epsilon = {\tilde \epsilon}=\pm 1$,
%as well as the CP phase $\alpha$. 
%\label{six201}}          
%\end{table}

%\begin{table}
%[htb]\footnotesize
%\renewcommand{\arraystretch}{1}
\begin{center}
\begin{tabular}{||c||c|c|c||}\hline\hline
\multicolumn{4}{c}{\small{RR tadpoles for 6-stack 
C6 quiver}}\\\hline
\hline
\hline
$N_i$ & $(n_i^1, m_i^1)$ & $(n_i^2, m_i^2)$ &
$(n_i^3, m_i^3)$\\\hline\hline
 $N_a=3$ & $(1/\b^1, 0)$  &
$(3,  \frac{1}{2}{\tilde \epsilon} \epsilon )$ & $\alpha^2 1_3$  \\
\hline
$N_b=2$  & $(n_b^1, -{\tilde \epsilon} \epsilon \b^1)$
& $(-1, -\frac{1}{2}{\epsilon}{\tilde \epsilon})$ &
$ 1_2$ \\
\hline
$N_c=1$ & $(n_c^1, -{\tilde \epsilon} \epsilon \b^1)$
&   $(0, {\epsilon}{\tilde \epsilon} )$  &
$1$ \\    
\hline
$N_d=1$ & $(1/\b^1, 0)$
&  $(1,   -\frac{1}{2}\epsilon{\tilde \epsilon})$  
  & $\alpha^2$  \\\hline
$N_e = 1$ & $(1/\b^1, 0)$ &  
$(-1,  \frac{1}{2}\epsilon 
{\tilde \epsilon} )$  
  & $\alpha$ 
\\\hline
$N_f = 1$ & $(1/\b^1, 0)$ &  
$(-1,  \frac{1}{2}\epsilon 
{\tilde \epsilon} )$  
  & $\alpha$ 
\\\hline
$N_h$ & $(\epsilon_h/ \b^1, 0)$ &  
$(2, 0 )$  
  & $1_{N_h}$ 
\\\hline
\end{tabular}
\end{center}
%\caption{\small General tadpole solutions for the six-stack 
%C6-type quiver of intersecting
%D5-branes, giving rise to an exotic structure
%with the left handed quarks
%of the 
%SM, and with the rest of SM fermions having exotic
%hypercharges. The solutions depend 
%on two integer parameters, 
%$n_b^1$, $n_c^1$, 
%the NS-background $\beta^1$ and
%the phase parameters $\epsilon = {\tilde \epsilon}=\pm 1$,
%as well as the CP phase $\alpha$. 
%\label{six301}}          
%\end{table}

%\begin{table}[htb]\footnotesize
%\renewcommand{\arraystretch}{1}
\begin{center}
\begin{tabular}{||c||c|c|c||}\hline\hline
\multicolumn{4}{c}{\small{RR tadpoles for 6-stack 
C7 quiver}}\\\hline
\hline
\hline
$N_i$ & $(n_i^1, m_i^1)$ & $(n_i^2, m_i^2)$ &
$(n_i^3, m_i^3)$\\\hline\hline
 $N_a=3$ & $(1/\b^1, 0)$  &
$(3,  \frac{1}{2}{\tilde \epsilon} \epsilon )$ & $\alpha^2 1_3$  \\
\hline
$N_b=2$  & $(n_b^1, -{\tilde \epsilon} \epsilon \b^1)$
& $(-1, -\frac{1}{2}{\epsilon}{\tilde \epsilon})$ &
$ 1_2$ \\
\hline
$N_c=1$ & $(n_c^1, -{\tilde \epsilon} \epsilon \b^1)$
&   $(0, {\epsilon}{\tilde \epsilon} )$  &
$1$ \\    
\hline
$N_d=1$ & $(1/\b^1, 0)$
&  $(1,   -\frac{1}{2}\epsilon{\tilde \epsilon})$  
  & $\alpha^2$  \\\hline
$N_e = 1$ & $(1/\b^1, 0)$ &  
$(1, - \frac{1}{2}\epsilon 
{\tilde \epsilon} )$  
  & $\alpha^2$ 
\\\hline
$N_f = 1$ & $(1/\b^1, 0)$ &  
$(-1, \frac{1}{2}\epsilon 
{\tilde \epsilon} )$  
  & $\alpha$ 
\\\hline
$N_h$ & $(\epsilon_h/ \b^1, 0)$ &  
$(2, 0 )$  
  & $1_{N_h}$ 
\\\hline
\end{tabular}
\end{center}
%\caption{\small General tadpole solutions for the six-stack 
%C7-type quiver of intersecting
%D5-branes, giving rise to the
%SM at low energies. The solutions depend 
%on two integer parameters, 
%$n_b^1$, $n_c^1$, 
%the NS-background $\beta^1$ and
%the phase parameters $\epsilon = {\tilde \epsilon}=\pm 1$,
%as well as the CP phase $\alpha$. 
%\label{six401}}          
%\end{table}

%\begin{table}[htb]\footnotesize
%\renewcommand{\arraystretch}{1}
\begin{center}
\begin{tabular}{||c||c|c|c||}\hline\hline
\multicolumn{4}{c}{\small{RR tadpoles for 6-stack 
C8 quiver}}\\\hline
\hline
\hline
$N_i$ & $(n_i^1, m_i^1)$ & $(n_i^2, m_i^2)$ &
$(n_i^3, m_i^3)$\\\hline\hline
 $N_a=3$ & $(1/\b^1, 0)$  &
$(3,  \frac{1}{2}{\tilde \epsilon} \epsilon )$ & $\alpha 1_3$  \\
\hline
$N_b=2$  & $(n_b^1, -{\tilde \epsilon} \epsilon \b^1)$
& $(1, \frac{1}{2}{\epsilon}{\tilde \epsilon})$ &
$ 1_2$ \\
\hline
$N_c=1$ & $(n_c^1, -{\tilde \epsilon} \epsilon \b^1)$
&   $(0, -{\epsilon}{\tilde \epsilon} )$  &
$1$ \\    
\hline
$N_d=1$ & $(1/\b^1, 0)$
&  $(1,   -\frac{1}{2}\epsilon{\tilde \epsilon})$  
  & $\alpha$  \\\hline
$N_e = 1$ & $(1/\b^1, 0)$ &  
$(-1,  \frac{1}{2}\epsilon 
{\tilde \epsilon} )$  
  & $\alpha^2$ 
\\\hline
$N_f = 1$ & $(1/\b^1, 0)$ &  
$(-1,  \frac{1}{2}\epsilon 
{\tilde \epsilon} )$  
  & $\alpha^2$ 
\\\hline
$N_h$ & $(\epsilon_h/ \b^1, 0)$ &  
$(2, 0 )$  
  & $1_{N_h}$ 
\\\hline
\end{tabular}
\end{center}
%\caption{\small General tadpole solutions for the six-stack 
%C8-type quiver of intersecting
%D5-branes, giving rise, at 
%low energy, to an exotic structure involving the SM left 
%handed quarks and with the rest of SM fermions having exotic 
%hypercharges. The solutions depend 
%on two integer parameters, 
%$n_b^1$, $n_c^1$, 
%the NS-background $\beta^1$ and
%the phase parameters $\epsilon = {\tilde \epsilon}=\pm 1$,
%as well as the CP phase $\alpha$. 
%\label{six801}}          
%\end{table}

%\begin{table}[htb]\footnotesize
%\renewcommand{\arraystretch}{1}
\begin{center}
\begin{tabular}{||c||c|c|c||}\hline\hline
\multicolumn{4}{c}{\small{RR tadpoles for 6-stack 
C9 quiver}}\\\hline
\hline
\hline
$N_i$ & $(n_i^1, m_i^1)$ & $(n_i^2, m_i^2)$ &
$(n_i^3, m_i^3)$\\\hline\hline
 $N_a=3$ & $(1/\b^1, 0)$  &
$(3,  \frac{1}{2}{\tilde \epsilon} \epsilon )$ & $\alpha 1_3$  \\
\hline
$N_b=2$  & $(n_b^1, -{\tilde \epsilon} \epsilon \b^1)$
& $(1, \frac{1}{2}{\epsilon}{\tilde \epsilon})$ &
$ 1_2$ \\
\hline
$N_c=1$ & $(n_c^1, -{\tilde \epsilon} \epsilon \b^1)$
&   $(0, -{\epsilon}{\tilde \epsilon} )$  &
$1$ \\    
\hline
$N_d=1$ & $(1/\b^1, 0)$
&  $(-1,   \frac{1}{2}\epsilon{\tilde \epsilon})$  
  & $\alpha^2$  \\\hline
$N_e = 1$ & $(1/\b^1, 0)$ &  
$(1, - \frac{1}{2}\epsilon 
{\tilde \epsilon} )$  
  & $\alpha$ 
\\\hline
$N_f = 1$ & $(1/\b^1, 0)$ &  
$(-1, \frac{1}{2}\epsilon 
{\tilde \epsilon} )$  
  & $\alpha^2$ 
\\\hline
$N_h$ & $(\epsilon_h/ \b^1, 0)$ &  
$(2, 0 )$  
  & $1_{N_h}$ 
\\\hline
\end{tabular}
\end{center}
%\caption{\small General tadpole solutions for the six-stack 
%C9-type quiver of intersecting
%D5-branes, 
%giving rise at
%low energy to an exotic structure involving the SM left 
%handed quarks and with the rest of SM fermions having exotic 
%hypercharges. The solutions depend 
%on two integer parameters, 
%$n_b^1$, $n_c^1$, 
%the NS-background $\beta^1$ and
%the phase parameters $\epsilon = {\tilde \epsilon}=\pm 1$,
%as well as the CP phase $\alpha$. 
%\label{six901}}          
%\end{table}

%\begin{table}
%[htb]\footnotesize
%\renewcommand{\arraystretch}{1}
\begin{center}
\begin{tabular}{||c||c|c|c||}\hline\hline
\multicolumn{4}{c}{\small{RR tadpoles for 6-stack 
C10 quiver}}\\\hline
\hline
\hline
$N_i$ & $(n_i^1, m_i^1)$ & $(n_i^2, m_i^2)$ &
$(n_i^3, m_i^3)$\\\hline\hline
 $N_a=3$ & $(1/\b^1, 0)$  &
$(3,  \frac{1}{2}{\tilde \epsilon} \epsilon )$ & $\alpha 1_3$  \\
\hline
$N_b=2$  & $(n_b^1, -{\tilde \epsilon} \epsilon \b^1)$
& $(1, \frac{1}{2}{\epsilon}{\tilde \epsilon})$ &
$ 1_2$ \\
\hline
$N_c=1$ & $(n_c^1, -{\tilde \epsilon} \epsilon \b^1)$
&   $(0, -{\epsilon}{\tilde \epsilon} )$  &
$1$ \\    
\hline
$N_d=1$ & $(1/\b^1, 0)$
&  $(-1,   \frac{1}{2}\epsilon{\tilde \epsilon})$  
  & $\alpha^2$  \\\hline
$N_e = 1$ & $(1/\b^1, 0)$ &  
$(-1,  \frac{1}{2}\epsilon 
{\tilde \epsilon} )$  
  & $\alpha^2$ 
\\\hline
$N_f = 1$ & $(1/\b^1, 0)$ &  
$(1, - \frac{1}{2}\epsilon 
{\tilde \epsilon} )$  
  & $\alpha$ 
\\\hline
$N_h$ & $(\epsilon_h/ \b^1, 0)$ &  
$(2, 0 )$  
  & $1_{N_h}$ 
\\\hline
\end{tabular}
\end{center}
%\caption{\small General tadpole solutions for the six-stack 
%C10-type quiver of intersecting
%D5-branes, giving rise  
%at
%low energy to an exotic structure involving the SM left 
%handed quarks and with the rest of SM fermions having exotic 
%hypercharges. The solutions depend 
%on two integer parameters, 
%$n_b^1$, $n_c^1$, 
%the NS-background $\beta^1$ and
%the phase parameters $\epsilon = {\tilde \epsilon}=\pm 1$,
%as well as the CP phase $\alpha$. 
%\label{six1001}}          
%\end{table}

%\newpage

\end{document}